\documentclass[fleqn,3p,onecolumn,11pt,nofootinbib,showkeys,showpacs,justification=justified]{revtex4-1}

\newcommand{\versionnumber}{v1.00}

\renewcommand{\today}{11.02.2014 (\versionnumber)}

\usepackage[small,justification=centerlast]{caption}
\usepackage{graphicx}
\usepackage{bm}
\usepackage{mathrsfs}
\usepackage[latin1]{inputenc}
\usepackage{stmaryrd}
\usepackage{array}
\usepackage{psfrag}
\usepackage{dsfont}
\usepackage{epsfig}
\usepackage{titletoc}
\usepackage{float}   
\usepackage{wrapfig} 
\usepackage{amsmath}\usepackage{amssymb,stmaryrd}
\usepackage{cancel}
\usepackage[dvips]{feynmp}
\usepackage{upgreek}
\usepackage{subfig}
\usepackage{color}
\usepackage{units}
\usepackage{tabularx}
\usepackage{multirow}

\usepackage{textfit} 

\usepackage{hyperref}

\renewcommand{\eqref}[1]{\mbox{Eq.~(\ref{#1})}}
\newcommand{\tabref}[1]{\mbox{Tab.~\ref{#1}}}
\newcommand{\figref}[1]{\mbox{Fig.~\ref{#1}}}
\newcommand{\secref}[1]{\mbox{Sec.~\ref{#1}}}

\newcommand{\appref}[1]{\mbox{App.~\ref{#1}}}

\newcommand{\Figref}[1]{\mbox{Figure \ref{#1}}}

\newcolumntype{C}[1]{>{\centering\arraybackslash}m{#1}}

\begin{document}

\title{Analytical bunch compression studies for FLUTE}

\author{M. Schreck} \email{marco.schreck@kit.edu} \thanks{Corresponding Author}
\affiliation{Laboratory for Applications of Synchrotron Radiation \\ Karlsruhe Institute of Technology --- Campus South \\
76131 Karlsruhe, Germany}
\author{P. Wesolowski} \email{pawel.wesolowski@kit.edu}
\affiliation{ANKA Synchrotron Radiation Facility \\ Karlsruhe Institute of Technology --- Campus North \\
76344 Eggenstein-Leopoldshafen, Germany}

\begin{abstract}
The current article deals with analytical bunch compression studies for FLUTE whose results are compared to simulations. FLUTE is a linac-based
electron accelerator with a design energy of approximately \unit[40]{MeV} currently being constructed at the Karlsruhe Institute of
Technology. One of the goals of FLUTE is to generate electron bunches with their length lying in the femtosecond regime. In the first phase
this will be accomplished using a magnetic bunch compressor. This compressor forms the subject of the studies presented.
The paper is divided into three parts. The first part deals with pure geometric investigations of the bunch compressor where space charge
effects and the back reaction of bunches with coherent synchrotron radiation (CSR) are neglected. The second part is dedicated to the
treatment of space charge effects and the third part gives some analytical results on the emission of CSR.
The upshot is that the results of the first and the third part agree quite well with what is obtained from simulations. However, the
space charge forces in the analytical model of the second part have the trend to be overestimated for large bunch charges.
With this paper we intend to form the basis for future analytical studies of the FLUTE bunch compressor and of bunch compression,
in general.
\end{abstract}

\hfill\today

\maketitle

\newpage
\setcounter{equation}{0}
\setcounter{section}{0}
\renewcommand{\theequation}{\arabic{section}.\arabic{equation}}

\section{Introduction}

\begin{figure}[t]
\centering
\includegraphics[scale=1]{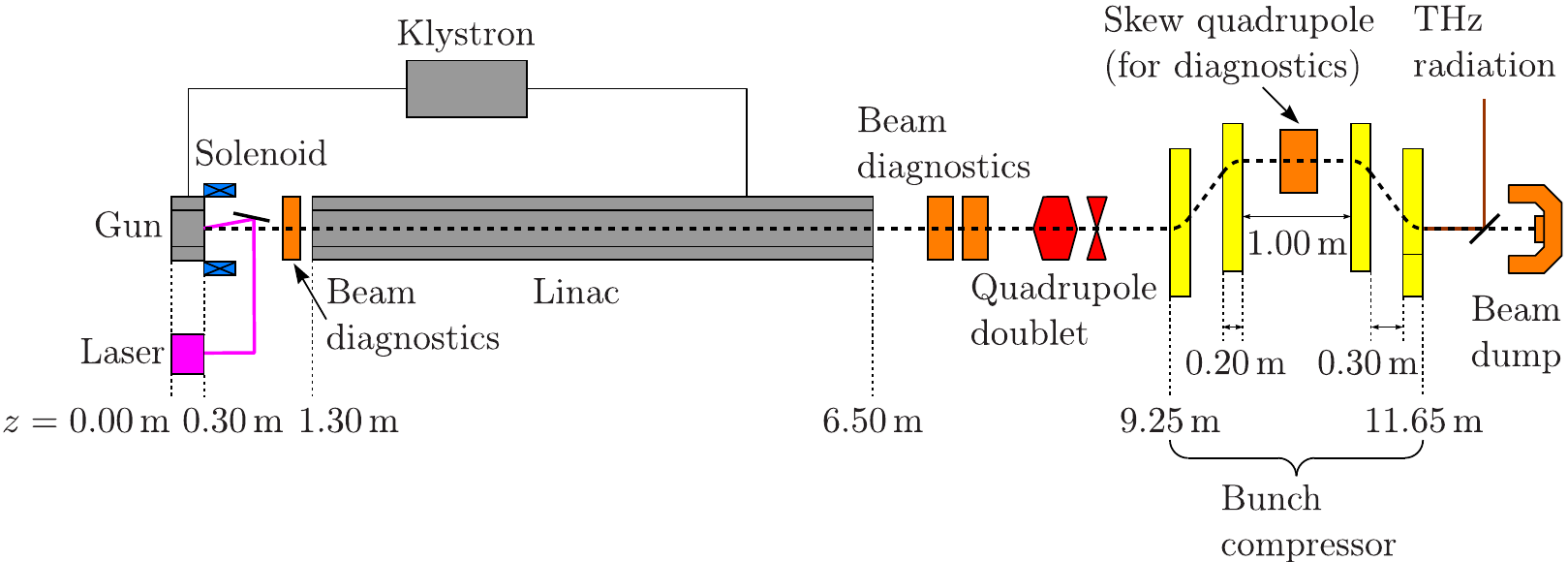}
\caption{Baseline layout of FLUTE in the first phase, where the position of the various parts of the machine are shown on the
$z$-axis. The dashed line is the trajectory of an electron bunch. Such bunches are produced in a photocathode gun
and accelerated by the linac to the design energy of  \unit[40]{MeV}. The rf of \unit[3]{GHz} for the
gun and the linac is delivered by a klystron. Solenoids and quadrupole magnets are used to focus the beam in the transverse
directions. We plan to place diagnostics at certain positions along the machine to extract information on the
transverse and longitudinal beam dimensions. Electron bunches are supposed to be compressed by a bunch compressor consisting
of four rectangular dipole magnets. After compressing, the bunches produce coherent THz radiation that is coupled out before
the electrons hit the beam dump.}
\label{fig:baseline-machine-layout}
\end{figure}

FLUTE is a linac-based electron accelerator which is presently being built at the ANKA Synchrotron Radiation Facility at
the Karlsruhe Institute of Technology. The acronym FLUTE stands for the German expression \textit{Ferninfrarot}
\textit{Linac- Und Test-Experiment} translated to English as ``Far-infrared Linac- and Test Experiment.'' FLUTE has a
design energy of approximately \unit[40]{MeV} where the baseline machine layout of the first phase is depicted in
\figref{fig:baseline-machine-layout}.

In the current design the electron source is a 2 1/2 cell photocathode radiofrequency (rf) gun with a maximum repetition rate of
\unit[10]{Hz}. Electrons are released by shooting a pulsed Ti:Sa laser with a fundamental wavelength of \unit[800]{nm} on a copper
cathode where its third frequency harmonic will be used. The released electrons are then accelerated to \unit[7]{MeV}. The charge
of the bunches produced by the gun is planned to range from \unit[1]{pC} to \unit[3]{nC}. Upon leaving the gun the beam is
transversally focused by a solenoid before entering the linac accelerating the electrons to the design energy of approximately
\unit[40]{MeV}. Behind the linac the beam is focused again by a doublet of quadrupole magnets before it enters the bunch
compressor consisting of four dipole magnets.

One goal of FLUTE is to produce coherent synchrotron radiation (CSR) in the terahertz (THz) range. To achieve this, sub-picosecond
bunch lengths will be necessary where the aim is to compress bunches to lengths in the femtosecond regime. For the past few
years there has been a growing interest in coherent THz sources due to the various possibilities of using this kind of radiation
both in research and in application. The following list does not claim to be complete but will give some representative
examples.

\begin{itemize}

\item In Ref.~\cite{Mikhailov:2007} it was shown theoretically that by applying an external oscillating electric field to a
sample of graphene, it is possible to produce higher harmonic modes. At room temperature this effect may occur for frequencies
in the THz regime. Therefore it could open the way to graphene devices in THz electronics.
\item In a cuprate superconductor a special kind of soliton was excited successfully by using intense and narrow-band THz
radiation \cite{Dienst:2013}. If the generation, acceleration, and stopping of such solitons is under control, these could
be exploited for transporting and storing information in such composites.
\item The chemical composition $\mathrm{BaTiO_3}$ is ferroelectric, i.e., below some critical temperature it exhibits domains
with a spontaneous electric dipole moment. These domains are separated by domain walls that can be manipulated by applying
a strong, external electric field. In Ref.~\cite{Shin:2007} the physical mechanisms occurring at microscopic scales are
investigated and the results are compared with experimental data. If the microscopic mechanisms of moving domain walls are
better understood, such ferroelectric materials could be the basis for ultra-fast computer memories.
\item By experiment it was shown that the magnetization direction of thin cobalt films can be reversed by short THz pulses, if
the magnetization vector lies in the plane of the film \cite{Back:2013}. Some (but not all) characteristics of the experimental
results can be described by a simple model based on the Landau--Lifshitz equation. A better understanding of the physics and
a further development of this method could lead to novel devices used for magnetic recording at high data rates.

\end{itemize}

The applications above have two characteristics in common: they need high electric and magnetic field strengths (in the order
of magnitude of MV/m and several hundred kA/m, respectively) and they happen on ultra-short time scales (picoseconds). These
properties can be provided by pulses of coherent synchrotron radiation in the THz regime (see, e.g., \cite{Huttel:2012zzb}).

In the first phase of FLUTE the compression of the electron bunches shall be achieved with a magnetic bunch compressor.
This compressor is a D-shape chicane consisting of four dipole magnets with each of them having the same magnetic field strength
value. The directions of the field in the first and fourth dipole magnet are opposite to the directions in the second
and third magnet. The distances between the first two and the last two magnets are supposed to be equal.

Since the electrons travel on curved trajectories inside this chicane they emit synchrotron radiation. If the bunch length
is much smaller than the wavelength of the radiation, wave trains emitted from different electrons are in phase with respect
to each other and they can interfere constructively. The radiation produced is then coherent and its intensity grows with
the number of radiating electrons squared. Hence, the FLUTE chicane serves the purpose of compressing the bunches and is the
place where the coherent radiation will be generated.

Due to space charge effects and the self-interaction of bunches with their own coherent radiation field a compression of bunches
to a length of several femtoseconds is a challenging task. That is why a better understanding of the chicane is of paramount importance.
Therefore, the scope of the current article is to provide a framework for analytical bunch compression studies for FLUTE. The
analytical results will also be compared to results obtained with the simulation tool Astra \cite{Floettmann:2013}.

\section{Bunch compression by path length differences}
\label{sec:bunch-compression-trajectory}
\setcounter{equation}{0}

\begin{figure}[b]
\centering
\includegraphics[scale=1]{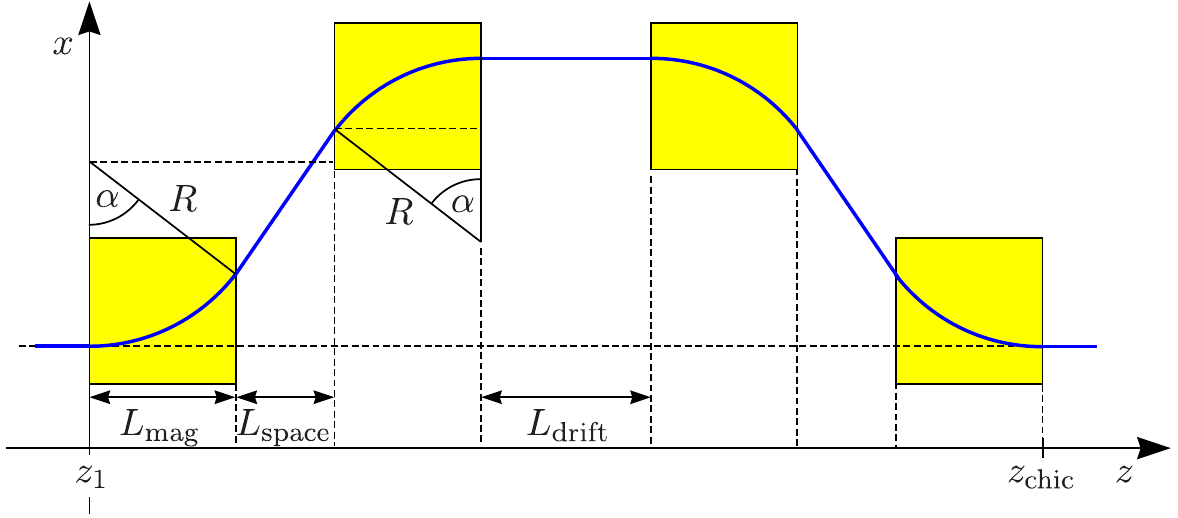}
\caption{Draft of the D-shape chicane that shall be constructed for FLUTE. A cartesian coordinate system is used where its labels
$x$ and $y$ (orthogonal to the drawing plane) correspond to the two transverse directions and the label $z$ corresponds to the
longitudinal direction. The chicane is assumed to lie in the $x$-$z$-place and the $z$-axis points along the direction of the
electron beam right before the chicane. The plain (blue) curve depicts one possible electron trajectory. The length of a single
chicane magnet is denoted as $L_{\mathrm{mag}}$. The distance between the first two and the last two magnets is called
$L_{\mathrm{space}}$, whereas the distance between the second and the third magnet is denoted as $L_{\mathrm{drift}}$. The angle
$\alpha$ is the bending angle of each magnet and $R$ is the bending radius.}
\label{fig:chicane-principle}
\end{figure}

In the current section analytical results on bunch compression in the FLUTE chicane are obtained, where a draft of the latter
is shown in \figref{fig:chicane-principle}. To make this approach feasible, first of all the D-shape chicane is
considered to consist of ideal dipole magnets. These are assumed to have a homogeneous magnetic field with flux density
$B$ inside the poles which immediately drops to zero outside. In the first and fourth magnet the field is to point along the
negative $y$-axis, whereas in the second and third magnet it points along the positive $y$-axis.

The bending radius in a chicane magnet is given by $R=p/(eB)$, where $p=\gamma(v) m_{\mathrm{e}}v$ is the relativistic electron
momentum with the Lorentz factor
\begin{equation}
\gamma(v)=\frac{1}{\sqrt{1-\beta^2}}\,,\quad \beta=\frac{v}{c}\,.
\end{equation}
Here $m_{\mathrm{e}}$ is the electron rest mass, $v$ the electron propagation velocity, and $c$ the speed of light. An electron
has the charge $q=-e$ with the elementary charge $e>0$. The bending angle can be computed as $\alpha=\arcsin(L_{\mathrm{mag}}/R)$.

First of all, space charge effects and the back reaction of the bunch with its CSR will be neglected. As a result, all considerations
of the current chapter are of geometrical nature.
The reduction of the bunch length within the chicane then essentially results from the path length difference of electrons with
different momenta. The length of the trajectory of an electron traveling with momentum $p$ is given by:
\begin{equation}
\label{eq:path-length-through-chicane}
L(p)=4R\arcsin\left(\frac{L_{\mathrm{mag}}}{R}\right)+\frac{2L_{\mathrm{space}}}{\sqrt{1-(L_{\mathrm{mag}}/R)^2}}
+L_{\mathrm{drift}}\,,\quad R=\frac{p}{eB}\,.
\end{equation}
Now the difference between the traveling lengths of two electrons is considered. The first electron is assumed to travel
with the design (reference) momentum $p$ and the second electron with a momentum that deviates from $p$ by $\Delta p$. For
$\Delta p\ll p$ a Taylor expansion can be performed with respect to the dimensionless normalized momentum deviation
$\delta\equiv \Delta p/p\ll 1$.
Due to the limited extension of the beam pipe, the bending angle $\alpha$ must be much smaller than $\pi/2$. This translates
to the necessary condition that $L_{\mathrm{mag}}\ll R$. Hence, it makes sense to perform a second expansion with
respect to the small ratio $L_{\mathrm{mag}}/R$. That leads to a transparent result for the path length difference:
\begin{equation}
\label{eq:path-length-difference}
\Delta L\equiv L(p+\Delta p)-L(p)=-2\left(\frac{L_{\mathrm{mag}}}{R}\right)^2\left[\frac{2}{3}L_{\mathrm{mag}}+L_{\mathrm{space}}\right]\delta
+\mathcal{O}\left[\delta^2,(L_{\mathrm{mag}}/R)^4\right]\,.
\end{equation}
It is evident that $\Delta L<0$ for $\delta>0$. This is clear since the bending angle of an electron with a larger momentum
is smaller resulting in a shorter path length traveled by the respective particle.

We decided to perform the following calculations throughout the paper for the two extreme cases that were simulated with Astra:
a bunch with the high charge of \unit[3]{nC} and a bunch with the very low charge of \unit[1]{pC}.

\subsection{Electron trajectory inside the chicane}
\label{ssec:electron-trajectory}

\begin{figure}[t!]
\centering
\subfloat[{$Q_b=\unit[3]{nC}$, $\sigma_s=\unit[2.30]{ps}$, $\sigma_p=1.87\cdot 10^{-2}$}]{\label{fig:bunch-before-chicane-3nc}
\includegraphics[scale=0.5]{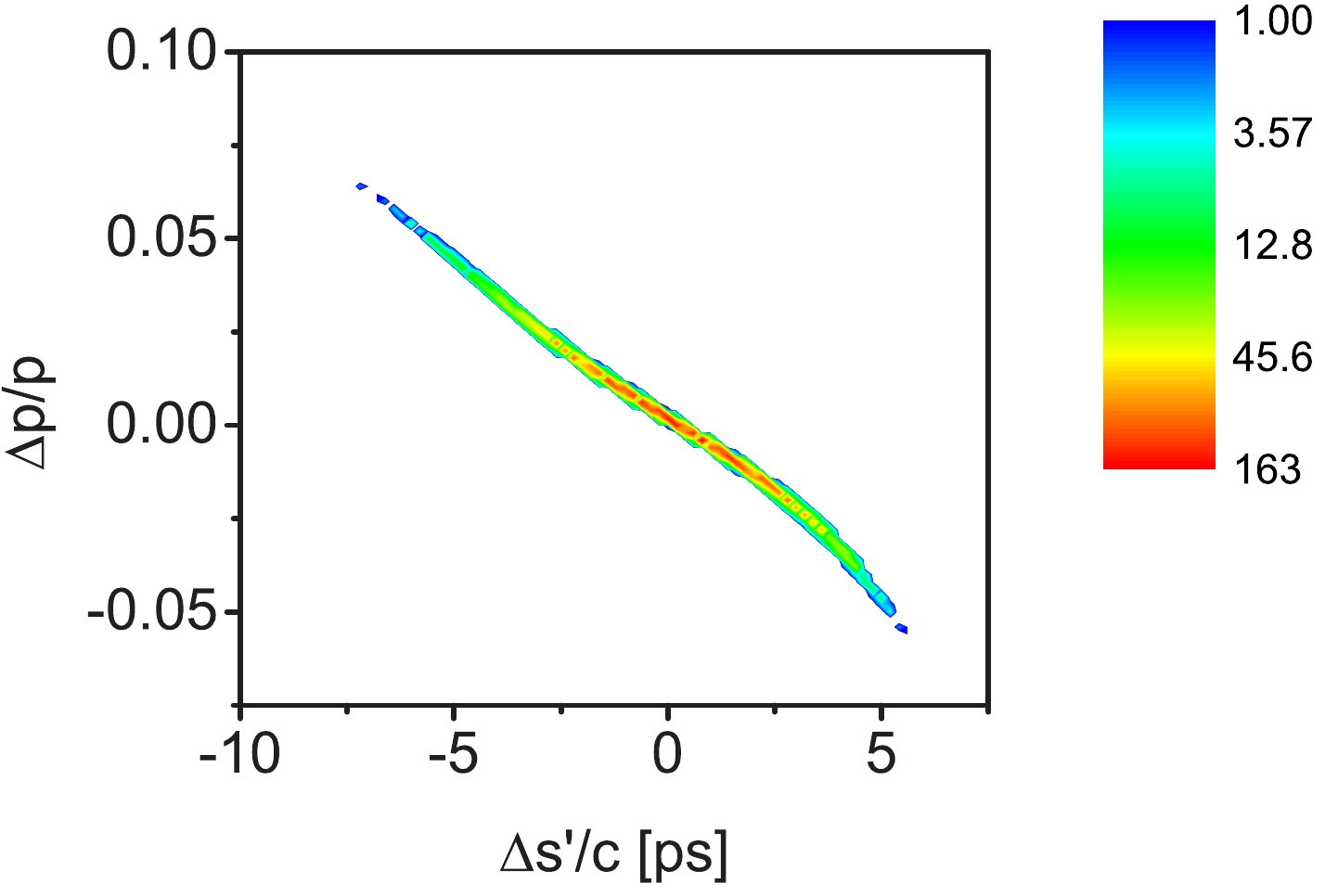}}\hspace{1.0cm}
\subfloat[{$Q_b=\unit[1]{pC}$, $\sigma_s=\unit[452]{fs}$, $\sigma_p=4.79\cdot 10^{-3}$}]{\label{fig:bunch-before-chicane-1pc}
\includegraphics[scale=0.5]{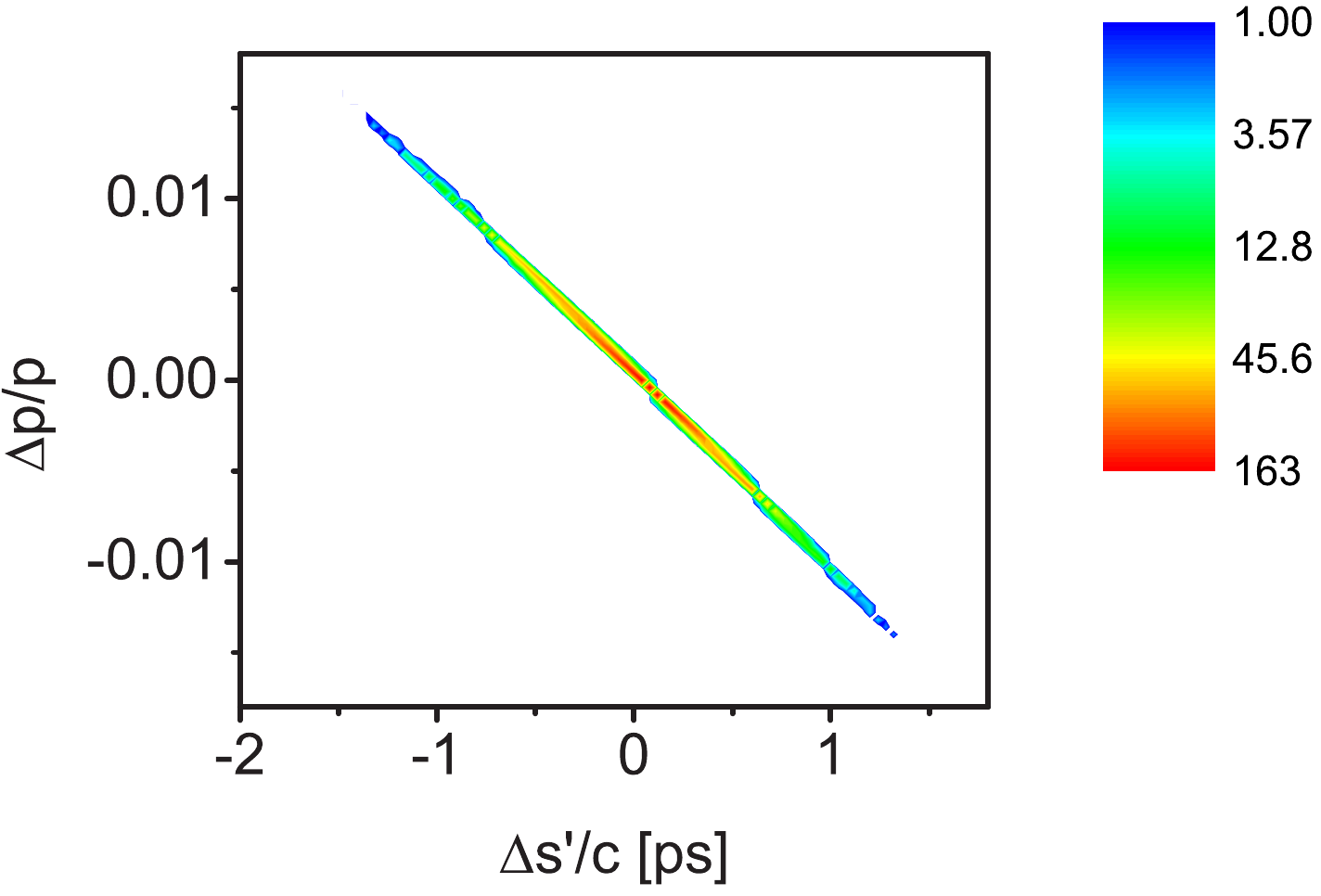}}
\caption{Longitudinal phase space plot of simulated \unit[3]{nC} and \unit[1]{pC} bunches at the position $z=\unit[8.19]{m}$ before the
chicane. Here the
normalized momentum spread is plotted against the distance $\Delta s'$ of a bunch particle with respect to the bunch center corresponding
to the mean of all distances. The spatial bunch coordinates are divided by the speed of light to convert them to the dimension
of time. Both distributions are centered on the mean relative momentum spread at the vertical axis as well. (This procedure is conducted
for all such distributions.) The rainbow color code represents the number of particles ranging from one (blue) to the maximum (red). The
substructures for the \unit[3]{nC} bunch, i.e., the two small superimposed bumps originate from the emission of the particles at the
cathode.}
\label{fig:bunch-before-chicane}
\end{figure}

The longitudinal phase space distribution of electron bunches produced at FLUTE, i.e., their longitudinal momentum spread $\Delta p/p$
as a function of the longitudinal particle position $\Delta s$ respective the reference particle has certain characteristics
directly after the linac. These are paramount for compression. In addition to a momentum spread based on statistical uncertainties,
the longitudinal phase space shows a correlated momentum spread (chirp). This means that the average momentum spread as a function
of $\Delta s$ is not zero but depends on $\Delta s$ (see \figref{fig:bunch-before-chicane-3nc} for a typical simulated \unit[3]{nC}
bunch and \figref{fig:bunch-before-chicane-1pc} for a \unit[1]{pC} bunch before the chicane).
In this paper the bunch length $\sigma_s$ is computed as the root mean square (rms) of the $\Delta s$-values where $(\Delta s)_n$
is related to the $n$-th particle:
\begin{equation}
\sigma_s\equiv \sqrt{\frac{1}{N} \sum_{n=1}^N \big[(\Delta s)_n-\overline{\Delta s}\,\big]^2}\,.
\end{equation}
The sum runs over all particles, i.e., $N$ is the number of particles in the bunch. The average particle position is denoted by $\overline{\Delta s}$.
From this formula directly follows the relation
\begin{equation}
\sigma_s=\sqrt{\overline{\Delta s^2}-\overline{\Delta s}^2}\,,
\end{equation}
meaning that the rms bunch length is given by the square root of the difference between the average of the squared particle positions and
the squared average position. The rms momentum spread $\sigma_p$ of a bunch is computed analogously. In fact, the Greek letter $\sigma$
will always indicate an rms quantity.

Since the momentum spread is negative at the head of the bunch the corresponding particles travel
with a lower velocity compared to the tail of the bunch where the momentum spread is positive. The distributions in \figref{fig:bunch-before-chicane}
were obtained by simulating electron bunches from their generation at the cathode to the linac exit with the help of Astra. These are the
bunches that we intend to use in the framework of the paper. Note that the typical length scale of a $\unit[3]{nC}$ bunch directly before
the FLUTE chicane lies in the picosecond regime, whereas the length of the $\unit[1]{pC}$ bunch is several hundred femtoseconds.

Now the phase space coordinates of these bunches
are tracked through the chicane analytically. To do so we need the parametric representation $\mathbf{r}(l,p)$ of the electron
trajectory with respect to the path length $l$ traveled. It consists of four parts of a circle and three straight lines (plus one
line directly before and one directly after the chicane) and it can be found in \appref{sec:parametric-representation-trajectory}.
Via the bending radius $R$ and the bending angle $\alpha$ it depends on the electron momentum $p$.
\begin{figure}[t!]
\centering
\subfloat[\,\,\,rms bunch length for ${Q_b=\unit[3]{nC}}$ as a function of the longitudinal distance]{
\label{bunch-length-in-chicane-3nc}\includegraphics[scale=1]{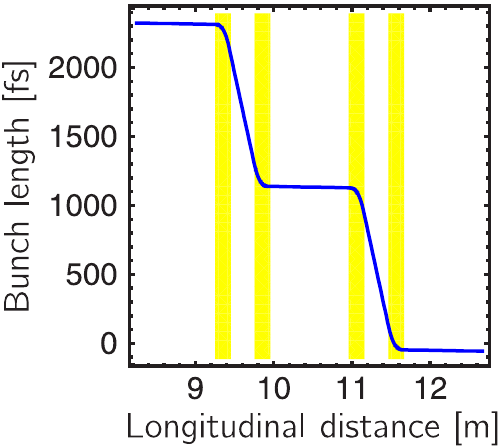}}\hspace{2.0cm}
\subfloat[\,\,\,the same as \protect\subref{bunch-length-in-chicane-3nc} with {${Q_b=\unit[1]{pC}}$}]{\label{bunch-length-in-chicane-1pc}\includegraphics[scale=1]{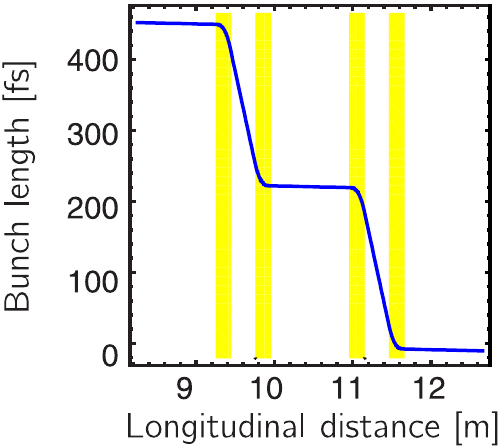}}
\caption{Reduction of the bunch length in the FLUTE chicane as a function of the longitudinal distance traveled. The horizontal axis
shows the distance traveled in meters and the vertical axis displays the bunch length. The yellow regions
indicate the positions of the chicane magnets along the longitudinal distance. The starting point of the chicane is chosen to be at
$z=\unit[9.25]{m}$. Besides, $L_{\mathrm{mag}}=\unit[0.2]{m}$, $L_{\mathrm{space}}=\unit[0.3]{m}$, and
$L_{\mathrm{drift}}=\unit[1.0]{m}$ are used (see \figref{fig:baseline-machine-layout}). In the left panel the reduction of the
bunch length for a bunch charge of $\unit[3]{nC}$ is presented and in the right panel for a bunch charge of $\unit[1]{pC}$.}
\label{fig:bunch-length-reduction}
\end{figure}%

What serves as an input are the bunch phase space coordinates one meter before the chicane (at $z=\unit[8.25]{m}$)\footnote{The six
centimeters difference from where the initial distributions are defined will be ignored leading to a modification of the bunch length
in the subfemtosecond regime.} that were obtained
from a simulation of the bunch from the cathode to this position. Every electron is then sent along its path through the chicane where
$\mathbf{r}(l,p+\Delta p)$ gives the spatial coordinates for an electron with momentum $p+\Delta p$ after travelling a distance $l$.
At the position $z=\unit[11.65]{m}$, which is one meter behind the bunch compressor, the $z$-coordinate of the reference electron
with momentum $p$ is subtracted from the $z$-coordinate of the electron with momentum $p+\Delta p$. This gives the position
$\Delta s^{(2)}$ of an electron with respect to the reference particle after the bunch compressor:
\begin{equation}
\Delta s^{(2)}=z(\unit[2]{m}+L(p+\Delta p),p+\Delta p)-z(\unit[2]{m}+L(p),p)\,.
\end{equation}
Herein, $L(p)$ is the traveling length through the chicane, which is given by \eqref{eq:path-length-through-chicane}. Hence, the bunch
coordinates after the compressor are projected on the $z$-axis. With this method particle velocity differences are taken into account
as well. Due to such a velocity difference a particle at the tail of the bunch with larger momentum may further catch up to a particle
at the head with lower momentum. This effect is called velocity bunching. An additional advantage of the technique described above is that
the spatial particle coordinates can be obtained after travelling an arbitrary length $l$ inside the chicane.

Figure~\ref{fig:bunch-length-reduction} shows how the bunch lengths for the \unit[3]{nC} and \unit[1]{pC} bunch evolve within the
bunch compressor. From the range of all electrons within the bunches we pick two with the initial distances \unit[2.30]{ps}
and \unit[452]{fs}, respectively (see the captions of \figref{fig:bunch-before-chicane}).
It is evident that the bunch length is mainly reduced in the regions between the first two and the last two magnets.
Hence, a large difference in traveling lengths is achieved with a large bending angle and a big $L_{\mathrm{space}}$. The drift
length between the second and the third magnet leads to a tiny reduction only originating from the velocity difference between
the electrons. Because of this, $L_{\mathrm{drift}}$ mainly decouples from the bunch length reduction, which can also be seen
from \eqref{eq:path-length-difference}.

The rate, which the reduction of the bunch length takes place with, increases in the first magnet until it reaches a constant value when
the bunch enters the drift between the first two magnets.\footnote{By ``rate'' we mean the amount of bunch length reduction per time
interval. This corresponds to the slope of the curves in \figref{fig:bunch-length-reduction}.}
In the second magnet the rate decreases by the same amount as it had increased in the first magnet. The same
pattern repeats in the last two magnets because of the symmetry of the chicane and due to the neglect of space charge forces and CSR
effects, which may become important for bunch lengths in the femtosecond regime. Comparing \figref{fig:bunch-length-reduction} to the Astra
output of figure 6 in \cite{Huttel:2012zza} (where space charge effects are taken into account) reveals that compression is suppressed in
the last two bending magnets because of the increasing space charge effects in the compressed bunch. Due to the neglect of space charge
effects in \figref{fig:bunch-length-reduction} the time evolutions for both bunch charges look the same. After all, the resulting curve
then scales only with the initial distance between the two electrons.

The overall bunch compression sensitively depends on the final rate of compression at the end of the first and third magnet, respectively.
If this rate is high it will
stay high during the propagation of the bunch between the first two or the last two magnets heavily reducing the bunch length. So these drift
spaces play a major role for compression. Note that the bunch length is decreased by a minor fraction in the drift lengths before, in the
middle, and after the chicane as a result of velocity bunching. Since these effects are rather small for a particle energy in the
$\unit[40]{MeV}$ range, $L_{\mathrm{drift}}$ does not play that much a role for compression at FLUTE. This space can be used rather for beam
diagnostics, e.g., a skew quadrupole, which can give information on the longitudinal bunch profile when combined with a fluorescent screen
placed downstream the chicane (see \figref{fig:baseline-machine-layout}).

Besides, the final distances between the two electrons considered are negative for both bunches indicating an overcompression.
However note that considering two electrons with the distance $\sigma_s$ is not equivalent to considering a bunch of many electrons
with the bunch length $\sigma_s$. Hence an overcompression does then not necessarily occur for the bunch as a whole, which can be
seen in the subsequent results. One can say that a minimum distance between two electrons and a minimum bunch length $\sigma_s$ are
different optimization criteria.
\begin{figure}[t!]
\centering
\subfloat[{final bunch profile with $Q_b=\unit[3]{nC}$ and $\sigma_s=\unit[211]{fs}$ obtained from path length differences}]{\label{fig:bunch-after-chicane-analytical-3nC}\includegraphics[scale=0.5]{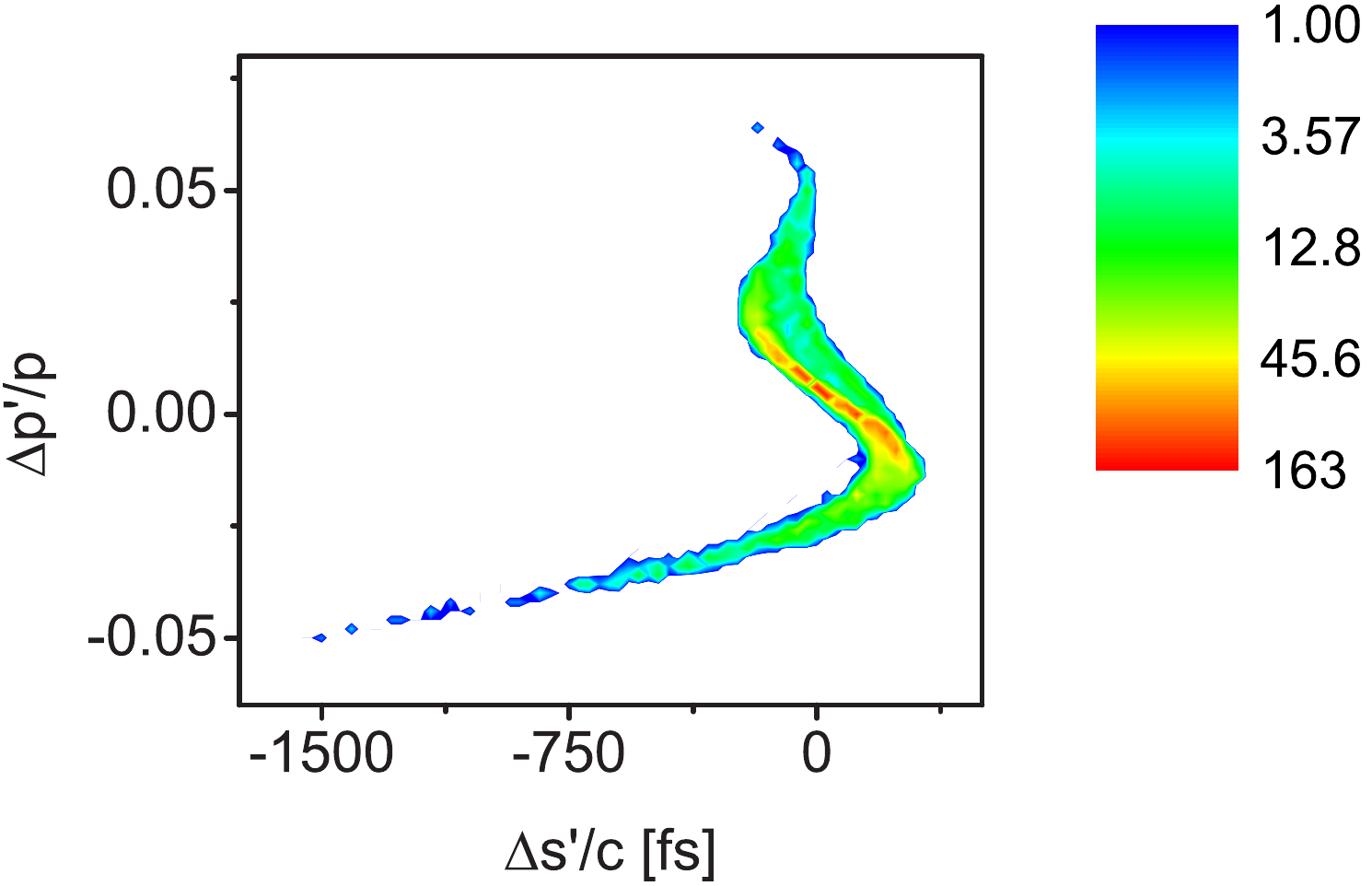}}\hspace{1.0cm}
\subfloat[the same as \protect\subref{fig:bunch-after-chicane-analytical-3nC} with {$Q_b=\unit[1]{pC}$} and {$\sigma_s=\unit[13]{fs}$}]{\label{fig:bunch-after-chicane-analytical-1pC}\includegraphics[scale=0.5]{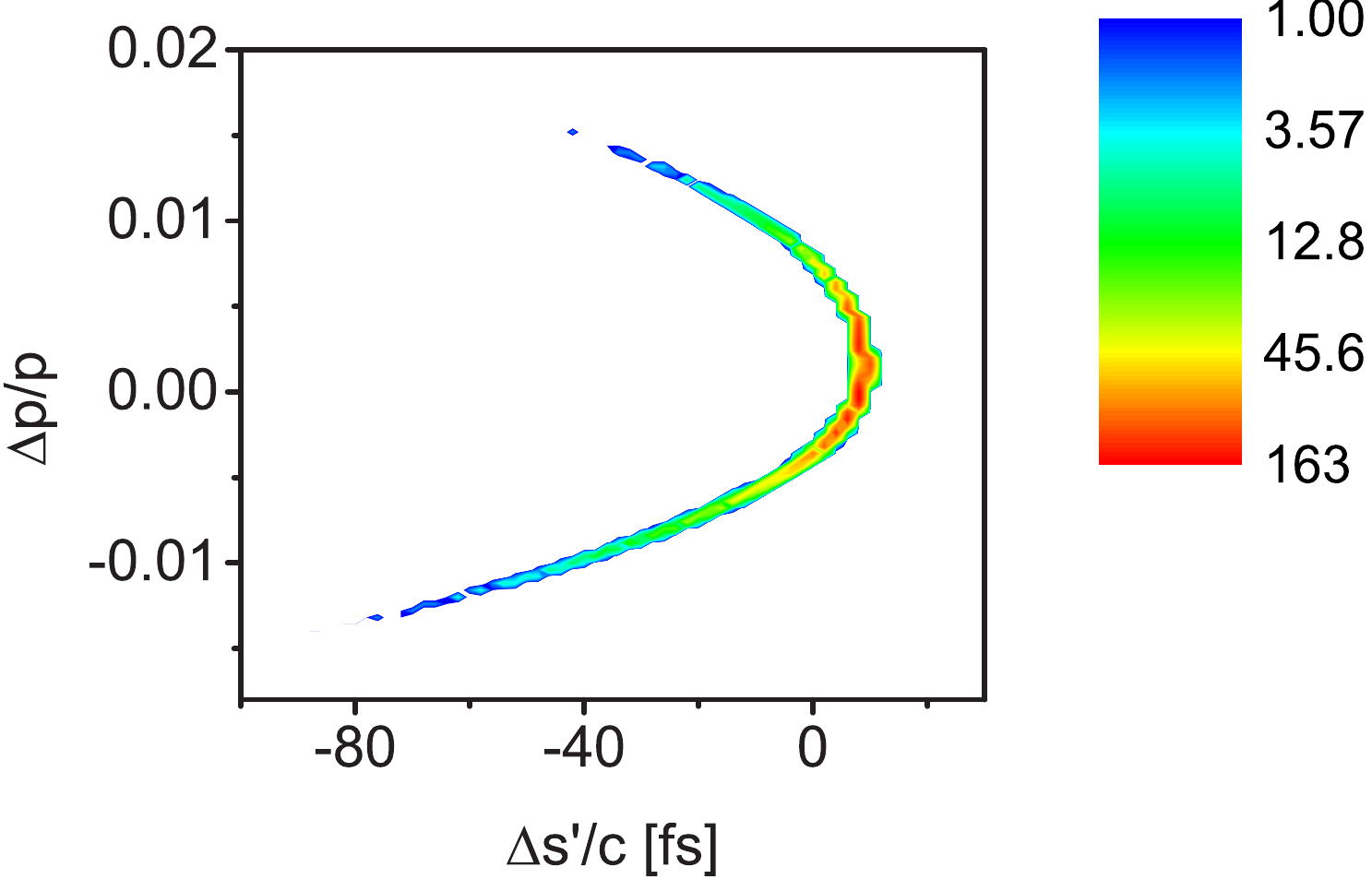}} \\
\subfloat[{final bunch profile with $Q_b=\unit[3]{nC}$ and $\sigma_s=\unit[201]{fs}$ obtained with Astra}]{\label{fig:bunch-after-chicane-astra-3nC}\includegraphics[scale=0.5]{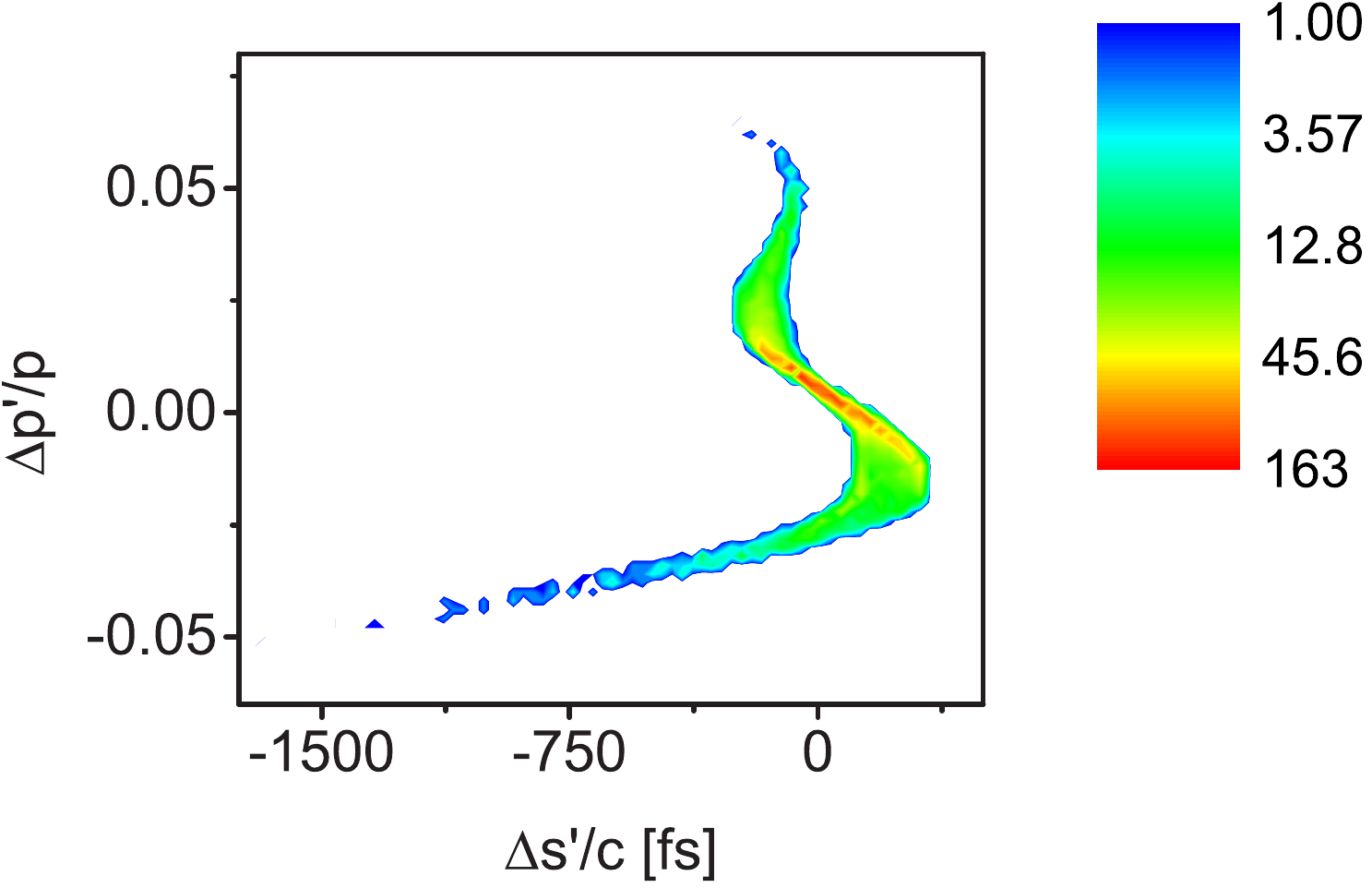}}\hspace{1.0cm}
\subfloat[the same as \protect\subref{fig:bunch-after-chicane-astra-3nC} with {$Q_b=\unit[1]{pC}$} and {$\sigma_s=\unit[13]{fs}$}]{\label{fig:bunch-after-chicane-astra-1pC}\includegraphics[scale=0.5]{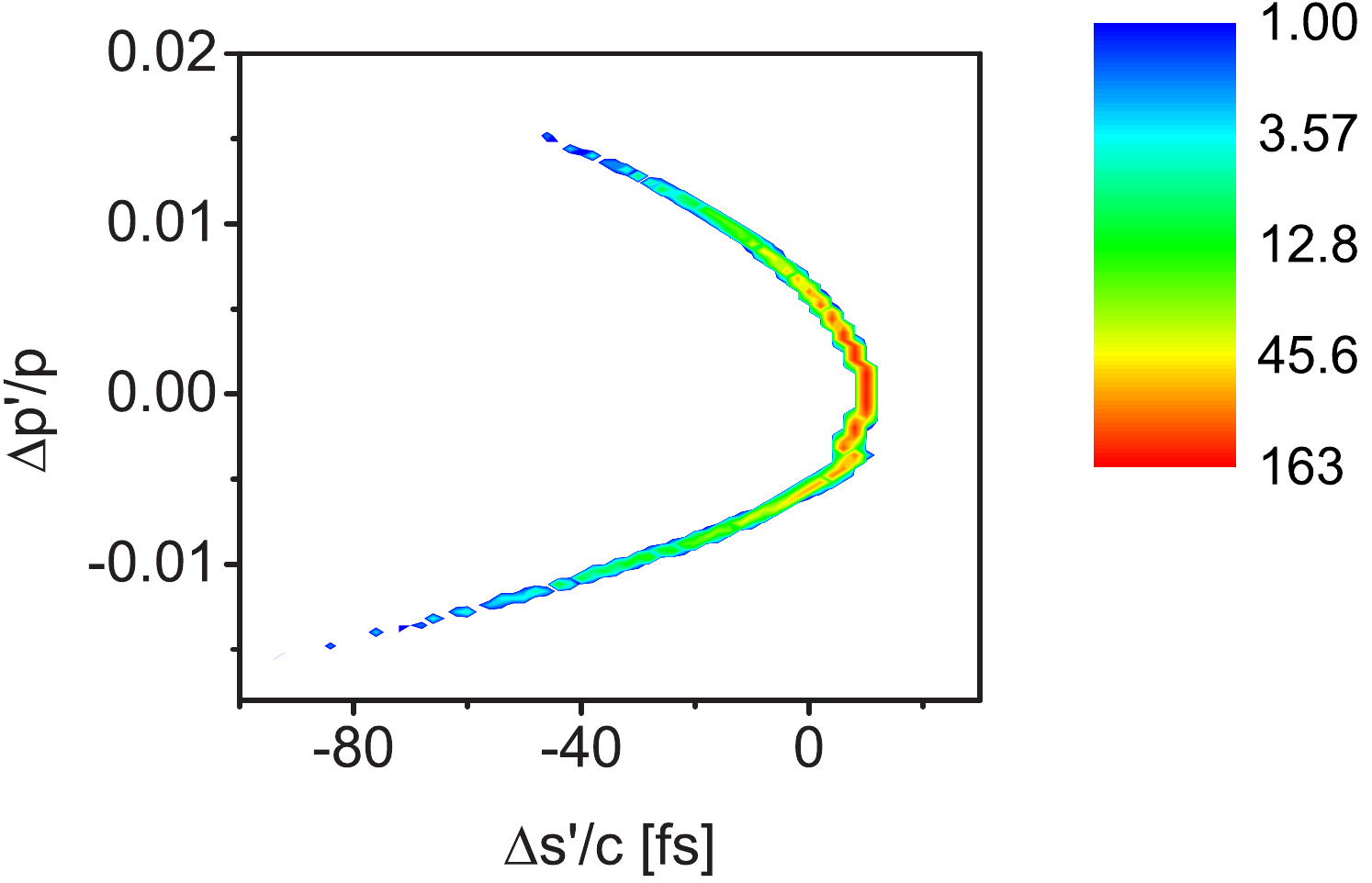}}
\caption{Longitudinal phase space plot of simulated \unit[3]{nC} and \unit[1]{pC} bunches at the position $z=\unit[12.65]{m}$ behind the chicane.
The chicane parameters are the same as in \figref{fig:bunch-length-reduction}.}
\label{fig:bunch-after-chicane}
\end{figure}%

Now we are interested in the longitudinal phase space after the chicane for the \unit[3]{nC} and the \unit[1]{pC} bunches used previously.
Sending each electron along its own trajectory leads to the results shown in \figref{fig:bunch-after-chicane}.
Note that the units used for the horizontal axis are now femtoseconds. The rms bunch length was reduced by a factor of 10.9 for the \unit[3]{nC}
bunch and a factor of 34.9 for the \unit[1]{pC} bunch. The double-s structure visible in \figref{fig:bunch-after-chicane-analytical-3nC} results
from the superimposed bumps in the initial distribution shown in \figref{fig:bunch-before-chicane}. Since all particle positions are reduced by
compression this structure is now more evident than it had been in the latter figure.

We see that both the final bunch lengths and the bunch profiles of the analytical calculation in Figs. \ref{fig:bunch-after-chicane-analytical-3nC},
\ref{fig:bunch-after-chicane-analytical-1pC} agree well with the Astra simulation results in Figs. \ref{fig:bunch-after-chicane-astra-3nC},
\ref{fig:bunch-after-chicane-astra-1pC}. For the \unit[3]{nC} bunch there is a deviation of the final bunch length of approximately 5\% and
for the \unit[1]{pC} bunch it is only 3\%.

\subsection{Transformation of an ideal phase space distribution}

Bunch distributions that are obtained from Astra simulations starting at the cathode usually are contaminated by substructures. These were
mentioned at the end of the previous section. Hence, for a theoretical understanding of the compression scheme it is more convenient to use an
ideal longitudinal phase space distribution. An ideal correlated energy spread is described by a straight line connecting the coordinates
$P_h$ of the head particle and $P_t$ the tail particle of the bunch. For simplicity the coordinate $P_r$ of the reference particle shall lie in
the center of the bunch. For the longitudinal phase space these are then given by:
\begin{subequations}
\begin{equation}
P_t\equiv \begin{pmatrix}
(\Delta s)_t \\
\delta_t \\
\end{pmatrix}=\begin{pmatrix}
-\sigma_s \\
\sigma_p \\
\end{pmatrix}\,,\quad P_r\equiv \begin{pmatrix}
(\Delta s)_r \\
\delta_r \\
\end{pmatrix}=\mathbf{0}\,,\quad
P_h\equiv \begin{pmatrix}
(\Delta s)_h \\
\delta_h \\
\end{pmatrix}=\begin{pmatrix}
\sigma_s \\
-\sigma_p \\
\end{pmatrix}\,,
\end{equation}
\end{subequations}
where $\sigma_s$ the rms bunch length, and $\sigma_p$ the rms (correlated) momentum spread. An
ideal chirp may then be parameterized by the following straight line:
\begin{equation}
\begin{pmatrix}
\Delta s \\
\delta \\
\end{pmatrix}=P_t+\begin{pmatrix}
\sigma_s \\
-\sigma_p \\
\end{pmatrix}\upsilon\,,\quad \upsilon\in [0,2]\,.
\end{equation}
The chicane transforms the chirp of the bunch. At linear order in the momentum spread this transformation can be written in matrix notation
such that a matrix $R$ acts on an initial longitudinal phase space vector $Z^{(1)}$ producing the final vector $Z^{(2)}$:
\begin{equation}
\label{eq:transformation-chirp-r56}
Z^{(2)}=RZ^{(1)}\,,\quad Z^{(1)}\equiv \begin{pmatrix}
\Delta s^{(1)} \\
\delta \\
\end{pmatrix}\,,\quad Z^{(2)}\equiv \begin{pmatrix}
\Delta s^{(2)} \\
\delta \\
\end{pmatrix}\,,\quad R=\begin{pmatrix}
1 & R_{56} \\
0 & 1 \\
\end{pmatrix}\,.
\end{equation}
Herein $\Delta s^{(1)}$ and $\Delta s^{(2)}$ are the longitudinal distances of a bunch particle before and after compression, respectively.
The matrix element $R_{56}$ relates the longitudinal distance to the normalized momentum spread. Note that in some papers instead of
$\Delta s$ the path length difference $\Delta L$ respective a reference particle is used directly.

Now consider, for instance, the head particle of the bunch. If $\Delta L^{(2)}-\Delta L^{(1)}>0$, the head particle has traveled a larger
distance through the chicane compared to the reference particle.
Then the distance with respect to the reference particle reduces by this amount, i.e., $\Delta s^{(2)}-\Delta s^{(1)}=-(\Delta L^{(2)}-\Delta L^{(1)})<0$.
From \eqref{eq:transformation-chirp-r56} it follows that $\Delta s^{(2)}-\Delta s^{(1)}=R_{56}\delta$. Then $R_{56}$ can be directly
extracted from \eqref{eq:path-length-difference}:
\begin{equation}
\label{eq:momentum-compaction-chicane-1}
R_{56}=2\left(\frac{L_{\mathrm{mag}}}{R}\right)^2\left(\frac{2}{3}L_{\mathrm{mag}}+L_{\mathrm{space}}\right)\,.
\end{equation}
In light of the previous arguments we have $R_{56}>0$.\footnote{In quite some papers, $R_{56}$ is defined to be negative (see, e.g.,
\cite{Kang:2004,Assmann:2013}). The reason is that these authors either use the path length difference $\Delta L$ instead of $\Delta s$ in
the phase space vector or they define $\Delta s^{(1)}-\Delta s^{(2)}\equiv R_{56}\delta$. None of these procedures will be followed in the
current paper.}
The parameter $R_{56}$ is also called momentum compaction factor and it contains the main information on bunch compression. The notation
used will be explained at a later point within this article. As the momentum spread is not transformed by the chicane, the matrix element at
the lower left corner of the matrix $R$ vanishes. This is clear since the bunch compressor solely consists of magnetic fields. In the
literature the following expression for $R_{56}$ for such a D-shape chicane is used as well (see, e.g.,
\cite{Castro:2003gx,Seletskiy:2009,Huttel:2012zza}):
\begin{equation}
\label{eq:momentum-compaction-chicane-2}
R_{56}'=2L_{\mathrm{space}}\frac{\tan^2\alpha}{\cos\alpha}-4L_{\mathrm{mag}}\left(\frac{\alpha-\tan\alpha}{\sin\alpha}\right)\,,\quad
\alpha=\arcsin\left(\frac{L_{\mathrm{mag}}}{R}\right)\,.
\end{equation}
The latter result is exact and valid for arbitrarily large bending angles $\alpha$. By a Taylor expansion with respect to $L_{\mathrm{mag}}/R$
it can be proven that $R_{56}=R_{56}'$ for $L_{\mathrm{mag}}\ll R$ besides corrections suppressed by $(L_{\mathrm{mag}}/R)^4$.
\begin{figure}[t]
\centering
\subfloat[\,\,\,linear part of transformation]{
\label{fig:phase-space-straight-3nC}\includegraphics[scale=1]{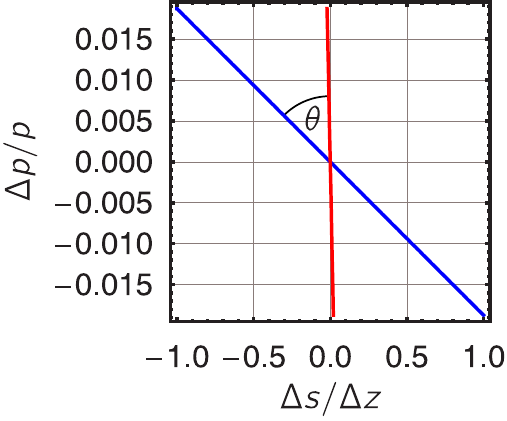}}\hspace{2.0cm}\subfloat[\,\,\,full transformation]{\includegraphics[scale=1]{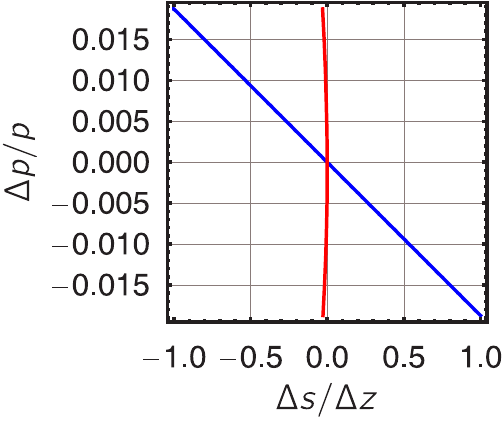}}
\caption{Comparison of ideal chirps before and after the FLUTE chicane. The horizontal axis gives the normalized distance of a particle with respect
to the reference particle. The vertical axis gives the normalized momentum spread. The phase space distribution before the chicane is shown as a
plain blue straight line, whereas the distribution after the chicane is illustrated by a plain red line. The angle between both lines in
\protect\subref{fig:phase-space-straight-3nC} is called $\theta$. The chicane parameters used are outlined in \figref{fig:baseline-machine-layout} and
the characteristic values $\sigma_s=\unit[2.30]{fs}$ and $\sigma_p=1.87\cdot 10^{-2}$ of the \unit[3]{nC} bunch were chosen for the plots.}
\label{fig:linear-and-full-transformation}
\end{figure}

\Figref{fig:linear-and-full-transformation} shows how an ideal chirp is transformed by the chicane. Both panels contain the initial chirp
(blue) and the final one (red). Here the longitudinal distance was normalized by the rms bunch length, i.e., it ranges from -1 to 1.
The left panel shows the result of the transformation given by \eqref{eq:transformation-chirp-r56}.
Since the transformation is linear the straight line is transformed to a straight line where the bunch length, i.e., the distance between the
head and the tail particle decreases. In case the final chirp is aligned along the vertical axis, the compression is optimal. This can be
characterized by the angle $\theta$ between the initial and the final chirp that can be computed as
\begin{equation}
\theta=\arccos\left(\frac{Z^{(1)}\cdot Z^{(2)}}{|Z^{(1)}||Z^{(2)}|}\right)\,.
\end{equation}
Best compression results for $\Delta s^{(2)}=0$ where the associated angle will be denoted as $\theta_b$. For the \unit[3]{nC} bunch we
obtain
\begin{equation}
\theta|_{\unit[3]{nC}}\approx 2.07^{\circ}\,,\quad \theta_b|_{\unit[3]{nC}} \approx 2.11^{\circ}\,,\quad (\theta/\theta_b)|_{\unit[3]{nC}} \approx 0.98\,,
\end{equation}
and for the \unit[1]{pC} bunch we compute
\begin{equation}
\theta|_{\unit[1]{pC}}\approx 1.61^{\circ}\,,\quad \theta_b|_{\unit[1]{pC}} \approx 1.62^{\circ}\,,\quad (\theta/\theta_b)|_{\unit[1]{pC}} \approx 0.99\,.
\end{equation}
Hence, the chicane parameters have evidently been chosen such that the angle $\theta$ lies in the vicinity of the optimum for both bunch charges.
This is even better for the \unit[1]{pC} bunch, since its initial chirp is closer to an ideal one --- in contrast to the \unit[3]{nC} bunch having
substructures (see \figref{fig:bunch-before-chicane}). The procedure proposed provides a good method of giving a first estimate on the optimum
parameters of the bunch compressor.

The right panel of \figref{fig:linear-and-full-transformation} shows the result of the transformation when the full particle trajectory is taken
into account as described in \secref{ssec:electron-trajectory}. Since this method also includes higher order terms in $\Delta p/p$, the transformation
of the chirp is no longer linear. Then the resulting chirp is not a straight line any more but it has a curvature.

Based on the transformation given by \eqref{eq:transformation-chirp-r56} every point of the chirp is transformed with the same matrix
element $R_{56}$. However this does not hold in general, but
\begin{equation}
\begin{pmatrix}
\Delta s^{(1)} \\
\delta \\
\end{pmatrix} \mapsto \begin{pmatrix}
\Delta s^{(2)} \\
\delta \\
\end{pmatrix}=f_{\mathrm{chic}}\left[\begin{pmatrix}
\Delta s^{(1)} \\
\delta \\
\end{pmatrix}\right]\,,
\end{equation}
where the function $f_{\mathrm{chic}}$ involves the phase space vector $(\Delta s,\delta)$ in a nonlinear way. So by talking about
nonlinearities in this context we mean that an initial ideal chirp being a straight line is not transformed into a straight line any more.
This is the case when the transformation contains higher-order polynomials in $\delta$. Such a situation cannot be described by a
matrix multiplication according to \eqref{eq:transformation-chirp-r56}.

\subsection{Transverse momentum components and magnetic field jitter}
\label{ssec:transverse-components-magnetic-jitter}

\begin{figure}[t]
\centering
\subfloat[\,\,\,modified chicane trajectory for electron with initial transverse momentum component]{
\label{fig:chicane-flute-modified-a}\includegraphics[scale=1]{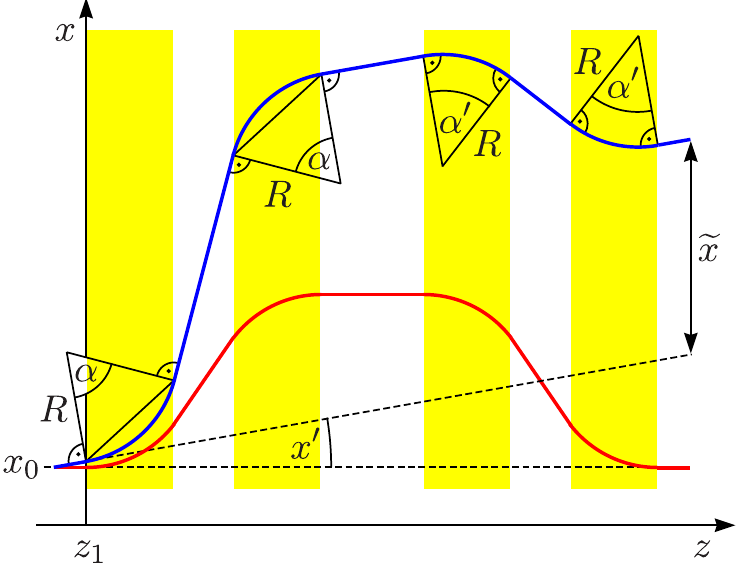}}\hspace{1.0cm}
\subfloat[\,\,\,chicane trajectory for different values of the magnetic fields in the chicane dipole magnets]{\label{fig:chicane-flute-modified-b}
\includegraphics[scale=1]{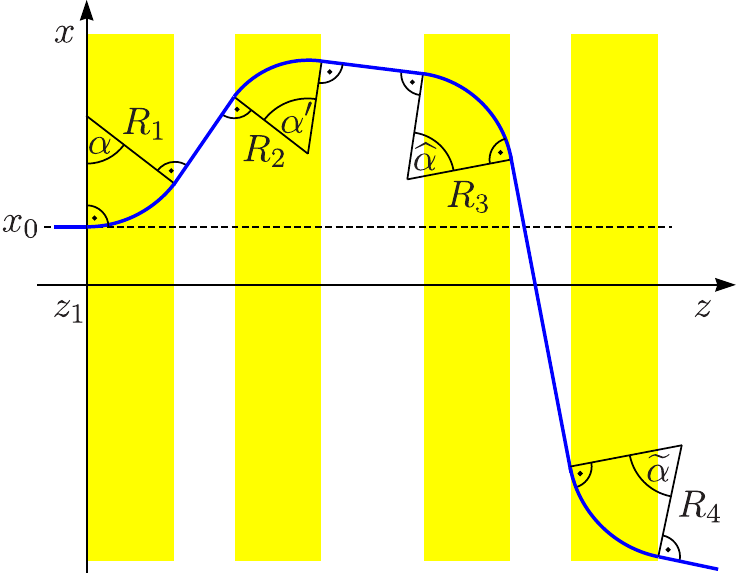}}
\caption{In the left panel the FLUTE chicane is considered with a particle trajectory travelled by an electron that has both an
initial transverse momentum and a longitudinal momentum component. Hence the angle $x'$ between the initial straight trajectory and
the horizontal axis is nonzero. The resulting modified trajectory is shown as a plain, blue curve. The trajectory for $x'=0$ is shown
as a plain, red curve and it is presented for comparison.
The right panel presents the chicane again for $x'=0$, but for different values of the bending radius for all four chicane magnets.
From both panels becomes evident that the electron after the chicane does not necessarily come back to the $z$-axis when such effects
are taken into consideration.}
\label{fig:chicane-flute-modified}
\end{figure}

In the previous considerations the incoming electron was assumed to only have a longitudinal momentum component. The resulting chicane
trajectory was shown in \figref{fig:chicane-principle}. However electrons entering the chicane are expected to have transverse
momentum components as well, which lie in the mrad regime for FLUTE. The transverse momentum components $p_x$ and $p_y$ can be
described by two angles $x'$ and $y'$ with respect to the $z$-axis:
\begin{equation}
\mathbf{p}=\begin{pmatrix}
p_x \\
p_y \\
p_z \\
\end{pmatrix}=|\mathbf{p}|\begin{pmatrix}
\sin x'\cos y' \\
\sin x'\sin y' \\
\cos x' \\
\end{pmatrix}\,.
\end{equation}
In case of nonvanishing angles $x'$ and $y'$ the trajectory of the electron through the chicane will be modified. The parametric representation
of the resulting particle trajectory was derived and is given in \appref{sec:parametric-representation-trajectory}. A draft for such
a modified trajectory restricted to the $x$-$z$-plane is shown in \figref{fig:chicane-flute-modified-a}, where the unmodified
trajectory is drawn as well. What becomes evident at first is that after the chicane the electron does not turn back to the path it
would have traveled without the chicane. Hence there is an offset transverse distance $\widetilde{x}$. The strong deviation off-axis is caused within
the distance $L_{\mathrm{space}}$ between the first two magnets, where the electron exits the first magnet under an additional angle.
This leads to a long drift in $x$-direction before the electron reaches the second magnet and this propagation is not compensated in
the last two magnets.

Furthermore, in practice the magnetic field strength values in the chicane magnets cannot be assumed to agree all the time. As
every power supply has a current jitter $\Delta I$ this will result in a jitter $\Delta B$ of the magnetic flux density:
\begin{equation}
\frac{\Delta I}{I}=\frac{\Delta B}{B}\,.
\end{equation}
The electron trajectory is modified by such a field jitter and a possible resulting curve is presented in
\figref{fig:chicane-flute-modified-b}. Here each dipole magnet has a different bending radius, which applies when the magnet
power supplies are operated in parallel.

\begin{figure}[t]
\centering
\subfloat[\,\,\,modified bunch lengths for transverse maximum angles ${|x'|=\unit[1]{mrad}}$ and transverse maximum
deviations ${|\Delta x|=\unit[10^{-2}]{m}}$]
{\label{fig:results-jitter-a}\includegraphics[scale=0.5]{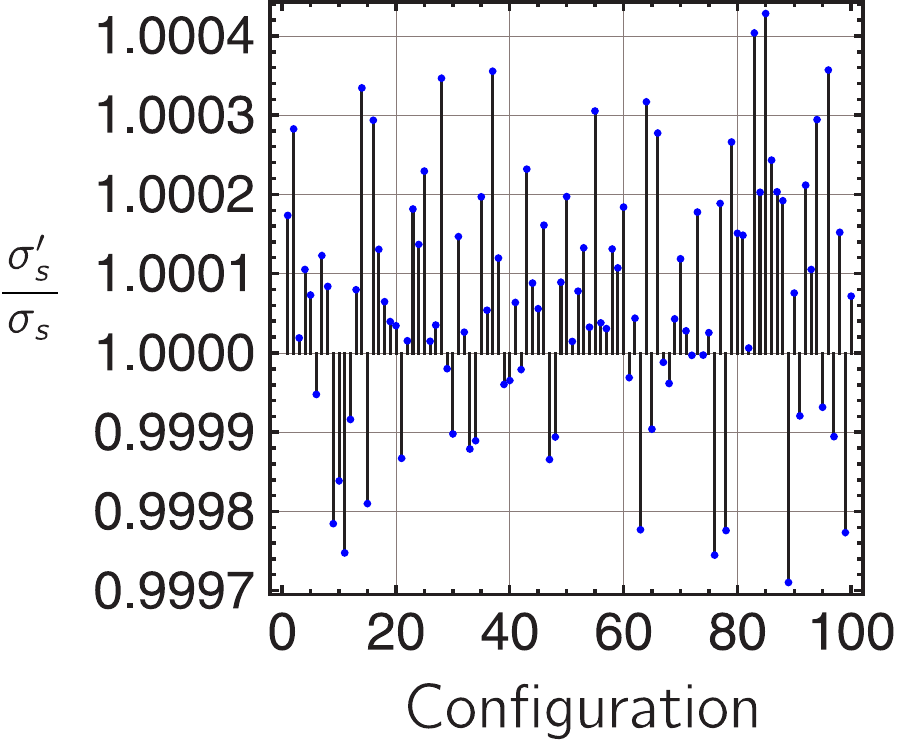}}\hspace{2cm}
\subfloat[\,\,\,modified bunch lengths for a magnetic field jitter $|\Delta B/B|=10^{-4}$]
{\label{fig:results-jitter-b}\includegraphics[scale=0.5]{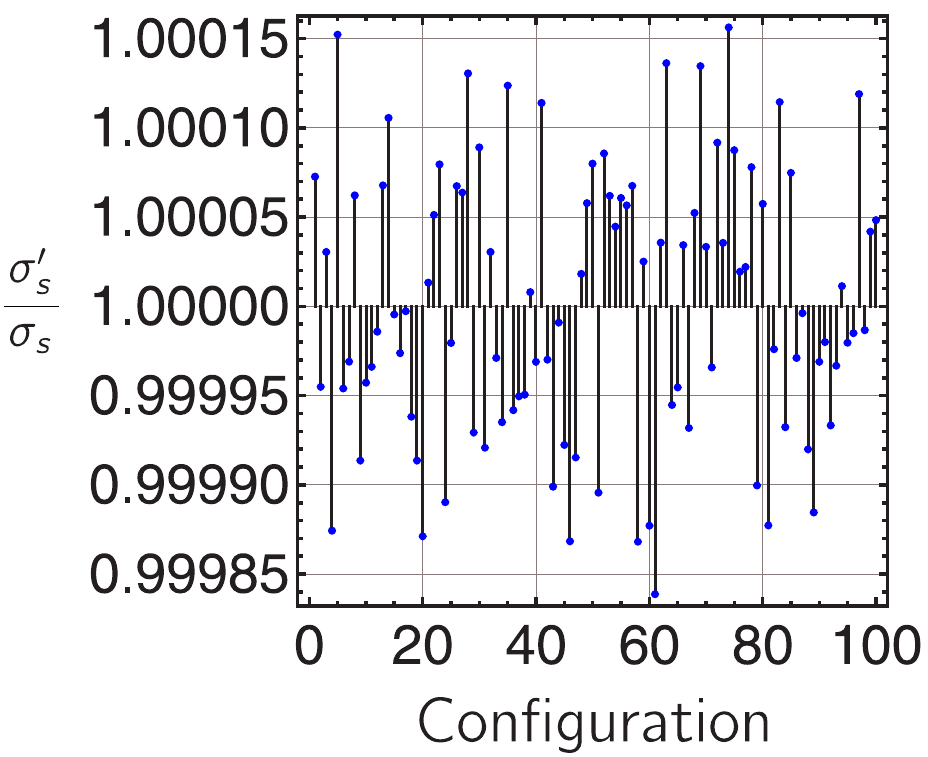}}
\caption{The current figure presents modified bunch lengths $\sigma_s'$ due to transverse angles and positions (left panel)
and due to a magnetic field jitter (middle panel). These are normalized by the final bunch length $\sigma_s$ without any
uncertainties. The first panel depicts the change of the bunch length when each initial particle trajectory encloses a random
transverse angle ${x'\in [-\unit[1]{mrad},\unit[1]{mrad}]}$ with the $z$-axis. Furthermore the particles have random
transverse positions ${\Delta x\in [-\unit[10^{-2}]{m},\unit[10^{-2}]{m}]}$. The second panel shows how the bunch length
changes when there is a magnetic field jitter. The rainbow color code in the third
panel illustrates the ratio $\sigma_s'/\sigma_s$ as a function of the mean transverse angle $(x')_{\mathrm{mean}}$ and the
mean transverse position $(\Delta x)_{\mathrm{mean}}$. The respective function values increase from violet to red.}
\label{fig:results-jitter}
\end{figure}

First of all the magnetic field values in all four dipole magnets are assumed to be constant. We investigate the influence on the
bunch length by initial transverse angles $x'$ with respect to the $z$-axis and initial transverse position deviations $\Delta x$.
The modified bunch length is obtained for 100 configurations. Talking about a ``configuration'' we mean that the angle $x'$ is
chosen randomly to lie in the interval $[-\unit[10^{-2}]{rad},\unit[10^{-2}]{rad}]$ for each particle. Furthermore, at the same
time this is done for the initial transverse deviation $\Delta x$ that is chosen within the range
$[-\unit[10^{-2}]{m},\unit[10^{-2}]{m}]$. The probability distributions are taken as uniform for both intervals. Note that
these intervals are hypothetical since realistic values are much smaller; this will be discussed below. The result is shown in
\figref{fig:results-jitter-a}.

As a second step a magnetic field jitter was considered without any transverse momentum components or offsets. The modified bunch
length due to this jitter can be computed and normalized by the final bunch length that is obtained without any jitter. This was
done for 100 configurations. In that case a ``configuration'' means that the flux density is chosen independently for each magnet
within the interval $[B-\Delta B,B+\Delta B]$, based on a uniform probability distribution. The result for $|\Delta B/B|=10^{-4}$
is presented in \figref{fig:results-jitter-b}.

\Figref{fig:results-jitter-a} indicates that the maximum modification of the bunch length due to transverse angles in the
$\unit[1]{mrad}$ regime and transverse offsets in the $\unit[10^{-2}]{m}$ range lies even below one per mill.
Realistic transverse deviations are much smaller. For a typical bunch, $\Delta x$ is in the order of $\unit[10^{-3}]{m}$
and $x'$ in the order of $\unit[10^{-1}]{mrad}$. Hence, the transverse electron coordinates do not seem to
have a large influence on bunch compression --- at least within this simplified analysis. Besides, it can be shown that for
angles $\gtrsim \unit[10^{-2}]{rad}$ the bunch length increases to a larger degree in comparison to the smaller angles
considered before. This may indicate that the transverse offset $\widetilde{x}$ behind the chicane starts playing a larger
role for these values. However such rather large angles are purely hypothetical and do not appear in typical bunches at
FLUTE.

\Figref{fig:results-jitter-b} reveals that the bunch length can be modified by a factor of $10^{-4}$ due to a jitter of the magnetic
flux density of $|\Delta B/B|=10^{-4}$. So the uncertainty in the stability of the flux density is directly imposed on the bunch
length without becoming larger or smaller. The conclusion is that there is a linear connection between this uncertainty and the
bunch length. Otherwise we would expect a heavily larger or a smaller modification.

\subsection{Sector chicane as a further (hypothetical) example}
\label{ssec:sector-chicane}

It is planned to construct the FLUTE bunch compressor using rectangular dipole magnets. However for theoretical reasons, in this paper we
additionally intend to consider the characteristics of a bunch compressor made up of sector dipole magnets. A sector dipole is characterized
by the property that the reference particle both enters and exits the magnet perpendicularly to its edges. This is not necessarily the
case for a rectangular magnet.

\begin{figure}[t]
\centering
\includegraphics[scale=1]{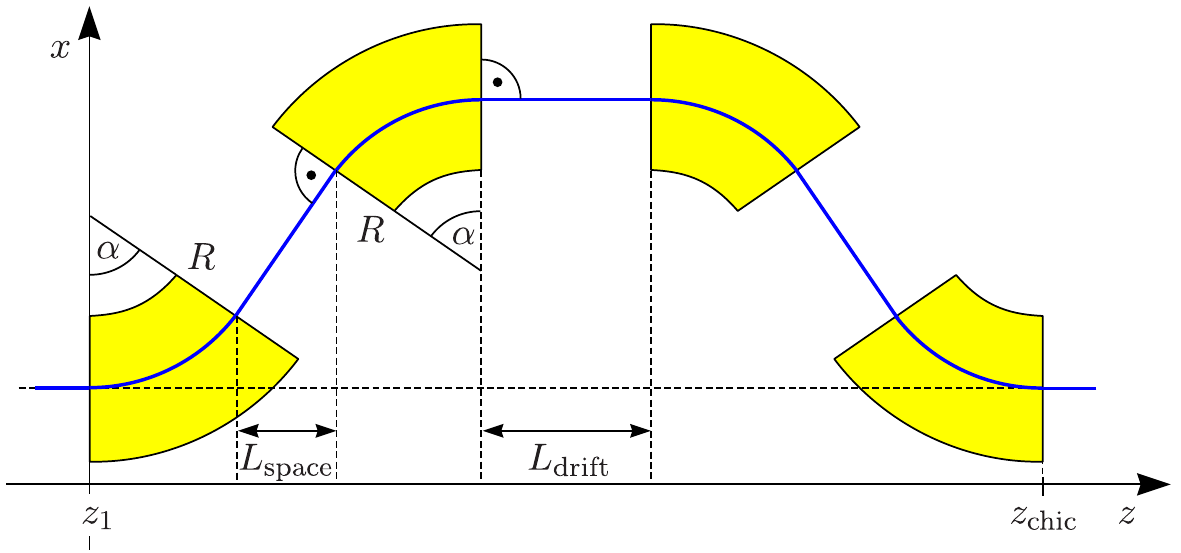}
\caption{Draft of a bunch compressor consisting of sector dipole magnets. The reference trajectory is shown as a plain, blue line.
The bending angle is given by $\alpha$ and the bending radius is denoted as $R$. The distance between the exit face of the first/third
dipole magnet and the entrance face of the second/fourth dipole magnet that is projected on the longitudinal axis is called
$L_{\mathrm{space}}$. The distance between the second and the third magnet is $L_{\mathrm{drift}}$.}
\label{fig:chicane-sector-principle}
\end{figure}

The principle of a chicane constructed with sector dipole magnets is shown in \figref{fig:chicane-sector-principle}. The free parameters
of such a chicane are the bending angle $\alpha$, the bending radius $R$, and the distances $L_{\mathrm{space}}$ and $L_{\mathrm{drift}}$.
We can then derive a parametric representation of the reference trajectory. The result can be found in \appref{ssec:chicane-sector-dipoles}.
Using this representation we compute the path length difference of two trajectories with normalized momentum spread $\delta=\Delta p/p$.
At first order in $\delta$ and for bending angles $\alpha\ll \pi/2$ we obtain:
\begin{equation}
\Delta L=-2\alpha^2\left(\frac{2}{3}R\alpha+L_{\mathrm{space}}\right)\delta+\mathcal{O}\left(\alpha^4,\delta^2\right)\,.
\end{equation}
If the chicane parameters $\alpha$, $R$, and $L_{\mathrm{space}}$ are chosen such that they correspond to the parameters of the chicane in
\figref{fig:chicane-principle}, the momentum compaction factor for $\delta\ll 1$ and $\alpha\ll \pi/2$ is the same for both types of
chicanes. However, note that effects from the fringes of the dipole magnets have been neglected in this derivation. We will come back
to the sector chicane at a later stage of the paper.

\section{Transfer matrix formalism applied on the FLUTE chicane}
\label{sec:transfer-matrix-formalism}
\setcounter{equation}{0}

In the previous chapters the FLUTE bunch compressor was investigated analytically by deriving parametric representations for particle
trajectories in the compressor. The advantage of this approach is that all geometrical effects are taken into account. However this
technique also has a number of disadvantages. First of all, the dipole field strength of the chicane magnets
has been assumed to fall off to zero directly outside the magnet, i.e., we have used a hard-edge model. This is not the case
for real magnets having a nonzero fringe field outside of the iron yoke. Secondly, the calculational time of this method is rather
large since the trajectory for each electron has to be computed separately. This may already take several minutes for 5000 particles,
which is the typical number of particles that we use.

For these reasons we are interested in considering an alternative approach in the current section. In general, each electron within
a bunch can be described by a six-dimensional phase space vector $Z$, which reads as follows:
\begin{equation}
\label{eq:phase-space-coordinates}
Z=\left(\Delta x,\frac{p_x}{p},\Delta y,\frac{p_y}{p},\Delta s,\delta\right)^T\simeq \left(\Delta x,x',\Delta y,y',\Delta s,\delta\right)^T\,.
\end{equation}
These components give positions in configuration space and momentum space with respect to a reference particle. The variables $\Delta x$
and $\Delta y$ are the two transverse offsets, $x'$ and $y'$ are the transverse angles, $\Delta s$ is the longitudinal distance,
and $\delta=\Delta p/p$ the normalized momentum spread. For typical angles in the mrad-regime it holds that $x'\simeq p_x/p$ and
$y'\simeq p_y/p$. This makes the description of the particle phase space by the first vector in \eqref{eq:phase-space-coordinates}
equivalent to the description by the second vector.

Solving the equations of motion of the reference particle is relatively straightforward in some cases. This is evident for the reference
trajectory of the FLUTE chicane that has a simple form. However the trajectories shown in, e.g., \figref{fig:chicane-flute-modified},
that are not those of a reference particle are more involved. In general, the equations
of motion for particles not being reference particles may be complicated to solve, even more when field inhomogeneities are taken into
account. For this reason the procedure is to not solve these equations exactly but perturbatively, i.e., as an expansion in the deviations
$\Delta x$, $x'$ etc. from the reference trajectory.

In the framework of perturbation theory each part of an accelerator such as a drift, dipole magnet, quadrupole magnet etc. modifies the
phase space distribution of a particle. Hence, it transforms an initial phase space vector $Z^{(1)}$ to a final vector $Z^{(2)}$.
Expanding this transformation to second order in the phase space vector it can be written with the help of a transfer matrix $R$
(a second-rank tensor) and a third-rank transfer tensor $T$ \cite{Brown:1984,Iselin:1992}:
\begin{equation}
\label{eq:corrections-phase-space-vector}
Z_j^{(2)}=Z_j^{(1)}+\sum_{k=1}^6 R_{jk}Z_k^{(1)}+\sum_{k,l=1}^6 T_{jkl}Z_k^{(1)}Z_l^{(1)}+\hdots\,.
\end{equation}
In this case a transfer matrix $R$ has $6^2=36$ components and a third-rank transfer tensor $T$ even $6^3=216$ components. Within this
formalism, two elements of an accelerator (e.g. a drift or a dipole magnet) can be combined at first order perturbation
theory by a simple matrix multiplication. If an electron propagates through an element designated by (a) and followed by an element (b)
the resulting transfer matrix is given by $R^c=R^bR^a$. The ordering of the matrices is such that the transfer matrix of the first
element, which the electron propagates through, is at the far right of the matrix product.

The third-rank tensor of a combination of two accelerator components (a) and (b) is given by~\cite{Iselin:1992}:
\begin{equation}
\label{eq:composition-third-rank-tensor}
T^c_{ijk}=\sum^6_{l=1} R_{il}^bT^a_{ljk}+\sum^6_{l=1}\sum^6_{m=1} T^b_{ilm}R^a_{lj}R^a_{mk}\,.
\end{equation}
This equation involves both the transfer matrices and the third-rank tensors of the respective accelerator components.

\subsection{First order perturbation theory}

First of all, we will concentrate simply on the transfer matrices $R$ that are taken from \cite{Iselin:1992}. The notation used in
\cite{Brown:1984,Iselin:1992} will be kept with some minor modifications that will be stated in the corresponding context. The
explicit matrices plus additional conventions are stated in \appref{sec:transfer-matrices-flute}. For the FLUTE chicane the 
transfer matrices of a drift and that of a rectangular dipole are needed.

The transfer matrix $R_{\mathrm{drift}}$ for a drift can be found in \eqref{eq:matrix-drift}. For a drifting particle both $\Delta x$
and $\Delta y$ increase with the length of the drift whereas the transverse angles $x'$ and $y'$ are not modified. The transfer
matrix $R_{\mathrm{sec}}$ of a sector dipole magnet is given by \eqref{eq:matrix-sector-dipole}. Since a particle travels on parts of
a circle through such a magnet the respective transfer matrix involves trigonometric functions.

If the particle does not enter or exit the dipole magnet perpendicularly to its surfaces, magnetic fringe fields have to be taken into
account. The action of these fringe fields on a particle are described by the matrix $R_{\mathrm{fringe}}$ of \eqref{eq:matrix-fringe}.
It involves the entrance and exit angle of the particle with respect to the magnet edges. Furthermore the magnetic fringe field profile
plays a great role in this context (see \appref{sec:transfer-matrices-flute}).
\begin{figure}[b]
\centering
\subfloat[magnet left turn]{\label{fig:magnet-left-turn}\includegraphics{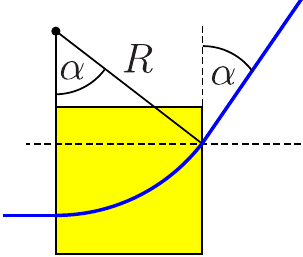}}\hspace{2cm}
\subfloat[magnet right turn]{\label{fig:magnet-right-turn}\includegraphics{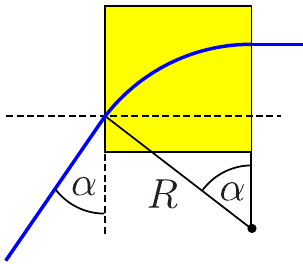}}
\caption{Left turn and right turn trajectory inside a rectangular dipole magnet of the FLUTE chicane.}
\label{fig:magnet-turns}
\end{figure}%

The transfer matrix $R_{\mathrm{rec}}$ for a rectangular dipole is not given directly in \cite{Brown:1984,Iselin:1992}. But it can be
constructed from the transfer matrices for the sector dipole and the magnet fringe:
\begin{equation}
R_{\mathrm{rec}}(L,h,\psi_1,\psi_2)=R_{\mathrm{fringe}}(\psi_2,h)R_{\mathrm{sec}}(L,h)R_{\mathrm{fringe}}(\psi_1,h)\,,
\end{equation}
with the entrance angle $\psi_1$, the exit angle $\psi_2$, and the curvature $h\equiv 1/R$ of the reference trajectory. For the sign
convention of $\psi_1$ and $\psi_2$ we refer to \figref{fig:sector-bend-sign-conventions}. In particular for the FLUTE chicane it holds
that
\begin{subequations}
\begin{equation}
R_{\mathrm{rec}}(L,-h,0,-\alpha)=R_{\mathrm{fringe}}(-\alpha,-h)R_{\mathrm{sec}}(L,-h)R_{\mathrm{fringe}}(0,-h)\,,
\end{equation}
for a left turn trajectory (see \figref{fig:magnet-left-turn}) and
\begin{equation}
R_{\mathrm{rec}}(L,h,0,\alpha)=R_{\mathrm{fringe}}(0,h)R_{\mathrm{sec}}(L,h)R_{\mathrm{fringe}}(\alpha,h)\,,
\end{equation}
\end{subequations}
for a right turn trajectory (see \figref{fig:magnet-right-turn}). The complete chicane can now be represented by the following matrix
$R$ as a function of the entrance and exit angles that go into $R_{\mathrm{fringe}}$:
\begin{align}
\label{eq:chicane-matrix}
R&=R_{\mathrm{rec}}(R\alpha,-h,0,-\alpha)R_{\mathrm{drift}}\left(\frac{L_{\mathrm{space}}}{\cos\alpha}\right)R_{\mathrm{rec}}(R\alpha,h,\alpha,0)R_{\mathrm{drift}}(L_{\mathrm{drift}}) \notag \\
&\phantom{{}={}}\times R_{\mathrm{rec}}(R\alpha,h,0,\alpha)R_{\mathrm{drift}}\left(\frac{L_{\mathrm{space}}}{\cos\alpha}\right)R_{\mathrm{rec}}(R\alpha,-h,-\alpha,0)\,.
\end{align}
Note that the curvature in the first and the fourth magnet has to be set to negative values since the magnetic field has a different
sign compared to the magnetic field in the second and the third magnet.
According to \eqref{eq:chicane-matrix} the transfer matrix of the whole FLUTE chicane is obtained by multiplying the appropriate
matrices. The element $R_{56}$ corresponds to the momentum compaction factor. For $L_{\mathrm{mag}}\ll R^2$
the following result is obtained:
\begin{equation}
\label{eq:transfer-matrices-r56}
R_{56}=2\left(\frac{L_{\mathrm{mag}}}{R}\right)^2\left[\frac{2}{3}L_{\mathrm{mag}}+L_{\mathrm{space}}\right]+\mathcal{O}\left[(L_{\mathrm{mag}}/R)^4\right]\,.
\end{equation}
It is equal to \eqref{eq:momentum-compaction-chicane-1} resulting from the path length difference of particle trajectories.

\begin{figure}[t!]
\centering
\subfloat[final bunch profile with {$Q_b=\unit[3]{nC}$} and {$\sigma_s=\unit[173]{fs}$} obtained with $R_{56}$ of \eqref{eq:transfer-matrices-r56}]{\label{fig:bunch-after-chicane-r56-3nC}\includegraphics[scale=0.5]{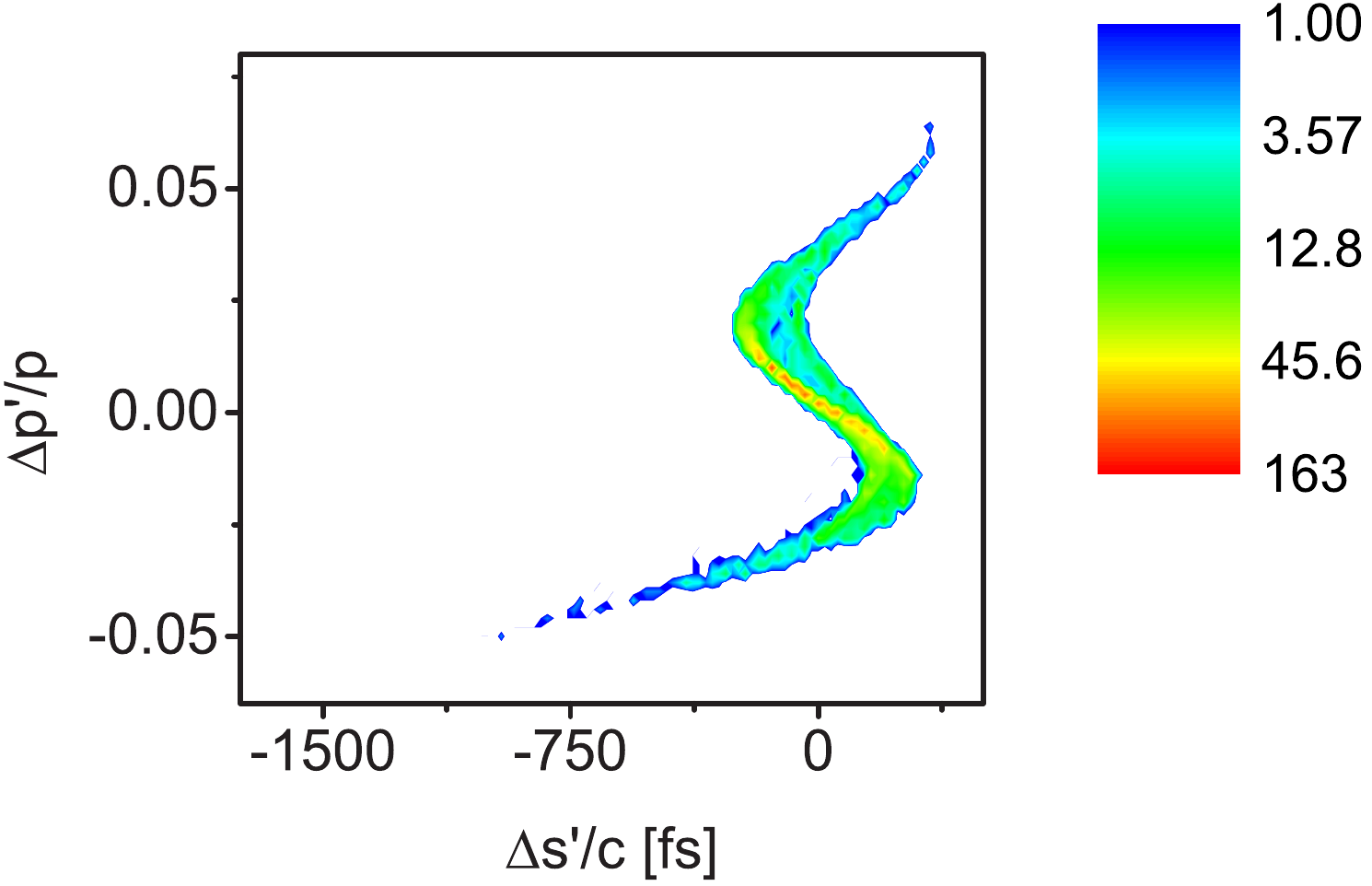}}\hspace{1.0cm}
\subfloat[the same as \protect\subref{fig:bunch-after-chicane-r56-3nC} with {$Q_b=\unit[1]{pC}$} and {$\sigma_s=\unit[7]{fs}$}]{\label{fig:bunch-after-chicane-r56-1pC}\includegraphics[scale=0.5]{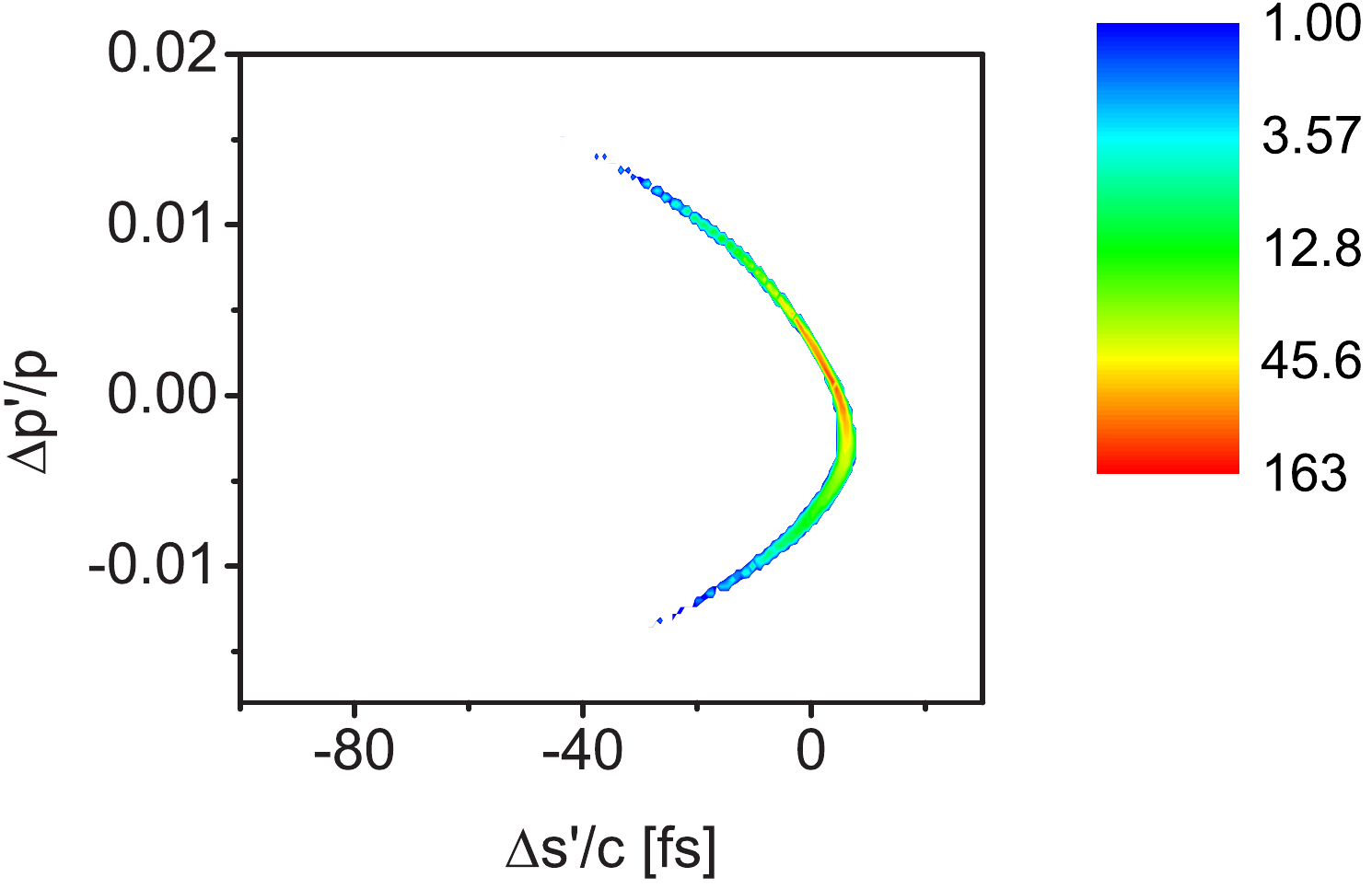}}
\caption{Longitudinal phase space plots of simulated \unit[3]{nC} and \unit[1]{pC} bunches after the chicane. The profiles shown were computed
by using the transfer matrix formalism at first order in $\Delta p/p$.}
\label{fig:bunch-after-chicane-r56}
\end{figure}

Now we would like to test the first order result on the \unit[3]{nC} and \unit[1]{pC} bunches used previously.
Unfortunately, the bunch profiles and final bunch length obtained by first order perturbation theory in Figs.~\ref{fig:bunch-after-chicane-r56-3nC},
\ref{fig:bunch-after-chicane-r56-3nC} differ from the Astra results
by quite some amount. For the \unit[3]{nC} bunch the deviation is 14\% and for the \unit[1]{pC} bunch it is even 43\%. This shows
that the transfer matrix formalism at first order in the momentum spread does not suffice to reproduce the Astra simulation output.

\subsection{Second order corrections}

The previous section dealt with the momentum compaction factor at first order perturbation theory. We are now interested to compute
the second order contribution of the path length difference in the chicane, i.e., the contribution proportional to $\delta^2$. It is
given by the tensor coefficient $T_{566}$ and can be obtained from Eqs.~(\ref{eq:path-length-through-chicane}), (\ref{eq:path-length-difference})
by including terms in the Taylor expansion up to second order in $\delta$. For $L_{\mathrm{mag}}\ll R$ is reads:
\begin{align}
\label{eq:path-length-difference-chicane-plus-second-order}
\Delta L &\equiv \Delta L^{(1)}+\Delta L^{(2)}+\dots \notag \\
&=-2\left(\frac{L_{\mathrm{mag}}}{R}\right)^2\left[\frac{2}{3}L_{\mathrm{mag}}+L_{\mathrm{space}}\right]\delta
+\left(\frac{L_{\mathrm{mag}}}{R}\right)^2\left[2L_{\mathrm{mag}}+3L_{\mathrm{space}}\right]\delta^2+\dots\,,
\end{align}
where $\Delta L^{(n)}$ denotes a correction proportional to $\delta^n$.
From the general relation $\Delta L=R_{56}\delta+T_{566}\delta^2+\dots$, the coefficient $T_{566}$ can be directly obtained by
comparison:
\begin{equation}
\label{eq:computation-566-from-r56}
T_{566}=\left(\frac{L_{\mathrm{mag}}}{R}\right)^2\left[2L_{\mathrm{mag}}+3L_{\mathrm{space}}\right]=-\frac{3}{2}R_{56}\,.
\end{equation}
Note that $T_{566}$ has the same order of magnitude as $R_{56}$ but it has a different sign.

Now let us compare the bunch profiles obtained from the transfer matrix formalism at second order in $\Delta p/p$ to the Astra
simulation output. In Figs.~\ref{fig:bunch-after-chicane-r56-3nC}, \ref{fig:bunch-after-chicane-r56-1pC} you see the bunch profiles
for the bunch charges \unit[3]{nC} and \unit[1]{pC}, respectively. The rms bunch length for \unit[3]{nC} is ca. 7\% larger than
the Astra result whereas the bunch length for \unit[1]{pC} is 10\% smaller. In comparison to perturbation theory at first order
in $\Delta p/p$ the final bunch profile at second order agrees much better with the simulations. The first order contribution has
the trend to underrate the final bunch length. This is corrected by the additional $T_{566}$ contribution having the opposite sign
as the $R_{56}$ term.
\begin{figure}[t!]
\centering
\subfloat[final bunch profile with {$Q_b=\unit[3]{nC}$} and {$\sigma_s=\unit[214]{fs}$} obtained with $R_{56}$ of
\eqref{eq:transfer-matrices-r56} and $T_{566}$ of \eqref{eq:computation-566-from-r56}]
{\label{fig:bunch-after-chicane-r56-t566-3nC}\includegraphics[scale=0.5]{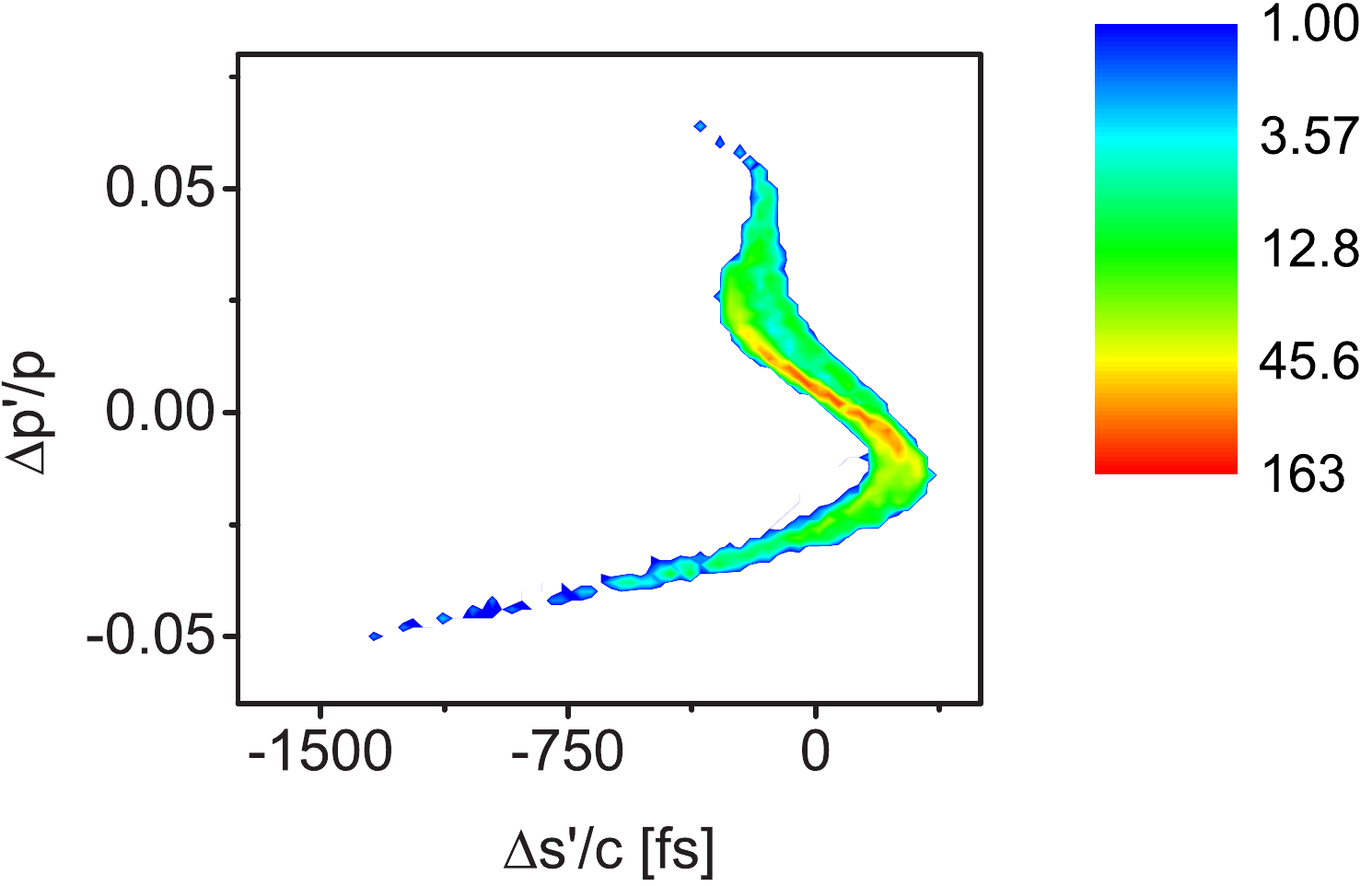}}\hspace{1.0cm}
\subfloat[the same as \protect\subref{fig:bunch-after-chicane-r56-t566-3nC} with {$Q_b=\unit[1]{pC}$} and {$\sigma_s=\unit[11]{fs}$}]{\includegraphics[scale=0.5]{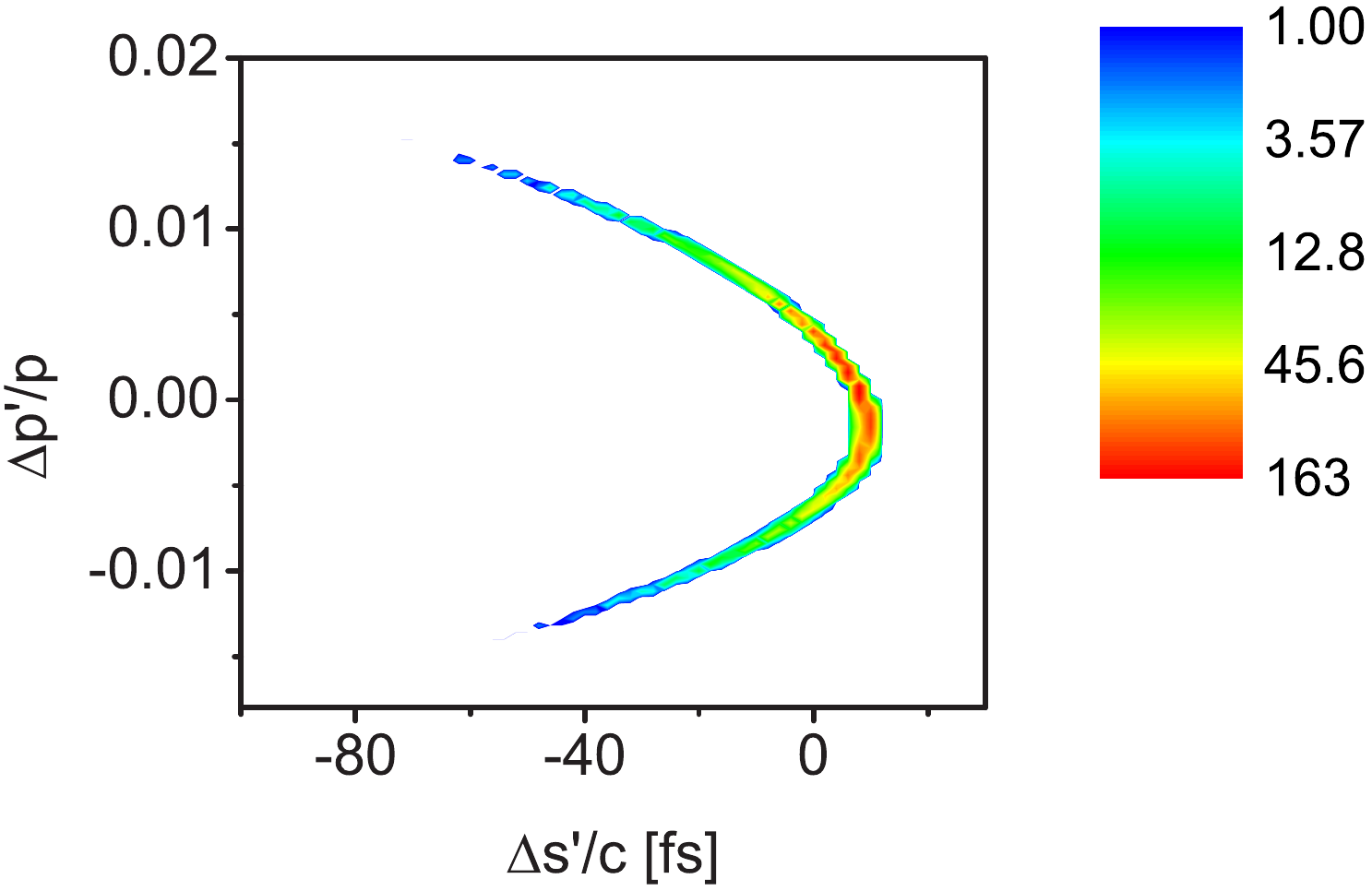}}
\caption{Longitudinal phase space plots of simulated \unit[3]{nC} and \unit[1]{pC} bunches after the chicane. The profiles shown were computed
by using the transfer matrix formalism at second order in $\Delta p/p$.}
\label{fig:bunch-after-chicane-r56}
\end{figure}%

As a consistency check, $T_{566}$ can be computed by using \eqref{eq:composition-third-rank-tensor}. Therefore we need the tensor
coefficients that relate path length differences to differences in angles and the momentum spread. Unfortunately, the respective
coefficients cannot be found in \cite{Brown:1984}. In \cite{Iselin:1992} some of these coefficients are stated, but a computation
according to \eqref{eq:composition-third-rank-tensor} showed that the result of \eqref{eq:computation-566-from-r56} cannot be
obtained with these coefficients alone. For this reason we conclude that the third-rank tensor coefficients given in \cite{Iselin:1992}
that relate path length difference to the five remaining phase space variables are not complete. Therefore they have to be derived
by ourselves.

For the derivation consider \figref{fig:chicane-trajectories-comparison}, which shows the particle trajectories in the first two
dipole magnets of a bunch compressor. Both the FLUTE chicane consisting of rectangular dipole magnets and a hypothetical chicane
of sector dipole magnets is considered. The regions where path length differences at second order in $\delta$ occur are encircled.
The method is to extract the relevant coefficients from the trajectories, i.e., from the solution of the equations of motion. It can
deliver results quite fast provided that the solution is on hand, which is the case here. Note that if the exact solutions are not
available the technique of Lie algebraic maps is more suitable \cite{Iselin:1992,Fartoukh:1997}. However, we will not follow the
latter approach in this paper.

The following two sections are rather technical. Readers who are only interested in the results may skip them and look at
\tabref{tab:path-length-contributions} where the results for rectangular dipole magnets are summarized.

\subsubsection{Rectangular dipole magnets}
\label{ssec:rectangular-magnets}

\begin{figure}[t]
\centering
\subfloat[rectangular dipole magnets]{\label{fig:rectangular-trajectories}\includegraphics[scale=1]{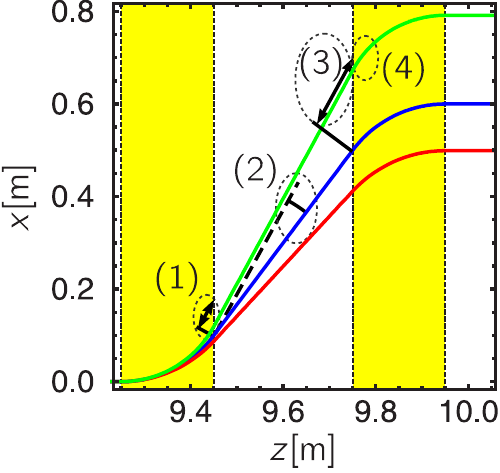}}\hspace{1cm}
\subfloat[sector dipole magnets]{\label{fig:sector-trajectories}\includegraphics[scale=1]{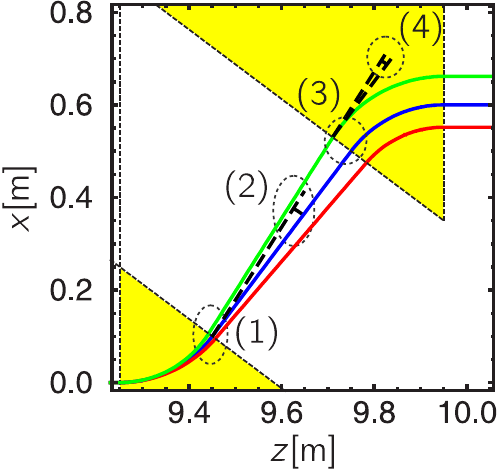}}
\caption{Particle trajectories inside the first two chicane magnets. The reference trajectory for a particle momentum of $p=\unit[40.66]{MeV}$
is shown in blue. The green trajectory is that for a particle with lower momentum $p+\Delta p=(1-0.09)p$. The particle travelling along the
red trajectory has a higher momentum $p+\Delta p=(1+0.09)p$. The left panel shows the trajectories in a chicane consisting of rectangular
dipole magnets with ${L_{\mathrm{mag}}=\unit[0.2]{m}}$, ${L_{\mathrm{space}}=\unit[0.3]{m}}$, $L_{\mathrm{drift}}=\unit[1.0]{m}$ and the
hypothetical bending radius $R=\unit[0.25]{m}$. The latter exaggerated value has been chosen such that the difference in the path lengths becomes
visible. The right panel shows the trajectories in a chicane of sector bending magnets. Here the chicane parameters are chosen such that the path
length of the reference trajectory is equal to the path length of the corresponding trajectory in the left panel. The regions where path length
differences proportional to $\delta^2$ originate from are encircled and marked by (1), (2), (3), and (4). The yellow areas show the dipole
magnets.}
\label{fig:chicane-trajectories-comparison}
\end{figure}

First of all we consider the rectangular D-shape bunch compressor that is planned for FLUTE (see \figref{fig:rectangular-trajectories}).
This chicane has a mirror symmetry with respect to an axis that is parallel to one of the transverse axes and has a distance
$2L_{\mathrm{mag}}+L_{\mathrm{space}}+L_{\mathrm{drift}}/2$ from the left edge of the first magnet. The path length difference of the chicane
from its start to the symmetry axis mentioned is 1/2 of the result given by \eqref{eq:path-length-difference-chicane-plus-second-order}.
Therefore it is sufficient to consider only the first two magnets. By doing so, we are interested in the origin of the terms that make up
$\Delta L^{(2)}$. The comparison of terms are understood to be based on the assumptions $\beta=v/c=1$ and $\alpha\ll \pi/2$, which will
not be mentioned for every instance.

\begin{itemize}

\item[1)] The first difference in path lengths at order $\delta^2$ comes from region (1) in \figref{fig:rectangular-trajectories}, i.e., it occurs
in the vicinity of the exit face of the first dipole magnet. Computing the difference in path length within the magnet as a function of
$\delta$ results in:
\begin{equation}
\label{eq:path-length-difference-rectangular}
\Delta L_{\mathrm{rec}}=R(\alpha-\tan\alpha)\delta+\frac{R}{2}(\tan^3\alpha)\,\delta^2+\mathcal{O}(\delta^3)\,.
\end{equation}
Note that the first order term in $\delta$ corresponds to the element $R_{56}$ of the sector dipole matrix of \eqref{eq:matrix-sector-dipole}
with $n=0$ (neglecting magnetic field inhomogeneities). The only difference is the occurrence of $\tan\alpha$ instead of $\sin\alpha$.
However both functions coincide for bending angles $\alpha\ll \pi/2$, which is the case for the FLUTE chicane. The second term of
\eqref{eq:path-length-difference-rectangular} then leads to
\begin{equation}
\Delta L^{(2)}_1=\Delta L_{\mathrm{rec}}^{(2)}=\frac{1}{2}\left(\frac{L_{\mathrm{mag}}}{R}\right)^2L_{\mathrm{mag}}\,\delta^2+\mathcal{O}\left[(L_{\mathrm{mag}}/R)^4\right]\,.
\end{equation}
Performing the analogue computation for a sector magnet we obtain the following result for the path length difference:
\begin{equation}
\label{eq:path-length-difference-sector}
\Delta L_{\mathrm{sec}}=R(\alpha-\sin\alpha)\delta-\frac{R}{6}(\sin^3\alpha)\,\delta^3+\mathcal{O}(\delta^4)\,.
\end{equation}
Contrary to \eqref{eq:path-length-difference-rectangular} there is no term proportional to $\delta^2$. Therefore the path length
difference at second order in $\delta$ in \eqref{eq:path-length-difference-rectangular} is not related to the body of the magnet. That
is why the magnet fringe must deliver a contribution to the path length difference proportional to $\delta^2$. This is described
by a tensor coefficient $T_{566}$ whose value can be obtained from \eqref{eq:path-length-difference-rectangular}:
\begin{equation}
T_{566}^{\text{exit fringe}}=\frac{1}{2h}\tan^3\alpha\,.
\end{equation}
Such a coefficient should be taken into account for the exit fringe of a rectangular dipole magnet with curvature $h=1/R$ and bending
angle $\alpha$.
\item[2)] The path length difference in the region between the first and second dipole magnet has two main contributions at second order in
$\delta$. The origin of the first contribution is given by region (2) in \figref{fig:rectangular-trajectories}. It is related to
the exit angle of the first dipole magnet with respect to the reference trajectory as a function of $\delta$. The latter results from
the scalar product of the respective tangent vectors $\widehat{\mathbf{t}}$ of the trajectories at the magnet exit:
\begin{align}
\label{eq:angles-tangent-vectors}
\Delta\phi&=\arccos\big\{\widehat{\mathbf{t}}[R+\Delta R,\alpha(R+\Delta R)]\cdot \widehat{\mathbf{t}}[R,\alpha(R)]\big\} \notag \\
&=(\tan\alpha)\delta-\tan\alpha\left(\frac{3+\cos(2\alpha)}{4\cos^2\alpha}\right)\delta^2+\mathcal{O}(\delta^3)\,.
\end{align}
From \eqref{eq:angles-tangent-vectors} we can read off the following transfer matrix and third-rank tensor coefficients that relate
$\Delta\phi$ to the momentum spread and to its square, respectively:
\begin{subequations}
\begin{align}
\label{eq:fringe-matrix-momentum-to-angle}
&R_{26}^{\text{exit fringe}}=\tan\alpha\,, \\
\label{eq:fringe-tensor-momentum-to-angle}
&T_{266}^{\text{exit fringe}}=-\tan\alpha\left(\frac{3+\cos(2\alpha)}{4\cos^2\alpha}\right)=-\tan\alpha+\mathcal{O}(\alpha^3)\,.
\end{align}
\end{subequations}
At the exit of the first dipole magnet the momentum spread $\delta$ is translated to an angle $\Delta\phi$ via
\eqref{eq:angles-tangent-vectors}. This is a contribution at first order perturbation theory in $\delta$. The path length
difference between two drifts that enclose an angle $\Delta\phi$ is of second order in this angle. That is why the aforementioned
$\Delta\phi$ then leads to a second order path length difference in the drift behind the first dipole magnet. Using the first term
of \eqref{eq:angles-tangent-vectors} we obtain:
\begin{align}
\label{eq:path-length-difference-2}
\Delta L^{(2)}_2&=\frac{L_{\mathrm{space}}}{\cos(\alpha+\Delta\phi)}-\frac{L_{\mathrm{space}}}{\cos\alpha}=\frac{(\Delta\phi^{(1)})^2}{2}\frac{L_{\mathrm{space}}}{\cos\alpha}(1+2\tan^2\alpha)+\mathcal{O}(\delta^3)= \notag \\
&=\frac{1}{2}\left(\frac{L_{\mathrm{mag}}}{R}\right)^2L_{\mathrm{space}}\,\delta^2+\mathcal{O}\left[(L_{\mathrm{mag}}/R)^4\delta^2,\delta^3\right]\,.
\end{align}
The latter equation relates a path length difference of a drift to the square of an angle with respect to the reference particle. This is
why it will be described by a product $T_{522}R_{26}^2$ where $R_{26}$ is given by \eqref{eq:fringe-matrix-momentum-to-angle}.
The coefficient $T_{522}$ must be that of a drift but these are not listed in \cite{Brown:1984,Iselin:1992}. However they are contained
in the \verb|MAD-X| \verb|Fortran| programming code \cite{Deniau:2013} and are given by:
\begin{equation}
\label{eq:coefficients-drift-madx}
T_{126}^{\mathrm{drift}}=\frac{L}{2\beta}=T_{162}^{\mathrm{drift}}=T_{346}^{\mathrm{drift}}=T_{364}^{\mathrm{drift}}=T_{522}^{\mathrm{drift}}=T_{544}^{\mathrm{drift}}\,,
\end{equation}
with $\beta=v/c$ and the length $L$ of the drift space. We see that for $L=L_{\mathrm{space}}$ the product
$T_{522}^{\mathrm{drift}}(R_{26}^{\text{exit fringe}})^2$ corresponds to \eqref{eq:path-length-difference-2}.
\item[3)] The second contribution for path length differences proportional to $\delta^2$ in the drift space behind the first dipole magnet
is related to region (3) in \figref{fig:rectangular-trajectories}. A trajectory enclosing an angle $\Delta\phi$ with the reference
trajectory has an additional length $\Delta L$ within the drift space because the trajectory encloses a nonzero angle with the
entrance edge of the second dipole magnet. With the second order term in \eqref{eq:angles-tangent-vectors} we obtain a second order
correction to the path length difference with respect to the momentum spread $\delta$.
\begin{align}
\label{eq:path-length-difference-3}
\Delta L^{(2)}_3&=\frac{L_{\mathrm{space}}}{\cos(\alpha+\Delta\phi)}-\frac{L_{\mathrm{space}}}{\cos\alpha}=\frac{L_{\mathrm{space}}}{\cos\alpha}\tan\alpha|\Delta\phi^{(2)}|+\mathcal{O}(\delta^3) \notag \displaybreak[0]\\
&=\frac{L_{\mathrm{space}}}{\cos\alpha}\left(\frac{3+\cos(2\alpha)}{4\cos^2\alpha}\right)(\tan\alpha)^2\,\delta^2+\mathcal{O}(\delta^3) \notag \displaybreak[0]\\
&=\left(\frac{L_{\mathrm{mag}}}{R}\right)^2L_{\mathrm{space}}\,\delta^2+\mathcal{O}\left[(L_{\mathrm{mag}}/R)^4\delta^2,\delta^3\right] \notag \displaybreak[0]\\
&=2\Delta L^{(2)}_2+\mathcal{O}\left[(L_{\mathrm{mag}}/R)^4\delta^2,\delta^3\right]\,.
\end{align}
Since \eqref{eq:path-length-difference-3} involves a second order angle the coefficient responsible for this path length contribution
must be of first order, i.e., an $R_{52}$. As it is related to the fringe of a rectangular dipole magnet we obtain:
\begin{equation}
\Delta L=(L\tan \alpha)\Delta\phi \Rightarrow R_{52}^{\text{entr. fringe}}=L\tan\alpha\,,
\end{equation}
where $L$ is the length of the drift space before the respective dipole magnet. With $L=L_{\mathrm{space}}$ the product
$(R_{52}^{\text{entr. fringe}})(T_{266}^{\text{exit fringe}})$ is equal to the result of \eqref{eq:path-length-difference-3}.
The sign of the angle in $T_{266}^{\text{exit fringe}}$ of \eqref{eq:fringe-tensor-momentum-to-angle} has to be chosen as
negative in the first magnet leading to the correct overall sign.
\item[4)] Finally, we end up with region (4) in \figref{fig:rectangular-trajectories} leading to a second order correction that corresponds to
the correction of region (1):
\begin{equation}
\Delta L^{(2)}_4=\Delta L^{(2)}_1=\frac{1}{2}\left(\frac{L_{\mathrm{mag}}}{R}\right)^2L_{\mathrm{mag}}\,\delta^2+\mathcal{O}\left[(L_{\mathrm{mag}}/R)^4\right]\,.
\end{equation}
Here it is related to the entrance fringe of the second dipole magnet. So it can only come from
\begin{equation}
T_{566}^{\text{exit fringe}}=\frac{1}{2h}\tan^3\alpha\,.
\end{equation}

\end{itemize}
Summing up $\Delta L^{(2)}_i$ for $i=1$ $\dots$ 4 and multiplying the result by 2 leads to $\Delta L^{(2)}$ of
\eqref{eq:path-length-difference-chicane-plus-second-order}.

\subsubsection{Sector dipole magnets}

In the current section we are interested in the path length difference at second order in $\delta$ for the hypothetical bunch compressor
made up of sector dipole magnets (see \secref{ssec:sector-chicane}). This further example will be studied for academic reasons
to understand the differences to the D-shape chicane of rectangular magnets. The path length difference at first and second order in
$\delta$ is given by:
\begin{equation}
\label{eq:path-length-difference-sector}
\Delta L=-2\alpha^2\left(\frac{2}{3}R\alpha+L_{\mathrm{space}}\right)\delta+\alpha^2\left(2R\alpha+3L_{\mathrm{space}}\right)\delta^2+\mathcal{O}(\alpha^4,\delta^3)\,.
\end{equation}
From the previous equation we can extract the third-rank tensor element for this chicane relating the momentum spread to
path length difference:
\begin{equation}
T_{566}=\alpha^2\left(2R\alpha+3L_{\mathrm{space}}\right)=-\frac{3}{2}R_{56}\,.
\end{equation}
We see that this is related to the matrix element $R_{56}$ in the same manner as for the chicane of rectangular magnets.
Analogous to \secref{ssec:rectangular-magnets} we now intend to derive the respective third-rank tensor coefficients for sector
dipole magnets such that this result can be reproduced.

For the sector chicane we were also able to identify four regions where path length differences originate from (see
\figref{fig:sector-trajectories}). As we saw in \eqref{eq:path-length-difference-sector} there is no path length difference
$\Delta L$ in a sector dipole magnet at second order in the normalized momentum spread $\delta$.  The major part of $\Delta L^{(2)}$
emerges at the second dipole magnet. Because of transverse displacements $\Delta x$, which emerge at several places, a particle
travels an approximate path length $(R+\Delta x)\alpha$ resulting in $\Delta L=\alpha\Delta x$.
\begin{itemize}

\item[1)] The first angular displacement $\Delta x_1$ already appears at the exit fringe of the first dipole magnet, i.e., at
region (1) in \figref{fig:sector-trajectories}. It is given by:
\begin{equation}
\label{eq:angular-displacement-1}
\Delta x_1=2R\sin^2\left(\frac{\alpha}{2}\right)\,\delta+\frac{1}{2}(R\sin^2\alpha)\,\delta^2+\mathcal{O}(\delta^3)\,.
\end{equation}
As indicated, this displacement leads to a longer path length in the second dipole magnet. Its contribution at second order in $\delta$
is:
\begin{equation}
\label{eq:path-length-from-transverse-displacement}
\Delta L^{(2)}_1=\Delta x^{(2)}_1 \alpha=\frac{1}{2}R\alpha\sin^2\alpha\,\delta^2=\frac{1}{2}R\alpha^3\,\delta^2+\mathcal{O}(\alpha^5)\,.
\end{equation}
From \eqref{eq:angular-displacement-1} we extract the respective transfer coefficients relating the first transverse coordinate with
the momentum spread:
\begin{equation}
R_{16}^{\mathrm{sec}}=2R\sin^2\left(\frac{\alpha}{2}\right)=R(1-\cos\alpha)\,,\quad T_{166}^{\mathrm{sec}}=\frac{1}{2}(R\sin^2\alpha)\,.
\end{equation}
Neglecting magnetic field homogeneities we obtain from \eqref{eq:matrix-sector-dipole} that $R_{51}^{\mathrm{sec}}=-\sin(\alpha)/\beta$
with $\beta=v/c$. For $\alpha\ll \pi/2$ and $\beta=1$ the product $(R_{51}^{\mathrm{sec}})(T_{166}^{\mathrm{sec}})$ corresponds to the
result of \eqref{eq:path-length-from-transverse-displacement}.
\item[2)] Any particle with normalized momentum spread $\delta$ exits the first dipole magnet with an angle $\Delta\phi$ with respect to
the reference particle:
\begin{equation}
\label{eq:angle-after-first-sector}
\Delta\phi=(\sin\alpha)\delta+(\sin\alpha)\delta^2+\frac{1}{12}[13-\cos(2\alpha)](\sin\alpha)\,\delta^3+\mathcal{O}(\delta^4)\,.
\end{equation}
From the latter equation we obtain:
\begin{equation}
\label{eq:sector-angle-momentum-spread}
R_{26}^{\mathrm{sec}}=\sin\alpha\,,\quad T_{266}^{\mathrm{sec}}=\sin\alpha\,.
\end{equation}
There is one contribution to $\Delta L$ at second order in $\delta$ that coincides with $\Delta L^{(2)}_2$ obtained for the
rectangular dipole magnet. Consider region (2) in the drift space between the first two magnets. A particle propagating along a trajectory
that encloses an angle $\Delta \phi$ with the reference trajectory travels a different path length at second order in $\delta$. It involves
the first order contribution of the angle $\Delta\phi$ of \eqref{eq:angle-after-first-sector}:
\begin{align}
\label{eq:angular-displacement-2}
\Delta L^{(2)}_2&=\frac{L_{\mathrm{space}}}{\cos(\alpha+\Delta\phi)}-\frac{L_{\mathrm{space}}}{\cos\alpha}=\frac{(\Delta\phi^{(1)})^2}{2}\frac{L_{\mathrm{space}}}{\cos\alpha}(1+2\tan^2\alpha)+\mathcal{O}(\delta^3)= \notag \\
&=\frac{(1+2\tan^2\alpha)\sin^2\alpha}{2\cos\alpha}L_{\mathrm{space}}\,\delta^2+\mathcal{O}(\delta^3)=\frac{1}{2}\alpha^2L_{\mathrm{space}}\,\delta^2+\mathcal{O}(\delta^3,\alpha^4)\,.
\end{align}
Using $T_{522}^{\mathrm{drift}}=L_{\mathrm{space}}/(2\beta)$ of \eqref{eq:coefficients-drift-madx} the product $T_{522}^{\mathrm{drift}}(R_{26}^{\mathrm{sec}})^2$
equals \eqref{eq:angular-displacement-2}.
\item[3)] Due to the second order contribution of $\Delta\phi$ the drift space between the first two magnets leads to a further transverse displacement
at the entrance of the second dipole magnet. This corresponds to region (3) in \figref{fig:sector-trajectories} and the displacement reads:
\begin{equation}
\Delta x_2=\frac{L_{\mathrm{space}}}{\cos\alpha}\tan \Delta\phi=\frac{L_{\mathrm{space}}}{\cos\alpha}\Delta\phi^{(2)}+\mathcal{O}(\delta^3)=\frac{L_{\mathrm{space}}}{\cos\alpha}\sin\alpha\,\delta^2+\mathcal{O}(\delta^3)\,.
\end{equation}
It again translates to a path length difference at second order in $\delta$ analogous to
\eqref{eq:path-length-from-transverse-displacement}:
\begin{equation}
\label{eq:angular-displacement-3}
\Delta L^{(2)}_3=\Delta x^{(2)}_2\alpha=\frac{\alpha\sin\alpha}{\cos\alpha}L_{\mathrm{space}}\,\delta^2=\alpha^2L_{\mathrm{space}}\,\delta^2+\mathcal{O}(\alpha^4)\,.
\end{equation}
\begin{table}[b!]
\centering
\begin{tabular}{ccccc}
\toprule
\multicolumn{2}{c}{Contribution}                                 & Composition                                               & Coefficient & Value \\
\colrule
$\Delta L_1^{(2)}$ & $L_{\mathrm{mag}}^2/(2R)$                   & $T_{566}^{\text{exit fringe}}$                            & $T_{566}^{\text{exit fringe}}$ & $\tan^2(\alpha)/(2h)$ \\
$\Delta L_2^{(2)}$ & $L_{\mathrm{mag}}^2L_{\mathrm{space}}/(2R)$ & $T_{522}^{\mathrm{drift}}(R_{26}^{\text{exit fringe}})^2$ & $T_{522}^{\mathrm{drift}}$ & $L/(2\beta)$ \\
                   &                                             &                                                           & $R_{26}^{\text{exit fringe}}$ & $\tan\alpha$ \\
$\Delta L_3^{(2)}$ & $L_{\mathrm{mag}}^2L_{\mathrm{space}}/R$    & $(R_{52}^{\text{entr. fringe}})(T_{266}^{\text{exit fringe}})$ & $R_{52}^{\text{entr. fringe}}$ & $L\tan\alpha$ \\
                   &                                             &                                                                & $T_{266}^{\text{exit fringe}}$ & $-\tan\alpha$ \\
$\Delta L_4^{(2)}$ & $L_{\mathrm{mag}}^3/(2R)$ & $T_{566}^{\text{exit fringe}}$ & $T_{566}^{\text{exit fringe}}$                  & $\tan^3(\alpha)/(2h)$ \\
\botrule
\end{tabular}
\caption{Path length differences at second order in $\delta$ for the chicane consisting of rectangular dipole magnets. The first two
columns show the contribution to the path length difference. The third column presents how each contribution can be expressed via the
transfer matrix and third-rank tensor coefficients. The last two columns list the individual matrix and tensor coefficients plus their
specific values.}
\label{tab:path-length-contributions}
\end{table}%
Computing the product $(R_{51}^{\mathrm{sec}})(R_{12}^{\mathrm{drift}})(T_{266}^{\mathrm{sec}})$ with $R_{51}^{\mathrm{sec}}=-\sin(\alpha)/\beta$,
$R_{12}^{\mathrm{drift}}=L_{\mathrm{drift}}$ (see \eqref{eq:matrix-drift}), and $T_{266}^{\mathrm{sec}}$ of \eqref{eq:sector-angle-momentum-spread}
results in $\Delta L^{(2)}_3$.
\item[4)] The fourth contribution to the whole $\Delta L$ proportional to $\delta^2$ comes from the fact that a particle enters the second magnet
under the angle $\Delta\phi$ with respect to the reference particle. That is marked as region (4) in \figref{fig:sector-trajectories}.
\begin{equation}
\Delta L^{(2)}_4=R(1-\cos\alpha)\Delta\phi^{(2)}=R(1-\cos\alpha)\sin\alpha\,\delta^2=\frac{1}{2}R\alpha^3\,\delta^2+\mathcal{O}(\alpha^5)\,.
\end{equation}
This result agrees with $(R_{52}^{\mathrm{sec}})(T_{266}^{\mathrm{sec}})$ where $T_{266}^{\mathrm{sec}}$ is taken from \eqref{eq:sector-angle-momentum-spread}. The matrix element $R_{52}^{\mathrm{sec}}=-R(1-\cos\alpha)/\beta$ is obtained from \eqref{eq:matrix-sector-dipole} again neglecting field inhomogeneities.

\end{itemize}
Summing up $\Delta L_i^{(2)}$ for $i=1 \dots 4$ and multiplying the result by 2 leads to the second order term in \eqref{eq:path-length-difference-sector}.

The results obtained are summarized in \tabref{tab:path-length-contributions}. We have shown that the path length difference
for both the rectangular and the sector chicane at second order in $\delta$ can be traced back to individual contributions. These may
originate from magnetic fringes, angles with respect to the reference trajectory or transverse displacements. The contributions are made
up of third-rank tensor coefficients or products of transfer matrix elements with tensor coefficients. Each of them must have a structure
``566'' of free indices relating the momentum spread square to a path length difference.

\section{Space charge effects}
\label{sec:space-charge-effects}
\setcounter{equation}{0}

So far, the FLUTE bunch compressor has been considered merely from the geometrical point of view. We investigated how a bunch evolves when
each particle is sent along its own trajectory through the chicane. The results agree well with what is obtained from Astra simulations
with the space charge routine switched off.
Furthermore the FLUTE chicane was examined with the transfer matrix formalism being a well-known tool in accelerator physics. Within this
perturbative method the first order is not sufficient to reproduce the simulation results, but the second order terms in the momentum spread
are necessary. Note that in the analytical calculations performed so far, both space charge effects and the back reaction of CSR on the bunch
were neglected.

The next step lies in taking space charge forces into account, i.e., the mutual interaction of bunch particles due to the attraction and repulsion
by their electromagnetic fields. We thereby follow the procedure described in the fourth chapter of \cite{reiser:2008}. This will be
applied to both the \unit[3]{nC} and the \unit[1]{pC} bunches considered before. As a starting point, the influence of space charge forces
on the bunch will be estimated by simple principles. Every charged particle beam can be considered as a plasma, i.e., as a gas of charged
particles. The space charge forces acting on a particle moving in transverse direction originate from the electric and magnetic fields.
Assuming a uniform, cylindric particle distribution, these forces depend linearly on the transverse coordinate $x$ and they are related to
what is known as the plasma frequency $\omega_p$. The latter is given by
\begin{equation}
\label{eq:plasma-frequency}
\omega_p=\sqrt{\frac{e^2n}{\varepsilon_0\gamma^3m}}\,,\quad n=\frac{Q_b}{e\pi \sigma_x\sigma_y \cdot (2\sigma_s)}\,,
\end{equation}
where $e$ is the elementary charge, $\varepsilon_0$ the vacuum permittivity, $m$ the electron mass, and $\gamma$ is the Lorentz factor of
the bunch. Furthermore, $n$ is the number density of electrons, $Q_b$ the bunch charge, $\sigma_i$ for $i=(x,y)$ is the rms transverse beam
size, and $\sigma_s$ the rms longitudinal bunch length. Note that for the cylinder length we use the double rms longitudinal bunch length
$2\sigma_s$ since $\sigma_s$ is the standard deviation from the mean and, therefore, it is a measure for one half of the width of the distribution.
Electrons in a plasma oscillate with the plasma frequency. While the plasma frequency describes a transverse
oscillation it nevertheless involves the Lorentz factor. The reason is that the relativistic mass and the relativistic electric and magnetic
fields go into the corresponding equation of motion. To get a feeling for the sizes of these values at FLUTE the \unit[3]{nC}
and \unit[1]{pC} bunches from above will be considered, in particular. We are interested in the behavior of the bunches right before
the fourth chicane magnet.\footnote{We obtain the respective distribution with the trajectory method described in \secref{sec:bunch-compression-trajectory}.
Thereby we assume that the change of transverse coordinates is negligible.} Space charge effects are expected to be most important in this
magnet as here the bunch has already been compressed by the largest fraction. The respective characteristic values of these bunches, e.g., the
bunch length are obtained with the trajectory method. Hence, we assume that space charge effects are negligible before the fourth magnet.
The results can be found in \tabref{tab:physical-parameters} and we then obtain:
\begin{table}[b!]
\centering
\begin{tabular}{cccc}
\toprule
Parameter & Unit & $Q_b=\unit[3]{nC}$ & $Q_b=\unit[1]{pC}$ \\
\colrule
$R$                 & m                & 1.006              & 1.135 \\
$B$                 & T                & 0.14               & 0.12 \\
$p$                 & MeV              & 41.2               & 41.2 \\
$\sigma_p$        &                  & $1.9\cdot 10^{-2}$ & $4.8\cdot 10^{-3}$ \\
$\sigma_x$          & m                & $2.4 \cdot 10^{-3}$ & $4.8 \cdot 10^{-4}$ \\
$\sigma_y$          & m                & $2.4\cdot 10^{-3}$ & $4.8 \cdot 10^{-4}$ \\
$\sigma_{p_x}$      &                  & $2.7\cdot 10^{-4}$ & $1.7\cdot 10^{-5}$ \\
$\sigma_{p_y}$      &                  & $2.6\cdot 10^{-4}$ & $1.7\cdot 10^{-5}$ \\
$\widetilde{v}_x/c$ &                  & $2.2\cdot 10^{-2}$ & $1.4\cdot 10^{-3}$ \\
$\widetilde{v}_y/c$ &                  & $2.1\cdot 10^{-2}$ & $1.4\cdot 10^{-3}$ \\
$\sigma_s$          & fs               & 286                 & 32 \\
\botrule
\end{tabular}
\caption{Physical parameters used for the \unit[3]{nC} and the \unit[1]{pC}, respectively, before the fourth chicane magnet (at $z=\unit[11.45]{m}$).
The momentum spread $\Delta p$, the bunch length $L$, and the beam sizes $\sigma_x$, $\sigma_y$ are rms values. The transverse velocities
$\widetilde{v}_x$, $\widetilde{v}_y$ are defined as the velocities corresponding to the rms values of the transverse momentum components $p_x$
and $p_y$, respectively.}
\label{tab:physical-parameters}
\end{table}%
\begin{equation}
n=\left\{\begin{array}{ll}
\unit[6.2\cdot 10^{18}]{1/m^3} & \text{for } \unit[3]{nC}\,, \\
\unit[4.5\cdot 10^{17}]{1/m^3} & \text{for } \unit[1]{pC}\,, \\
\end{array}
\right.\quad \omega_p=\left\{\begin{array}{ll}
\unit[1.9\cdot 10^8]{1/s} & \text{for } \unit[3]{nC}\,, \\
\unit[5.2\cdot 10^7]{1/s} & \text{for } \unit[1]{pC}\,. \\
\end{array}
\right.
\end{equation}
Although these frequencies seem to be very high, they are heavily suppressed by the Lorentz factor --- contrary to a nonrelativistic
plasma with these particle densities.

A characteristic quantity for the behavior of space charge forces in a particle beam is the Debye length $\lambda_D$ being the ratio of the
rms transverse velocity $\widetilde{v}_x$ and the plasma frequency:
\begin{equation}
\lambda_D=\frac{\widetilde{v}_x}{\omega_p}=\sqrt{\frac{\varepsilon_0\gamma^2 k_BT}{e^2n}}\,,
\end{equation}
with Boltzmann's constant $k_B$. The Debye length emerges as a length scale in the Poisson equation of a distribution of charged particles.
It is a measure of the influence that each particle has on the other particles within a plasma. If the Debye length lies in the order of
the beam dimensions, the smeared-out behavior of the particle distribution will be more important than the interaction of single particles.
For a Debye length in the order of the distances between the individual particles the interaction between nearest neighbors will dominate \cite{reiser:2008}.
This may contribute to the effect of emerging of grainy substructures in a bunch whereby microbunching (see \cite{Saldin:2003mk}, amongst others)
is the most prominent of those effects.

Due to the motion of particles a beam can be considered as a thermal distribution. Via $\gamma m\widetilde{v}_x^2=k_BT$ we can then assign
a transverse temperature $T$ to it. Whether we choose $\widetilde{v}_x$ or $\widetilde{v}_y$ as the transverse velocity does not matter if
$(\widetilde{v}_x-\widetilde{v}_y)/\widetilde{v}_x\ll 1$. The latter is the case for the \unit[3]{nC} and the \unit[1]{pC} distribution that
are considered. We then obtain:
\begin{equation}
T=\left\{\begin{array}{ll}
\unit[2.3\cdot 10^8]{K} & \text{for } \unit[3]{nC}\,, \\
\unit[8.8\cdot 10^5]{K} & \text{for } \unit[1]{pC}\,,
\end{array}
\right.\quad
\lambda_D=\left\{\begin{array}{ll}
\unit[3.4\cdot 10^{-2}]{m} & \text{for } \unit[3]{nC}\,, \\
\unit[7.8\cdot 10^{-3}]{m} & \text{for } \unit[1]{pC}\,, \\
\end{array}
\right.
\end{equation}
for the temperature $T$ and the Debye length $\lambda_D$. The average inter-particle distance $l_p$ and the number $N_p$ of particles inside
a sphere with radius $\lambda_D$ is given by:
\begin{equation}
l_p=\left\{\begin{array}{ll}
\unit[5.4\cdot 10^{-7}]{m} & \text{for } \unit[3]{nC}\,, \\
\unit[1.3\cdot 10^{-6}]{m} & \text{for } \unit[1]{pC}\,,
\end{array}
\right.\quad
N_p=\left\{\begin{array}{ll}
9.8\cdot 10^{14} & \text{for } \unit[3]{nC}\,, \\
8.8\cdot 10^{11} & \text{for } \unit[1]{pC}\,. \\
\end{array}
\right.
\end{equation}
We see that the Debye length is one order of magnitude larger than the beam radius (compare to $\sigma_x$ or $\sigma_y$ in \tabref{tab:physical-parameters})
directly before the fourth magnet. Besides, $\lambda_D\gg l_p$ and $N_p\gg 1$. Under these conditions the interaction of a single
particle with other particles due to space charge effects can be described by considering a smooth particle distribution.

Furthermore, due to $k_BT\ll 1$ the transverse beam density profile is expected to be uniform with respect to the radial distance $r$ from
the beam center, i.e., it is assumed to have a sharp radius $r_m$
\cite{reiser:2008}:
\begin{equation}
n(r)=\left\{\begin{array}{ll}
n_0=\mathrm{const.} & \text{for } r\leq r_m\,, \\
0 & \text{for } r>r_m\,. \\
\end{array}
\right.
\end{equation}
Because of the reasons given we intend to describe a particle bunch before the fourth magnet of the FLUTE bunch compressor as a
uniformly charged distribution within one sigma in all three spatial dimensions. Besides, it is assumed to have a sharp edge according
to the latter formula.

Finally, as a measure for the net radial force on particles in a uniform cylindric beam without any external fields the dimensionless
generalized perveance $K$ can be introduced. For $K>0$ the beam particles are pushed outwards in radial direction, which leads to an
increase of the beam radius. For $K<0$ the opposite happens and the beam size becomes smaller. The latter can only occur when there
are particles inside the beam of opposite charge that neutralize themselves. Especially for FLUTE the generalized perveance is given
by:
\begin{equation}
\label{eq:generalized-perveance}
K=\frac{\omega_p^2r_m^2}{2\beta^2c^2}=\left\{\begin{array}{ll}
1.2 \cdot 10^{-6}  & \text{for } \unit[3]{nC}\,, \\
3.5 \cdot 10^{-9} & \text{for } \unit[1]{pC}\,. \\
\end{array}
\right.
\end{equation}
We see that for both types of bunches $K\ll 1$ indicating that space charge forces are expected to be weak. To summarize, all the
previous simple estimates demonstrate that space charge forces are of minor influence right before the fourth chicane magnet. However
one has to keep in mind that this conclusion results from a rough and simple estimate, where external electric and magnetic fields are
neglected. The estimate gives a first idea on the importance of space charge forces within a typical bunch at FLUTE, though.

\begin{figure}[t!]
\centering
\includegraphics{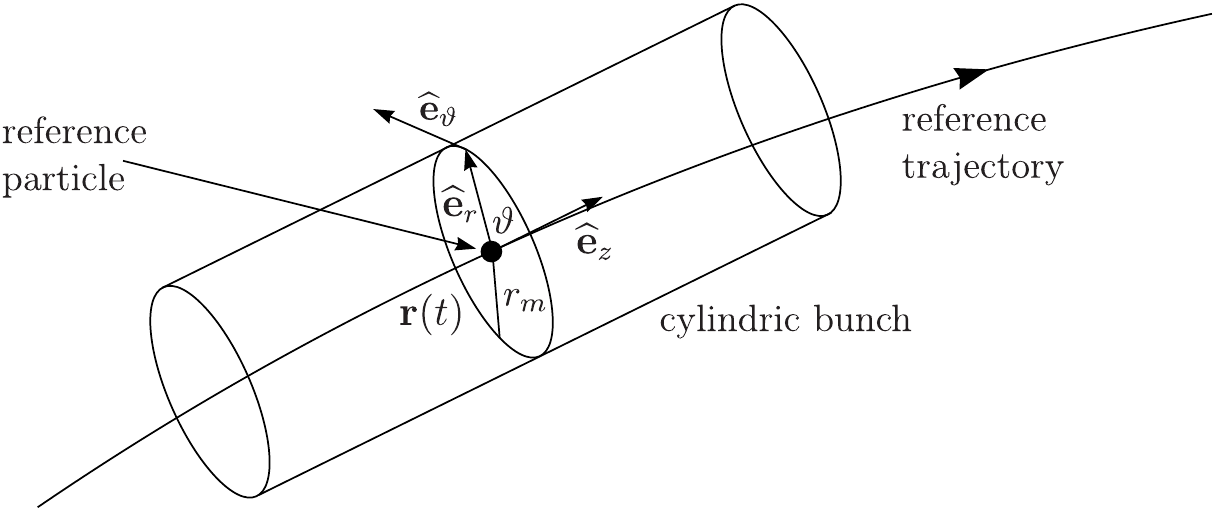}
\caption{Bunch traveling along a reference trajectory parameterized by $\mathbf{r}(t)$. We assume the bunch to be of cylindric shape.
The coordinates of a bunch particle are described by a cylindric, orthogonal coordinate system whose origin corresponds to the position
of the reference particle. The coordinate system is spanned by the basis vectors $\widehat{\mathbf{e}}_r$,
$\widehat{\mathbf{e}}_{\varphi}$, and $\widehat{\mathbf{t}}$. The first points in radial direction, the second in circular direction,
and the third tangentially to the reference trajectory. The beam radius is called $r_m$.}
\label{fig:space-charge-coordinate-system}
\end{figure}%
In what follows, the behavior of a particle bunch inside the FLUTE chicane shall be examined in more detail. To do so we
describe the shape of a particle bunch again by a cylinder (see \figref{fig:space-charge-coordinate-system}).
In general, particles moving inside the beam pipe are subject to the Lorentz force that originates both from internal and external
electromagnetic fields. Internal fields are those that are generated by the charged particles themselves, whereas the external fields
are generated by the accelerator, e.g., cavities, dipole magnets etc. The relativistic equations of motion for an electron moving along
a trajectory $\mathbf{r}(t)$ is given by:
\begin{equation}
\frac{\mathrm{d}}{\mathrm{d}t}(\gamma(t)m\dot{\mathbf{r}})_i=\dot{\gamma}m\dot{\mathbf{r}}_i+\gamma m\ddot{\mathbf{r}}_i
=q(\mathbf{E}+\dot{\mathbf{r}}\times \mathbf{B})_i\,,
\end{equation}
with the Lorentz factor $\gamma$, the electric field vector $\mathbf{E}$, and the magnetic field vector $\mathbf{B}$.
To set up the coordinate system shown in \figref{fig:space-charge-coordinate-system} we need the Frenet trihedron $\{\widehat{\mathbf{t}},\widehat{\mathbf{b}},\widehat{\mathbf{n}}\}$ of a general curve. This is made up of the tangent vector
$\widehat{\mathbf{t}}$, the normal vector $\widehat{\mathbf{n}}$, and the binormal vector $\widehat{\mathbf{b}}$. These vectors
are unit vectors. For their derivatives with respect to time $t$ the Frenet equations hold:
\begin{subequations}
\begin{equation}
\dot{\widehat{\mathbf{t}}}=|\dot{\mathbf{r}}|\kappa\widehat{\mathbf{n}}\,,\quad
\dot{\widehat{\mathbf{n}}}=|\dot{\mathbf{r}}|(\tau\widehat{\mathbf{b}}-\kappa\widehat{\mathbf{t}})\,,\quad
\dot{\widehat{\mathbf{b}}}=-|\dot{\mathbf{r}}|\tau\widehat{\mathbf{n}}\,,
\end{equation}
where $\kappa=\kappa(t)$ is the curvature and $\tau=\tau(t)$ the torsion of the curve:
\begin{equation}
\kappa(t)=\frac{|\dot{\widehat{\mathbf{t}}}(t)|}{|\dot{\mathbf{r}}(t)|}=\frac{|\dot{\mathbf{r}}(t)\times \ddot{\mathbf{r}}(t)|}{|\dot{\mathbf{r}}(t)|^3}\,,\quad \tau(t)=\frac{\left[\dot{\mathbf{r}}(t)\times \ddot{\mathbf{r}}(t)\right]\cdot \dddot{\mathbf{r}}(t)}{|\dot{\mathbf{r}}(t)\times \ddot{\mathbf{r}}(t)|^2}\,.
\end{equation}
\end{subequations}

We now consider the propagation of an electron bunch inside a dipole magnet with constant magnetic field pointing in positive
$y$-direction. We split the trajectories of the bunch particles in the reference trajectory $\mathbf{r}_0(t)$ plus the
coordinates $\mathbf{r}_b(t)$ of each particle with respect to the reference particle:
\begin{equation}
\mathbf{r}(t)=\mathbf{r}_0(t)+\mathbf{r}_b(t)\,.
\end{equation}
The reference particle is supposed to be situated in the center of the bunch. The equations of motion can then be written in the
following form:
\begin{equation}
\frac{\mathrm{d}}{\mathrm{d}t}(\gamma m\dot{\mathbf{r}}_0)+\frac{\mathrm{d}}{\mathrm{d}t}(\gamma m\dot{\mathbf{r}}_b)=q\left[\mathbf{E}+(\dot{\mathbf{r}}_0+\dot{\mathbf{r}}_b)\times \mathbf{B}\right]\,.
\end{equation}
Writing the electric and magnetic field as a sum of an internal and an external contribution according to
\begin{subequations}
\begin{align}
\mathbf{E}&=\mathbf{E}^{\mathrm{int}}+\mathbf{E}^{\mathrm{ext}}\,, \\
\mathbf{B}&=\mathbf{B}^{\mathrm{int}}+\mathbf{B}^{\mathrm{ext}}\,,
\end{align}
\end{subequations}
we obtain:
\begin{align}
\label{eq:equations-of-motion-compact}
\underbrace{\frac{\mathrm{d}}{\mathrm{d}t}(\gamma m\dot{\mathbf{r}}_0)-q(\mathbf{E}^{\mathrm{ext}}+\dot{\mathbf{r}}_0\times \mathbf{B}^{\mathrm{ext}})}_{=0}&+\frac{\mathrm{d}}{\mathrm{d}t}(\gamma m\dot{\mathbf{r}}_b) \notag \\
&=q\left[\mathbf{E}^{\mathrm{int}}+\dot{\mathbf{r}}_b\times (\mathbf{B}^{\mathrm{int}}+\mathbf{B}^{\mathrm{ext}})+\dot{\mathbf{r}}_0\times \mathbf{B}^{\mathrm{int}}\right]\,.
\end{align}
On the left-hand side of the latter equation the equations of motion of the reference particle can be found, which is assumed to
be fulfilled by the trajectory $\mathbf{r}_0$.

We now intend to consider the behavior of the particles that move with a velocity with respect to the reference particle. To
derive the equations of motion, the reference trajectory is needed. In a dipole magnet with a constant magnetic field strength
vector pointing along the positive $y$-axis it holds that
\begin{equation}
\mathbf{r}(t)=R\begin{pmatrix}
\cos(\omega_0t) \\
0 \\
\sin(\omega_0t) \\
\end{pmatrix}\,,\quad \omega_0=\frac{qB}{\gamma m}\,,
\end{equation}
where $\omega_0$ the cyclotron frequency and $B$ the magnetic flux density. For this particular curve the Frenet trihedron is
given by:
\begin{equation}
\widehat{\mathbf{t}}(t)=\begin{pmatrix}
-\sin(\omega_0t) \\
0 \\
\cos(\omega_0t) \\
\end{pmatrix}\,,\quad \widehat{\mathbf{n}}(t)=-\begin{pmatrix}
\cos(\omega_0 t) \\
0 \\
\sin(\omega_0 t) \\
\end{pmatrix}\,,\quad \widehat{\mathbf{b}}(t)=\begin{pmatrix}
0 \\
-1 \\
0 \\
\end{pmatrix}\,,
\end{equation}
and we obtain $\kappa(t)=1/R$, $\tau(t)=0$, $\dot{\kappa}(t)=0$, and $\dot{\tau}(t)=0$ for the curvature, torsion, and their derivatives.
The modulus of the velocity of a bunch particle with respect to the reference particle is
\begin{equation}
\label{eq:velocity-bunch-particles}
v_b\equiv |\dot{\mathbf{r}}_b|=\sqrt{\dot{r}^2+r^2\dot{\vartheta}^2+\dot{z}^2}\,.
\end{equation}
The acceleration $a_b$ yields then
\begin{equation}
\label{eq:acceleration-bunch-particles}
a_b\equiv \frac{\mathrm{d}|\dot{\mathbf{r}_b}|}{\mathrm{d}t}=\frac{\mathrm{d}v_b}{\mathrm{d}t}=\frac{1}{v_b}\left[\dot{r}\ddot{r}+r\dot{r}\dot{\vartheta}^2+r^2\dot{\vartheta}\ddot{\vartheta}+\dot{z}\ddot{z}\right]\,.
\end{equation}
Please note that $v_b\neq v$ where $v$ is the velocity of the reference particle, i.e., $v_b\ll v$. Using this information
the equations of motion for an electron moving inside the magnetic field of a dipole magnet can be obtained where the calculational details 
are relegated to \appref{sec:space-charge-cylindrical-bunch}. They read as follows:
\begin{subequations}
\begin{align}
\dot{\gamma}m&\left(\dot{r}+\frac{v_b}{R}z\cos\vartheta\right)+\gamma m\left[\ddot{r}+\left(\frac{a_b}{R}z+\frac{2v_b}{R}\dot{z}\right)\cos\vartheta-r\left(\dot{\vartheta}^2+\frac{v_b^2}{R^2}\cos^2\vartheta\right)\right] \notag \\
&=-e\left[E^{\mathrm{int}}_r-vB^{\mathrm{int}}_{\vartheta}+\left(\frac{v_b}{R}r\cos\vartheta-\dot{z}\right)(B^{\mathrm{int}}_{\vartheta}+B^{\mathrm{ext}}_{\vartheta})+\left(r\dot{\vartheta}-\frac{v_b}{R}z\sin\vartheta\right)B^{\mathrm{int}}_t\right]\,,
\end{align}
\begin{align}
\dot{\gamma}m&\left(r\dot{\vartheta}-\frac{v_b}{R}z\sin\vartheta\right)+\gamma m\left[r\left(\ddot{\vartheta}+\frac{v_b^2}{2R^2}\sin(2\vartheta)\right)+2\dot{r}\dot{\vartheta}-\left(\frac{2v_b}{R}\dot{z}+\frac{a_b}{R}z\right)\sin\vartheta\right] \notag \\
&=-e\left[E^{\mathrm{int}}_{\vartheta}+vB^{\mathrm{int}}_r+\left(\dot{z}-\frac{v_b}{R}r\cos\vartheta\right)(B^{\mathrm{int}}_r+B^{\mathrm{ext}}_r)-\left(\dot{r}+\frac{v_b}{R}z\cos\vartheta\right)B^{\mathrm{int}}_t\right]\,,
\end{align}
\begin{align}
\dot{\gamma}m&\left(\dot{z}-\frac{v_b}{R}r\cos\vartheta\right)+\gamma m\left[\ddot{z}-\frac{2v_b}{R}\dot{r}\cos\vartheta-\frac{v_b^2}{R^2}z+r\left(\frac{2v_b}{R}\dot{\vartheta}\sin\vartheta-\frac{a_b}{R}\cos\vartheta\right)\right] \notag \\
&=-e\left[E^{\mathrm{int}}_t+\left(\dot{r}+\frac{v_b}{R}z\cos\vartheta\right)(B^{\mathrm{int}}_{\vartheta}+B^{\mathrm{ext}}_{\vartheta})-\left(r\dot{\vartheta}-\frac{v_b}{R}z\sin\vartheta\right)(B^{\mathrm{int}}_r+B^{\mathrm{ext}}_r)\right]\,.
\end{align}
\end{subequations}
Note that no approximations have been made so far, i.e., the latter three equations are exact. Since there is no
external electric field accelerating the particles we use $\dot{\gamma}=0$ and $\dot{v}=0$.

According to \cite{reiser:2008} we introduce dimensionless functions as follows:
\begin{equation}
\label{eq:definition-functions}
r(t)=r_0\varrho(\xi)\,,\quad z(t)=L\zeta(\xi)\,,\quad l=l_0\xi\,,\quad l_0=\frac{r_0}{\sqrt{2K}}\,,\quad K=\frac{eI}{2\pi\varepsilon_0 m(c\beta\gamma)^3}\,.
\end{equation}
Here $r_0$ is the initial radial particle distance to the cylinder axis and $L$ the initial cylinder length,\footnote{With $L$ we mean the full length of the cylinder.}
which both are characteristic length scales of the problem considered. We express the traveled distance $l$ of the bunch via $r_0$ as well.\footnote{This
choice is in accordance with \cite{reiser:2008}; in principle $L$ could also be used.} $K$ is the dimensionless generalized perveance. Taking $\dot{l}=v$
into account with the velocity $v$ of the reference particle, the derivatives of the functions can be expressed via dimensionless derivatives
and the length scales previously introduced. Furthermore, we use the notation $\widetilde{\mathbf{E}}(\xi)\equiv \mathbf{E}(t(\xi))$,
$\widetilde{\mathbf{B}}(\xi)\equiv \mathbf{B}(t(\xi))$, $\widetilde{v}(\xi)\equiv v(t(\xi))$, and $\widetilde{\gamma}(\xi)=\gamma(t(\xi))$
for the respective functions in terms of the dimensionless variable $\xi$. We then obtain:
\begin{subequations}
\begin{align}
\dot{r}&=\frac{\mathrm{d}r}{\mathrm{d}t}=\dot{l}\frac{\mathrm{d}r}{\mathrm{d}l}=\frac{\dot{l}r_0}{l_0}\frac{\mathrm{d}\varrho}{\mathrm{d}\xi}=v\sqrt{2K}\varrho'(\xi)\,,\quad
\ddot{r}=\dot{v}\sqrt{2K}\varrho'(\xi)+2K\frac{v^2}{r_0}\varrho''(\xi)\,, \\[2ex]
\dot{\vartheta}&=\frac{v}{l_0}\varphi'(\xi)=\frac{v}{r_0}\sqrt{2K}\varphi'(\xi)\,,\quad \ddot{\vartheta}=\frac{\dot{v}}{r_0}\sqrt{2K}\varphi'(\xi)+2K\left(\frac{v}{r_0}\right)^2\varphi''(\xi)\,, \\[2ex]
\label{eq:velocity-z}
\dot{z}&=\frac{vL}{l_0}\zeta'(\xi)=v\sqrt{2K}\frac{L}{r_0}\zeta'(\xi)\,,\quad \ddot{z}(t)=\frac{\dot{v}L}{r_0}\sqrt{2K}\zeta'(\xi)+2K\left(\frac{v}{r_0}\right)^2L\zeta''(\xi)\,, \\[2ex]
\dot{v}&=\dot{l}\frac{\mathrm{d}v}{\mathrm{d}l}=\sqrt{2K}\frac{\widetilde{v}}{r_0}\widetilde{v}'(\xi)\,,\quad \dot{\gamma}=\dot{l}\frac{\mathrm{d}\widetilde{\gamma}}{\mathrm{d}l}=\sqrt{2K}\frac{\widetilde{v}}{r_0}\widetilde{\gamma}'(\xi)\,.
\end{align}
\end{subequations}
The dimensionless equations of motion containing the general internal and external electric and magnetic fields can be found in
Eqs.~\ref{eq:space-charge-eq-general-1} -- \ref{eq:space-charge-eq-general-3}.

We now employ the following assumptions for a first simplification of the equations of motion. A cylindric bunch with length $L$,
homogeneous charge $Q=-Q_b$ with $Q_b>0$, and velocity $v\geq 0$ can be associated with the current $I=-I_b=-Q_bv/L$ (with $I_b>0$).
Such a bunch current produces an electric field pointing in radial direction and a magnetic field pointing in circular direction.
They are given by (see, e.g., \cite{reiser:2008}):
\begin{equation}
\label{eq:internal-field-components-important}
\mathbf{E}^{\mathrm{int}}(r)=-\frac{I_b}{2\pi\varepsilon_0v}\frac{r}{r_m^2}\widehat{\mathbf{e}}_r\,,\quad
\mathbf{B}^{\mathrm{int}}(r)=-\frac{\mu_0I_b}{2\pi}\frac{r}{r_m^2}\widehat{\mathbf{e}}_{\varphi}\,,\quad r\leq r_m\,,
\end{equation}
where $\varepsilon_0$ is the vacuum permittivity, $\mu_0$ the vacuum permeability, and $r_m$ the radius of the cylinder. The distance
from the symmetry axis of the bunch is given by $r$. The unit vector pointing in radial direction is $\widehat{\mathbf{e}}_r$
and the unit vector in circular direction is $\widehat{\mathbf{e}}_{\varphi}$. From \eqref{eq:cross-product-velocity-magnetic-field}
we see that the internal fields are mainly involved in the $r$-component of the Lorentz force. With $\dot{\mathbf{r}}_0=v\widehat{\mathbf{t}}$
this leads to the following Lorentz force acting on an electron with charge $q=-e$:
\begin{align}
\label{eq:cancelation-internal-fields}
-e\left(\mathbf{E}^{\mathrm{int}}-\dot{\mathbf{r}}_0\times \mathbf{B}^{\mathrm{int}}\right)_r&=-e(E^{\mathrm{int}}_r-vB^{\mathrm{int}}_{\varphi})=e\left(\frac{I_b}{2\pi\varepsilon_0v}\frac{r}{r_m^2}-v \cdot \frac{\mu_0I_b}{2\pi}\frac{r}{r_m^2}\right) \notag \\
&=\frac{eI_b}{2\pi\varepsilon_0v}\frac{r}{r_m^2}\left(1-\frac{v^2}{c^2}\right)=\frac{eI_b}{2\pi\varepsilon_0v}\frac{r}{r_m^2}\frac{1}{\gamma^2}\,.
\end{align}
Hence, the space charges forces in radial direction that a particle feels in a homogeneous cylindric bunch are suppressed
by a factor $1/\gamma^2$. As a next step we assume that the remaining internal field components are negligible, i.e.,
\begin{equation}
\widetilde{E}^{\mathrm{int}}_{\varphi}=\widetilde{E}^{\mathrm{int}}_t=\widetilde{B}^{\mathrm{int}}_r=\widetilde{B}^{\mathrm{int}}_t=0\,.
\end{equation}
The velocity and acceleration of bunch particles in dimensionless coordinates result from Eqs.~(\ref{eq:velocity-bunch-particles}), (\ref{eq:acceleration-bunch-particles}) and
read as follows:
\begin{subequations}
\begin{align}
&\widetilde{v}_b=\sqrt{2K}\widetilde{v}f[\varrho,\varphi,\zeta]\,,\quad \widetilde{a}_b=\frac{2K\widetilde{v}\widetilde{v}'}{r_0}f[\varrho,\varphi,\zeta]+\frac{2K\widetilde{v}^2}{r_0}g[\varrho,\varphi,\zeta]\,, \displaybreak[0]\\[2ex]
&f[\varrho,\varphi,\zeta]\equiv \sqrt{\varrho'^2+\varrho^2\varphi'^2+(L/r_0)^2\zeta'^2}\,, \displaybreak[0]\\[2ex]
&g[\varrho,\varphi,\zeta]\equiv \frac{\varrho'\varrho''+\varrho\varrho'\varphi'^2+\varrho^2\varphi'\varphi''+(L/r_0)^2\zeta'\zeta''}{\sqrt{\varrho'^2+\varrho^2\varphi'^2+(L/r_0)^2\zeta'^2}}\,.
\end{align}
\end{subequations}
The notation $f=f[\bullet]$, $g=g[\bullet]$ shall indicate that $f$, $g$ contain the functions given as arguments plus additional derivatives
of these respective functions.

We now express the equations of motion solely using dimensionless functions. All physical parameters then do not appear in the
functions or their derivatives any more but in quantities that are denoted as Greek letters. Furthermore these are numbered according
to their order in the differential equations. The differential equation describing the motion of bunch particles in radial direction of
the cylinder in \figref{fig:space-charge-coordinate-system} is given by:
\begin{subequations}
\label{eq:eqom-simplified-1}
\begin{align}
\eta_1\left(\varrho'+\eta_2f\zeta\cos\varphi\right)&+\varrho''+\eta_3\varrho'+\left[\eta_4(\eta_3f+g)\zeta+\eta_5f\zeta'\right]\cos\varphi-\varrho(\varphi'^2+\eta_6f^2\cos^2\varphi) \notag \\
&=\eta_7\left[\eta_8\varrho+\left(\eta_9f\varrho\cos\varphi-\eta_{10}\zeta'\right)(\eta_{11}\varrho-\widetilde{B}_{\varphi}^{\mathrm{ext}})\right]\,,
\end{align}
\begin{align}
\eta_1&=\frac{\widetilde{\gamma}'}{\widetilde{\gamma}}\,,\quad \eta_2=\frac{L}{R}\,,\quad \eta_3=\frac{\widetilde{v}'}{\widetilde{v}}\,,\quad \eta_4=\eta_2\,,\quad \eta_5=2\eta_2\,, \\[1ex]
\eta_6&=\frac{r_0^2}{R^2}\,,\quad \eta_7=\frac{er_0^2}{2K\widetilde{v}^2\widetilde{\gamma}m}\,,\quad \eta_8=\frac{I_b}{2\pi\varepsilon_0 \widetilde{v}\widetilde{\gamma}^2r_m^2}\,, \\[1ex]
\eta_9&=\frac{\sqrt{2K}\widetilde{v}}{R}\,,\quad \eta_{10}=\frac{\sqrt{2K}\widetilde{v}L}{r_0^2}\,,\quad \eta_{11}=\frac{\mu_0I_br_0}{2\pi r_m^2}\,.
\end{align}
\end{subequations}
Note that both $\eta_1$ and $\eta_3$ are exactly equal to zero when the norm of the particle velocity is constant. Quantities containing
only bunch dimensions or velocities are merely related to kinematics, whereas quantities containing the elementary charge $e$ have to do
with space charge forces. Furthermore, the occurrence of the bunch current $I_b$ indicates internal electric and magnetic fields that are generated
by the bunch itself. The parameter $\eta_8$ shows the cancelation of the internal radial electric field and the internal circular magnetic
field proportional to $1/\gamma^2$. This was already indicated in \eqref{eq:cancelation-internal-fields}.

The differential equation describing the circular motion of bunch particles is as follows:
\begin{subequations}
\label{eq:eqom-simplified-2}
\begin{align}
\chi_1(\varrho\varphi'-\chi_2f\zeta\sin\varphi)&+\varrho\left[\varphi''+\chi_3\varphi'+\chi_4f^2\sin(2\varphi)\right]+2\varrho'\varphi'-\left[\chi_5(\chi_3f+g)\zeta+\chi_6f\zeta'\right]\sin\varphi \notag \\
&=\chi_7(\chi_9f\varrho\cos\varphi-\chi_{10}\zeta')\widetilde{B}_{\varrho}^{\mathrm{ext}}\,,
\end{align}
\begin{align}
\chi_1&=\frac{\widetilde{\gamma}'}{\widetilde{\gamma}}\,,\quad \chi_2=\frac{L}{R}\,,\quad \chi_3=\frac{\widetilde{v}'}{\widetilde{v}}\,,\quad \chi_4=\frac{r_0^2}{2R^2}\,,\quad\chi_5=\chi_2\,, \\[1ex]
\chi_6&=2\chi_2\,,\quad \chi_7=\frac{er_0^2}{2K\widetilde{v}^2\widetilde{\gamma}m}\,,\quad \chi_9=\frac{\sqrt{2K}\widetilde{v}}{R}\,,\quad \chi_{10}=\frac{\sqrt{2K}\widetilde{v}L}{r_0^2}\,.
\end{align}
\end{subequations}
Contrary to \eqref{eq:eqom-simplified-1} this equation of motion involves the radial external magnetic field component instead of the circular
one. Furthermore, the internal electric and magnetic fields do not play a role for the circular motion of the particle.

Finally, the differential equation for the motion of the bunch particles in axial direction of the cylindric bunch reads
\begin{subequations}
\label{eq:eqom-simplified-3}
\begin{align}
\psi_1(\zeta'-\psi_2f\varrho\cos\varphi)&+\zeta''+\psi_3\zeta'-\psi_4f^2\zeta-\psi_5f\varrho'\cos\varphi+\varrho\left[\psi_5f\varphi'\sin\varphi-\psi_6(\psi_3f+g)\cos\varphi\right] \notag \\
&=\psi_7\left[(\psi_9\rho'+\psi_{10}f\zeta\cos\varphi)(\psi_{11}\rho-\widetilde{B}_{\varphi}^{\mathrm{ext}})+(\psi_9\rho\varphi'-\psi_{10}f\zeta\sin\varphi)\widetilde{B}_{\varrho}^{\mathrm{ext}}\right]\,,
\end{align}
\begin{align}
\psi_1&=\frac{\widetilde{\gamma}'}{\widetilde{\gamma}}\,,\quad \psi_2=\frac{r_0^2}{RL}\,,\quad \psi_3=\frac{\widetilde{v}'}{\widetilde{v}}\,,\quad \psi_4=\frac{r_0^2}{R^2}\,,\quad \psi_5=2\psi_2\,, \\[1ex]
\psi_6&=\psi_2\,,\quad \psi_7=\frac{er_0^2}{2K\widetilde{v}^2\widetilde{\gamma}m}\,,\quad \psi_9=\frac{\sqrt{2K}\widetilde{v}}{L}\,,\quad \psi_{10}=\frac{\sqrt{2K}\widetilde{v}}{R}\,,\quad \psi_{11}=\frac{\mu_0I_br_0}{2\pi r_m^2}\,.
\end{align}
\end{subequations}
The quantities $\{\eta_1,\dots,\eta_6,\eta_7\times \{\eta_8,B\eta_9,B\eta_{10},\eta_9\eta_{11},\eta_{10}\eta_{11}\}\}$,
$\{\chi_1,\dots,\chi_6,B\chi_7\times \{\chi_9,\chi_{10}\}\}$, and $\{\psi_1,\dots,\psi_6,\psi_7\times \{B\psi_9,B\psi_{10},\psi_9\psi_{11},\psi_{10}\psi_{11}\}\}$
with the modulus of the external magnetic flux density $B$ are dimensionless. The numbering of the coefficients has been performed such that a correspondence
between coefficients of different equations of motion is evident. The first six coefficients of each differential equation are related to the
kinematics; they only involve kinematic quantities such as beam dimensions and velocities. The product of the seventh and eight coefficient
describes the space charge effects due to the internal electric and magnetic field. The fact that no $\chi_8$ appears in \eqref{eq:eqom-simplified-2}
and no $\psi_8$ in \eqref{eq:eqom-simplified-3} demonstrates that this special kind of force does not appear in the circular and the longitudinal
equation of motion.

\subsection{Space charge effects in the FLUTE bunch compressor}
\label{sec:space-charge-effects-flute}

In the calculations of the previous section none of the terms in the equations of motion were neglected a priori. We will now estimate the order
of magnitude of the related quantities for the FLUTE chicane such that they can be compared with each other. First of all, certain physical values, e.g.,
the beam size or the beam current depend on the bunch charge considered. We decided to compare the two extremal cases that were simulated with Astra:
a bunch with the high charge of \unit[3]{nC} and a bunch with the very low charge of \unit[1]{pC}.
\begin{figure}[b!]
\centering
\subfloat[Current for a simulated {\unit[3]{nC}} bunch right before the fourth chicane magnet as a function of the longitudinal coordinate
divided by $c$ (centered on the mean). A binning of {\unit[10]{fs}} has been chosen resulting in {$I_{\mathrm{peak}}=\unit[5250]{A}$}.]{\label{fig:current-3nc}\includegraphics[scale=0.5]{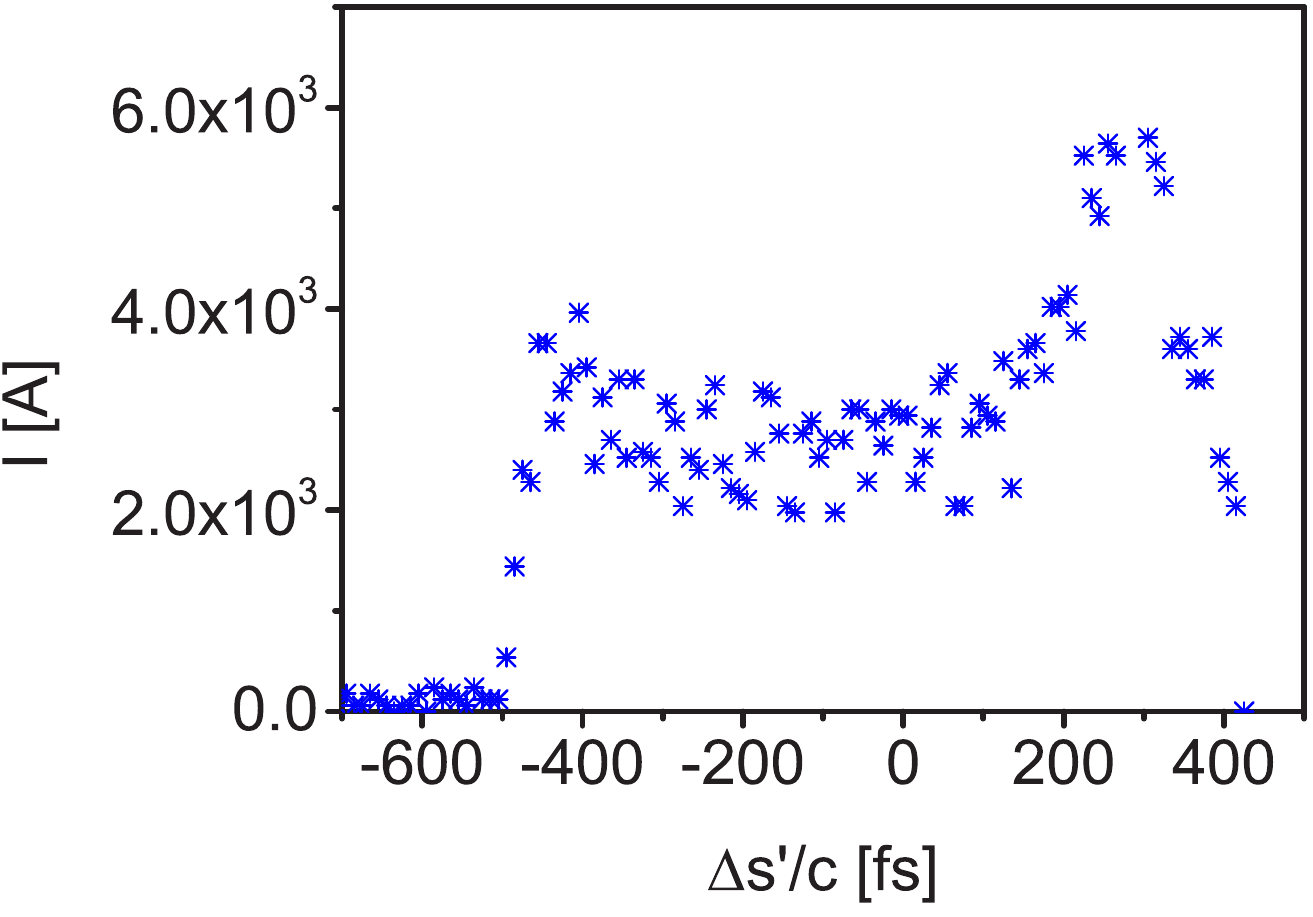}}
\hspace{1cm}
\subfloat[The same as in \protect\subref{fig:current-3nc} for a bunch charge of {\unit[1]{pC}} and a binning of {\unit[2]{fs}} leading to {$I_{\mathrm{peak}}=\unit[16.0]{A}$}.]{\label{fig:current-1pc}
\includegraphics[scale=0.5]{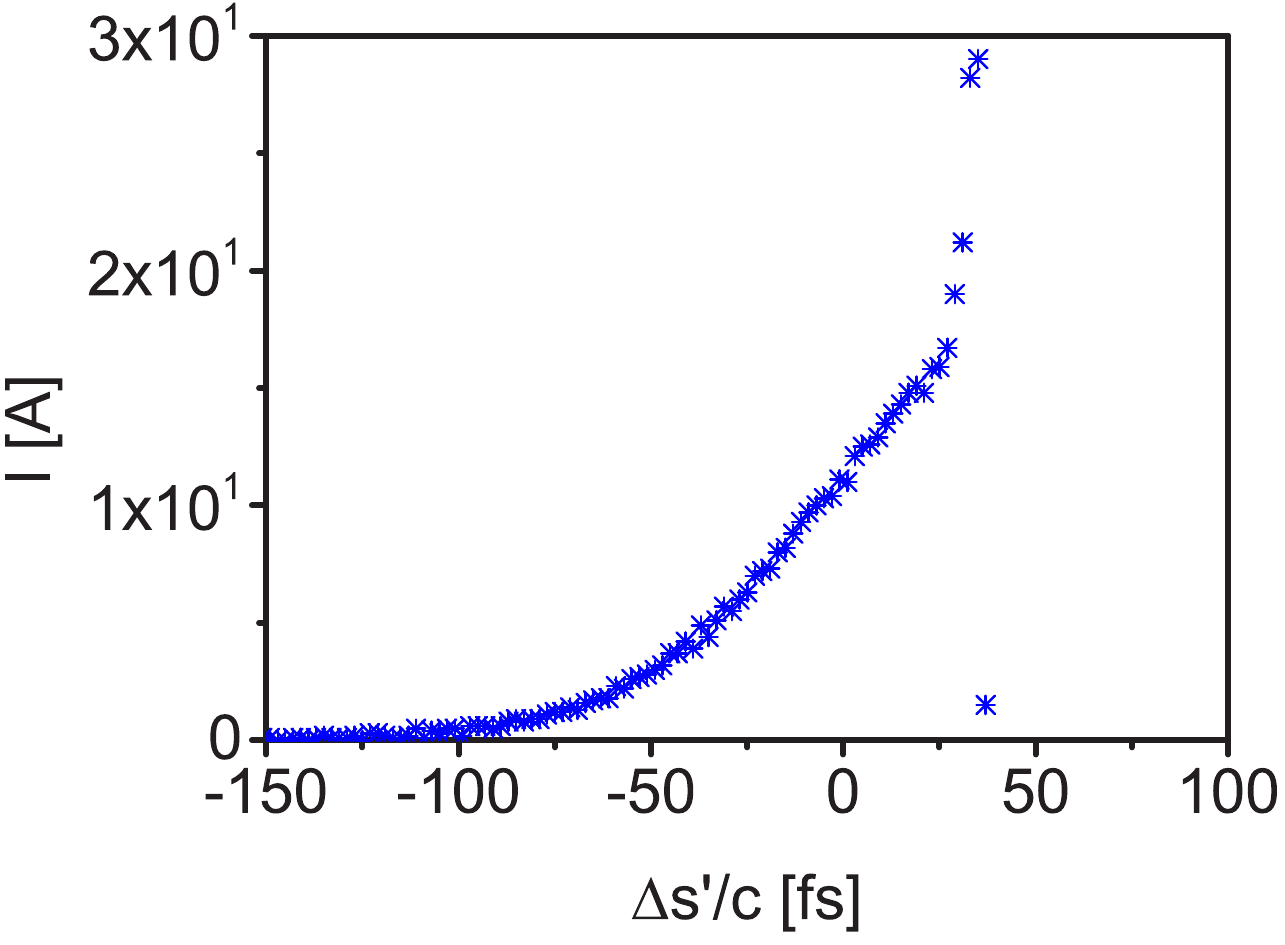}}
\caption{Current for different simulated bunches as a function of the longitudinal coordinate that is understood to be projected on the
$z$-axis.}
\label{fig:bunch-both-bunches}
\end{figure}%

Furthermore one has to keep in mind that the bunch properties are not constant in the chicane. For example during the process of bunch
compression the peak current will increase. That is why as a simple estimate of the behavior of the bunch due to space charge forces we
take the initial values right before the fourth chicane magnet. Another important point is that each bunch is a smeared-out particle distribution.
Hence, it has no sharp edges opposite to the pictorial representation of the cylindric bunch in \figref{fig:space-charge-coordinate-system}.
For this reason we take the respective rms values, e.g., the rms beam size\footnote{According to the charge density $n$ given in
\eqref{eq:plasma-frequency} it makes sense to obtain the beam size as the geometric average of the transverse beam sizes $\sigma_x$ and $\sigma_y$.}
for the radius $r_m$ and two times the rms bunch length $\sigma_s^{\mathrm{(4th)}}$ (before the fourth magnet) for the cylinder length $L$.
\begin{equation}
r_m\equiv \sqrt{\sigma_x\sigma_y}\,,\quad L\equiv 2\sigma_s^{\mathrm{(4th)}}\,.
\end{equation}
The bending radius is chosen from the design values in \cite{Assmann:2013}. The current directly follows from the simulated bunch data using
an appropriate binning (see \figref{fig:bunch-both-bunches}). Such a bunch consists of $N_p=5\cdot 10^4$ macroparticles. Counting the number of
macroparticles inside a bin, multiplying with $Q_b/N_p$ (where $Q_b$ is the bunch charge) and dividing the product by the bin size leads
to the current in terms of the longitudinal coordinate of the bunch. We then define the peak current $I_{\mathrm{peak}}$ of a bunch as
\begin{equation}
I_{\mathrm{peak}}\equiv \frac{Q_b}{L}=\frac{Q_b}{2\sigma_s^{\mathrm{(4th)}}}\,.
\end{equation}
The Alfv\'{e}n current $I_A$ is the maximum current possible for a collimated, cylindrical beam of charged particles
under the influence of space charge effects. It can be written with the characteristic current $I_0$ as follows \cite{diver:2001}:
\begin{equation}
I_A=I_0\beta\gamma\,,\quad I_0=\frac{4\pi\varepsilon_0mc^3}{e}\,.
\end{equation}
The characteristic current is the part of the Alfv\'{e}n current that is not related to the kinematics of the beam. The peak current
of the bunch normalized by $I_0$ approximately corresponds to the Budker parameter $\nu_B$ for relativistic particles
\cite{reiser:2008,diver:2001}. According to the peak current obtained in \figref{fig:bunch-both-bunches} the Budker parameter
is given by
\begin{equation}
\label{eq:budker-parameter}
\nu_B\equiv \frac{I_{\mathrm{peak}}}{I_0\beta}\approx \frac{I_{\mathrm{peak}}}{I_0}=\left\{\begin{array}{ll}
3.1\cdot 10^{-1} & \text{for } \unit[3]{nC}\,, \\
9.2\cdot 10^{-4} & \text{for } \unit[1]{pC}\,. \\
\end{array}
\right.
\end{equation}
We see that for the \unit[1]{pC} bunch at FLUTE the peak current is much smaller than the characteristic current and even more than the
Alv\'{e}n current (because of the Lorentz factor). So we are far away from the regime where the beam may become unstable due to space
charge forces. This is what happens only for currents that lie in the vicinity of $I_A$. However, for the \unit[3]{nC} bunch the peak
current is, indeed, smaller than $I_0$ but not negligibly small. This may have some influence on the treatment of space charge effects
and we will come back to this issue at the end of the current chapter. Note that also a geometrical factor due to the beam shape may
shift the effective Budker parameter, what will not be considered further, though. Using the definition of $\nu_B$ in \eqref{eq:budker-parameter},
the generalized perveance $K$ can also be computed as follows:
\begin{equation}
K=\frac{I_{\mathrm{peak}}}{I_0}\frac{2}{\beta^3\gamma^3}=\frac{2\nu_B}{\beta^2\gamma^3}\,,
\end{equation}
giving values that are in accordance with \eqref{eq:generalized-perveance}.

Bear in mind that the terms in the equations of motion \ref{eq:eqom-simplified-1} -- \ref{eq:eqom-simplified-3} that do not appear
together with a dimensionless physical quantity such as $\eta_2$ are multiplied with 1. In this context also the term including the
prefactor $\eta_7\eta_8=1/2$ is characteristic. We now simplify the equations of motion such that all terms multiplied by a number
much smaller than 1 according to \tabref{tab:physical-parameters-eqom} are neglected. This leads to a set of simplified differential equations
given as follows:
\begin{subequations}
\begin{align}
\varrho''&=\eta_7\left[\eta_8\varrho-\eta_{10}\zeta'(\eta_{11}\varrho-\widetilde{B}^{\mathrm{ext}}_{\varphi})\right]\,, \\[5pt]
\varrho\varphi''+2\varrho'\varphi'&=-\chi_7\chi_{10}\zeta'\widetilde{B}^{\mathrm{ext}}_{\varrho}\,, \\[5pt]
\zeta''&=\psi_7\psi_9\left[\varrho'(\psi_{11}\varrho-\widetilde{B}^{\mathrm{ext}}_{\varphi})+\varrho\varphi'\widetilde{B}^{\mathrm{ext}}_{\varrho}\right]\,.
\end{align}
\end{subequations}
Setting $\widetilde{B}^{\mathrm{ext}}_{\varrho}=\widetilde{B}^{\mathrm{ext}}_{\varphi}=0$, the resulting set of equations holds for the
drift spaces of the FLUTE chicane. In this case the first of these simplified equations of motion partially decouples from the other
two, i.e., the angular variable $\varphi$ does not appear any more. This shows that for mere drifts the circular motion of particles
inside the bunch due to the magnetic fields can be neglected when considering the increase of the transverse beam dimensions.

\begin{figure}[t]
\centering
\includegraphics[scale=1]{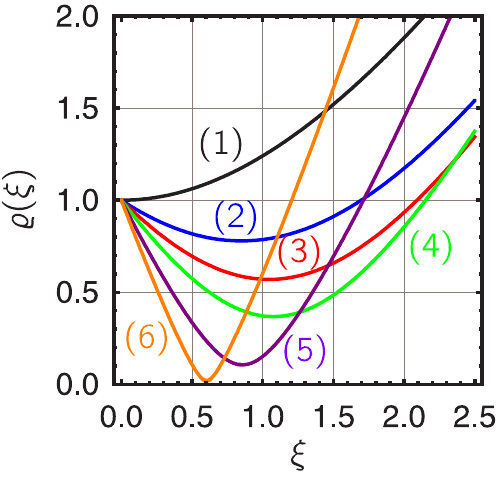}
\caption{Behavior of the dimensionless function $\varrho(\xi)$ defined by \eqref{eq:definition-functions} (amongst others). The different
curves follow by solving \eqref{eq:space-charge-final-dgl-1} for $\zeta'=0$ and $B=0$ numerically for different initial slopes. The curves
are numbered according to the initial conditions: $\varrho'(\xi=0)=(0,-0.5,-0.75,-1.0,-1.5,-2.0)$.}
\label{fig:space-charge-solutions}
\end{figure}

The (constant) external magnetic flux density in the dipole magnet along the positive $y$-direction can be decomposed in a radial
and a circular component:
\begin{equation}
\widetilde{\mathbf{B}}^{\mathrm{ext}}=B\left[(\widehat{\mathbf{e}}_y\cdot \widehat{\mathbf{e}}_{\varrho})\widehat{\mathbf{e}}_{\varrho}+(\widehat{\mathbf{e}}_y\cdot \widehat{\mathbf{e}}_{\varphi})\widehat{\mathbf{e}}_{\varphi}+(\widehat{\mathbf{e}}_y\cdot \widehat{\mathbf{t}})\widehat{\mathbf{t}}\right]=-B\left[\widehat{\mathbf{e}}_{\varrho}\sin\varphi+\widehat{\mathbf{e}}_{\varphi}\cos\varphi\right]\,.
\end{equation}
\begin{table}[t]
\centering
\setlength{\extrarowheight}{2pt}
\begin{tabular}{C{0.05\textwidth}C{0.095\textwidth}C{0.095\textwidth}C{0.095\textwidth}C{0.095\textwidth}C{0.095\textwidth}C{0.095\textwidth}C{0.095\textwidth}C{0.095\textwidth}C{0.075\textwidth}}
\toprule
\multirow{1}{*}{$Q_b$}          & \multirow{1}{*}{$\eta_2$} & \multirow{1}{*}{$\eta_4$} & \multirow{1}{*}{$\eta_5$} & \multirow{1}{*}{$\eta_6$} & \multirow{1}{*}{$\eta_7\widetilde{\eta}_8$} & \multirow{1}{*}{$\eta_7\eta_9B$} & \multirow{1}{*}{$\eta_7\eta_9\widetilde{\eta}_{11}$} & \multirow{1}{*}{$\eta_7\eta_{10}B$} & \multirow{1}{*}{$\eta_7\eta_{10}\widetilde{\eta}_{11}$} \\
\colrule
\multirow{1}{*}{$\unit[3]{nC}$} & \multirow{1}{*}{$1.70\cdot 10^{-4}$} & \multirow{1}{*}{$1.70\cdot 10^{-4}$} & \multirow{1}{*}{$3.41\cdot 10^{-4}$} & \multirow{1}{*}{$5.52\cdot 10^{-6}$} & \multirow{1}{*}{0.5} & \multirow{1}{*}{$3.61\cdot 10^{-3}$} & \multirow{1}{*}{$1.17\cdot 10^{-2}$} & \multirow{1}{*}{0.111} & \multirow{1}{*}{0.361} \\
\multirow{1}{*}{$\unit[1]{pC}$} & \multirow{1}{*}{$1.68\cdot 10^{-5}$} & \multirow{1}{*}{$1.68\cdot 10^{-5}$} & \multirow{1}{*}{$3.36\cdot 10^{-5}$} & \multirow{1}{*}{$1.80\cdot 10^{-7}$} & \multirow{1}{*}{0.5} & \multirow{1}{*}{$2.14\cdot 10^{-3}$} & \multirow{1}{*}{$1.16\cdot 10^{-4}$} & \multirow{1}{*}{0.200} & \multirow{1}{*}{$1.08\cdot 10^{-2}$} \\
\colrule
               & \multirow{1}{*}{$\chi_2$} & \multirow{1}{*}{$\chi_4$} & \multirow{1}{*}{$\chi_5$} & \multirow{1}{*}{$\chi_6$} & \multirow{1}{*}{$\chi_7\chi_9B$} & \multirow{1}{*}{$\chi_7\chi_{10}B$} & & & \\
\colrule
\multirow{1}{*}{$\unit[3]{nC}$} & \multirow{1}{*}{$1.70\cdot 10^{-4}$} & \multirow{1}{*}{$2.76\cdot 10^{-6}$} & \multirow{1}{*}{$1.70\cdot 10^{-4}$} & \multirow{1}{*}{$3.51\cdot 10^{-4}$} & \multirow{1}{*}{$3.61\cdot 10^{-3}$} & \multirow{1}{*}{0.111} & & & \\
\multirow{1}{*}{$\unit[1]{pC}$} & \multirow{1}{*}{$1.68\cdot 10^{-5}$} & \multirow{1}{*}{$9.00\cdot 10^{-8}$} & \multirow{1}{*}{$1.68\cdot 10^{-5}$} & \multirow{1}{*}{$3.36\cdot 10^{-5}$} & \multirow{1}{*}{$2.14\cdot 10^{-3}$} & \multirow{1}{*}{0.200} & & & \\
\colrule
               & \multirow{1}{*}{$\psi_2$} & \multirow{1}{*}{$\psi_4$} & \multirow{1}{*}{$\psi_5$} & \multirow{1}{*}{$\psi_6$} & \multirow{1}{*}{$\psi_7\psi_9B$} & \multirow{1}{*}{$\psi_7\psi_9\widetilde{\psi}_{11}$} & \multirow{1}{*}{$\psi_7\psi_{10}B$} & \multirow{1}{*}{$\psi_7\psi_{10}\widetilde{\psi}_{11}$} \\
\colrule
\multirow{1}{*}{$\unit[3]{nC}$} & \multirow{1}{*}{$3.24\cdot 10^{-2}$} & \multirow{1}{*}{$5.52\cdot 10^{-6}$} & \multirow{1}{*}{$6.48\cdot 10^{-2}$} & \multirow{1}{*}{$3.24\cdot 10^{-2}$} & \multirow{1}{*}{21.2} & \multirow{1}{*}{68.7} & \multirow{1}{*}{$3.61\cdot 10^{-3}$} & \multirow{1}{*}{$1.17\cdot 10^{-2}$} \\
\multirow{1}{*}{$\unit[1]{pC}$} & \multirow{1}{*}{$1.07\cdot 10^{-2}$} & \multirow{1}{*}{$1.80\cdot 10^{-7}$} & \multirow{1}{*}{$2.15\cdot 10^{-2}$} & \multirow{1}{*}{$1.07\cdot 10^{-2}$} & \multirow{1}{*}{128}  & \multirow{1}{*}{6.89} & \multirow{1}{*}{$2.14\cdot 10^{-3}$} & \multirow{1}{*}{$1.16\cdot 10^{-4}$} \\
\botrule
\end{tabular}
\caption{Dimensionless physical parameters as they appear in the equations of motion \ref{eq:eqom-simplified-1} --
\ref{eq:eqom-simplified-3}. Each pair of columns gives the respective parameters plus their values for FLUTE using
\tabref{tab:physical-parameters} and \figref{fig:bunch-both-bunches}.}
\label{tab:physical-parameters-eqom}
\end{table}%
The next important issue to mention is that the internal electric and magnetic field components given by
\eqref{eq:internal-field-components-important} themselves depend on the cylinder radius $r_m$. Since we are interested in the evolution of
$r_m$ as a function of time we cannot take it as a constant. Under the assumption that the particle trajectories are laminar, i.e., they do not
intersect each other it suffices to consider the envelope particles. Because of this we set $r_m=r_0\varrho(\xi)$ with
with $r_0=\sqrt{\sigma_x\sigma_y}$ being the initial radial distance of an envelope particle to the cylinder axis. This procedure is
followed in \cite{reiser:2008} as well and leads to the final system of differential equations
\begin{subequations}
\begin{align}
\label{eq:space-charge-final-dgl-1}%
\varrho''&=\eta_7\left[\frac{\widetilde{\eta}_8}{\varrho}-\eta_{10}\zeta'\left(\frac{\widetilde{\eta}_{11}}{\varrho}+B\cos\varphi\right)\right]\,, \displaybreak[0]\\[4pt]
\label{eq:space-charge-final-dgl-2}%
\varrho\varphi''+2\varrho'\varphi'&=\chi_7\chi_{10}\zeta'B\sin\varphi\,, \displaybreak[0]\\[4pt]
\label{eq:space-charge-final-dgl-3}%
\zeta''&=\psi_7\psi_9\left[\varrho'\left(\frac{\widetilde{\psi}_{11}}{\varrho}+B\cos\varphi\right)-\varrho\varphi'B\sin\varphi\right]\,,
\end{align}
with the following definitions where the cylinder radius $r_m$ corresponds to the initial radial particle distance $r_0$:
\begin{equation}
\widetilde{\eta}_8=\eta_8|_{r_m=r_0}\,,\quad \widetilde{\eta}_{11}=\eta_{11}|_{r_m=r_0}\,,\quad \widetilde{\psi}_{11}=\psi_{11}|_{r_m=r_0}\,.
\end{equation}
\end{subequations}
If in \eqref{eq:space-charge-final-dgl-1} we set the external magnetic field $B$ equal to zero and neglect particle motions along the
$z$-direction of the coordinate system (resulting in $\zeta'=0$) we obtain:
\begin{equation}
\varrho''=\frac{\eta_7\widetilde{\eta}_8}{\varrho}\,.
\end{equation}
This differential equation is discussed at the beginning of the fourth chapter in \cite{reiser:2008}.
The numerical solutions for different initial conditions are presented in \figref{fig:space-charge-solutions}. They correspond to the
plots given in the latter reference, which is a good crosscheck for the method used here. In the figure we see that space charge effects
always blow up the radial beam dimension. If the beam is focused, e.g., by magnetic quadrupoles the beam size first decreases until a certain
minimum value and then it starts increasing again.

The model considered here is more general in the sense that it does not neglect certain effects at the start of the calculations. The
differential equations given by \ref{eq:space-charge-final-dgl-1} -- \ref{eq:space-charge-final-dgl-3} consider the motion of particles
in radial, angular, and $z$-direction with respect to the reference particle. Furthermore external magnetic fields can be taken into account.
The equations for a drift space follow by setting $B=0$. Using the values of \tabref{tab:physical-parameters-eqom}, the system of
differential equations can be solved numerically. This is done for both the \unit[3]{pC} and the \unit[1]{pC} bunch right before the
fourth magnet of the FLUTE bunch compressor.

We intend to solve the system of differential equations for the following initial conditions:
\begin{subequations}
\begin{align}
r(t=0)&=r_0\,,\quad \dot{r}(t=0)=0\,,\quad \vartheta(t=0)=\vartheta_0\,,\quad \dot{\vartheta}(t=0)=0\,, \\
z(t=0)&=L\,,\quad \dot{z}(t=0)=\Delta v\,.
\end{align}
\end{subequations}
The first two conditions mean that the initial beam size is $r_0$ and the change of the beam size vanishes, which makes sense in case
that no focusing or defocusing is taken into account. The subsequent two conditions state that an arbitrary initial angle $\varphi_0$ is chosen
that initially does not change as well. By the fifth condition a head particle is considered and the sixth condition takes the velocity
difference $\Delta v$ of this particle with respect to the reference particle into account. Now these initial conditions have to be translated to
the dimensionless variables.

The first five can be translated directly by using \eqref{eq:definition-functions}. The last one is a bit more involved. Here we first
need the velocity difference $\Delta v$ of the head particle with respect to the reference particle for a bunch traveling through the last
bending magnet. In the following, this difference is assumed to be constant. Let $\sigma_s^{\text{(4th)}}$ be the bunch length
directly before the fourth chicane magnet, $\sigma_s^{\mathrm{(fin)}}$ the final bunch length, and $v$ the (constant) velocity of the reference particle.
It then makes sense to state that both the head and the tail particle will travel half of the distance $\sigma_s^{\mathrm{(fin)}}-\sigma_s^{\text{(4th)}}$
during compression. Such a distance will be traveled in the time period $\Delta t=R\arcsin(L_{\mathrm{mag}}/R)/v$. Then the velocity
difference of the head particle with respect to the reference particle can be obtained as follows:
\begin{equation}
\Delta v=\frac{\sigma_s^{\mathrm{(fin)}}-\sigma_s^{\text{(4th)}}}{2\Delta t}=\frac{(\sigma_s^{\mathrm{(fin)}}-\sigma_s^{\text{(4th)}})v}{2R\arcsin(L_{\mathrm{mag}}/R)}<0\,.
\end{equation}
Now $\Delta v$ has to be expressed via the prefactor in $\dot{z}(t)$ of \eqref{eq:velocity-z}. This then leads to a
dimensionless quantity. Finally we end up with the following dimensionless initial conditions:
\begin{subequations}
\begin{align}
\varrho(\xi=0)&=1\,,\quad \varrho'(\xi=0)=0\,,\quad \varphi(\xi=0)=\varphi_0\,,\quad \varphi'(\xi=0)=0\,, \\[2pt]
\zeta(\xi=0)&=1\,,\quad \zeta'(\xi=0)=\Delta v\left(v\sqrt{2K}\frac{\sigma_s^{\text{(4th)}}}{r_0}\right)^{-1}\,.
\end{align}
\end{subequations}
Via \eqref{eq:definition-functions} the dimensionless variable $\xi$ is related to the dimensionful traveled length $l$. The maximum
traveling length $l_m$ of the reference particle inside the fourth bending magnet connects to the following $\xi_m$:
\begin{equation}
\xi_m\equiv \sqrt{2K}\frac{l_m}{r_0}=\sqrt{2K}\frac{R\arcsin(L_{\mathrm{mag}}/R)}{r_0}\,.
\end{equation}
The bunch lengths for both bunch charges right before the fourth bending magnet are obtained using the particle trajectory described
in \secref{ssec:electron-trajectory}. They are corrected by a factor $1/\cos\alpha$ with the bending angle $\alpha$ since the bunch
length obtained with this procedure is understood to be projected on the longitudinal axis. Finally, for the \unit[3]{nC} bunch we
get with the choice $\varphi_0=1$:
\begin{equation}
\varrho(\xi_m)|_{\unit[3]{nC}}=1.00750\,,\quad \varphi(\xi_m)|_{\unit[3]{nC}}=0.999281\,,\quad \zeta(\xi_m)|_{\unit[3]{nC}}=0.895620\,.
\end{equation}
The corresponding values for the \unit[1]{pC} bunch are given by:
\begin{equation}
\varrho(\xi_m)|_{\unit[1]{pC}}=1.00092\,,\quad \varphi(\xi_m)|_{\unit[1]{pC}}=0.999129\,,\quad \zeta(\xi_m)|_{\unit[1]{pC}}=0.705793\,.
\end{equation}
The dependence of these values on the initial angle were tested as well. For the \unit[1]{pC} bunch the results vary in the per mill
regime, whereas for \unit[3]{nC} the maximum variations are 2\%. Note that the problem is not completely cylindrically symmetric.

How the space charge forces influence bunch compression can be deduced from $\zeta(\xi_m)$. Twice this value corresponds to the amount
of bunch compression if it is assumed that the head particle travels the same distance as the tail particle. So we have
\begin{equation}
\sigma_s|_{\begin{subarray}{l}\unit[3]{nC} \\ \text{with space charge}\end{subarray}}=\left[2\zeta(\xi_m)|_{\unit[3]{nC}}-1\right]\sigma_s^{\text{(4th)}}|_{\unit[3]{nC}}=1.07\sigma_s|_{\unit[3]{nC}}\,,
\end{equation}
for the \unit[3]{nC} bunch and
\begin{equation}
\sigma_s|_{\begin{subarray}{l}\unit[1]{pC} \\ \text{with space charge}\end{subarray}}=\left[2\zeta(\xi_m)|_{\unit[1]{pC}}-1\right]\sigma_s^{\text{(4th)}}|_{\unit[1]{pC}}=1.009\sigma_s|_{\unit[1]{pC}}\,,
\end{equation}
for the \unit[1]{pC} bunch.
\begin{figure}[t!]
\centering
\subfloat[final bunch profile with {$Q_b=\unit[3]{nC}$} and {$\sigma_s=\unit[205]{fs}$} obtained with Astra]{\label{fig:bunch-after-chicane-astra-sc-3nC}\includegraphics[scale=0.5]{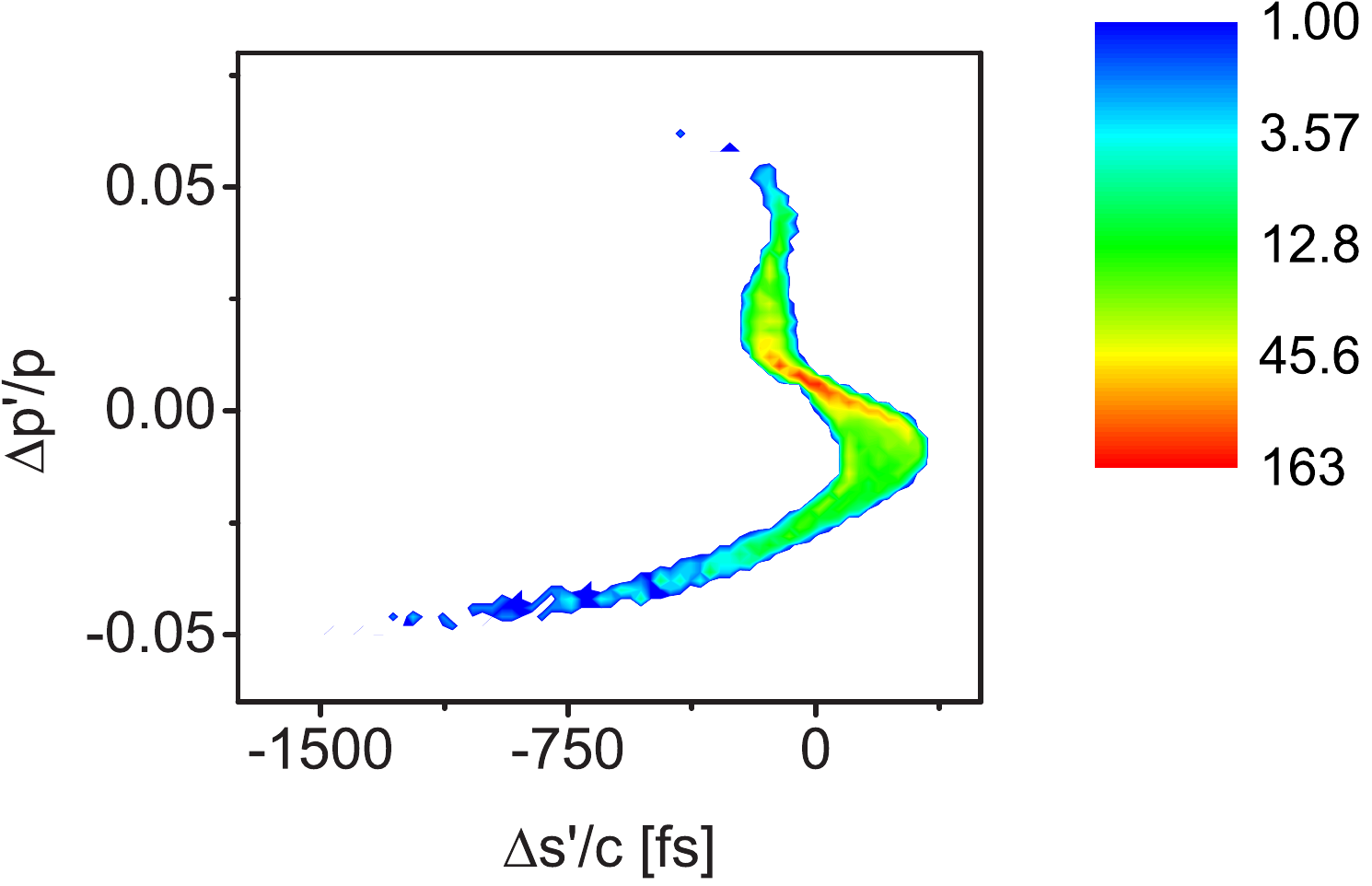}}\hspace{1.0cm}
\subfloat[the same as \protect\subref{fig:bunch-after-chicane-astra-sc-3nC} with {$Q_b=\unit[1]{pC}$} and {$\sigma_s=\unit[13]{fs}$}]{\label{fig:bunch-after-chicane-astra-sc-1pC}\includegraphics[scale=0.5]{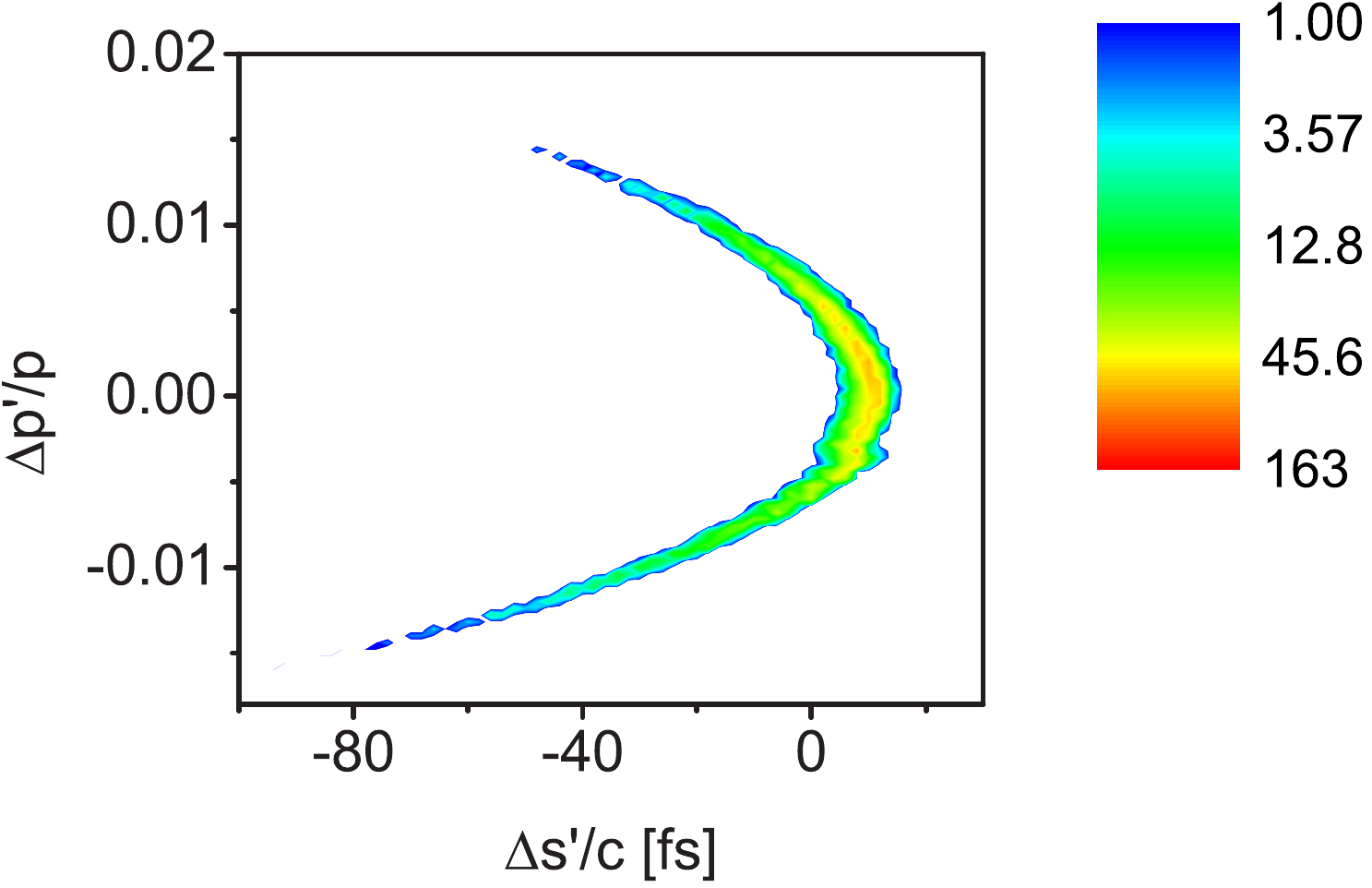}}
\caption{Longitudinal phase space plots of simulated \unit[3]{nC} and \unit[1]{pC} bunches after the chicane. The profiles shown were computed
with the Astra space charge routine.}
\label{fig:bunch-after-chicane-spacecharge}
\end{figure}%

Now we compare these results to the output of the Astra space charge routine that is shown in \figref{fig:bunch-after-chicane-astra-sc-3nC}
for the \unit[3]{nC} bunch and in \figref{fig:bunch-after-chicane-astra-sc-1pC} for the \unit[1]{pC} bunch. In comparison to the Astra
results without space charges the bunch length increases by approximately 2\% for \unit[3]{nC} and 3\% for \unit[1]{pC}. Hence, the
relative increase for the smaller bunch charge is larger. This may be related to the fact that the \unit[1]{pC} bunch is compressed
by an additional factor of 16 compared to the \unit[3]{nC} bunch.

For the simple, analytical model presented in this chapter we see that the increase of the beam size due to bunch compression lies in the
regime of few per mill for both bunch charges, where this increase is larger for the \unit[3]{nC} bunch. Furthermore the bunch length of
the \unit[1]{pC} bunch rises by approximately 1\% in comparison to the case without space charges. It becomes immediately evident that the
bunch length in case of the \unit[3]{nC} bunch is affected much more, i.e., it grows by 7\%, which is not in accordance with the simulations.
This relatively strong increase originates from the large value $\psi_7\psi_9\widetilde{\psi}_{11}=68.7$ in \tabref{tab:physical-parameters-eqom}
that is approximately by a factor of 10 larger compared to the corresponding value of the \unit[1]{pC} bunch. The combination $\psi_7\psi_9\widetilde{\psi}_{11}$
describes the size of the force in the circular magnetic field experienced by an electron moving outwards. This force works against bunch
compression. The bunch length of the \unit[1]{pC} bunch is, indeed, smaller by a factor of 9 versus the \unit[3]{nC} bunch. However note
that the bunch charges differ by a factor of 3000 enlarging the space charge forces for the \unit[3]{nC} bunch.

The behavior of the \unit[3]{nC} bunch indicates that the validness of the simple space charge model presented in this chapter
breaks down for high bunch charges. This may also have to do with the fact that the Budker parameter lies in the vicinity of 1
for this bunch, cf. \eqref{eq:budker-parameter}.

\section{Coherent synchrotron radiation}
\setcounter{equation}{0}

In the previous section we investigated the amount of space charge effects that may play a role for bunch compression in the FLUTE
chicane. The upshot was that the analytical model presented overestimates the space charge forces for the \unit[3]{nC} bunch. The Astra
simulations show that space charge effects are negligible for both the \unit[3]{nC} and the \unit[1]{pC} bunch --- even in
the fourth chicane magnet.

Unlike the space charges, the back reaction of the emitted coherent synchrotron radiation on a bunch plays a major role in the fourth
bending magnet. This is evident from simulations performed with the tool CSRtrack \cite{dohlus:2013} (see \cite{Assmann:2013}).
In this last section of the paper we intend to understand this back reaction better. Therefore we first would like to review the
most important formulas of synchrotron radiation and the properties of coherent radiation.

\begin{figure}[t]
\centering
\includegraphics[scale=1]{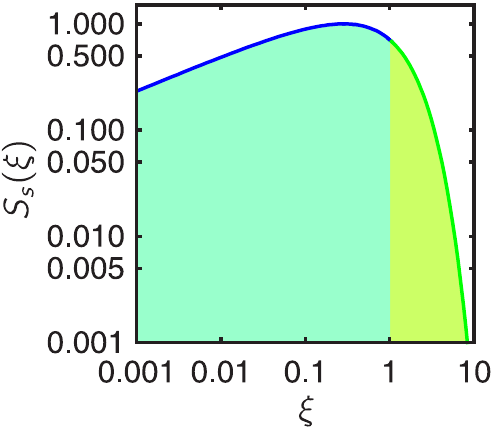}
\caption{Characteristic function $S_s(\xi)$ of the synchrotron radiation spectrum as defined in \eqref{eq:power-spectrum-synchrotron}
shown in a double-logarithmic plot. The function is normalized by its maximum value. The area under the curve is divided by two at
$\xi=1$.}
\label{fig:characteristic-function}
\end{figure}

The power radiated of an electron undergoing a circular motion at a
given time $t$ in a unit frequency interval was first computed by Schwinger within a purely classical framework. It is given by
Eq.~(II.16) in \cite{schwinger:1949}. Taking an additional factor of $(4\pi\varepsilon_0)^{-1}$ into account it reads in SI-units:
\begin{subequations}
\begin{align}
\label{eq:power-spectrum-synchrotron}
\frac{\mathrm{d}P(\omega,t)}{\mathrm{d}\omega}&=\frac{P_0}{\omega_c}S_s\left(\frac{\omega}{\omega_c}\right)\,,\quad
S_s(\xi)=\frac{9\sqrt{3}}{8\pi}\xi\int_{\xi}^{\infty} K_{5/3}(\zeta)\,\mathrm{d}\zeta\,, \\[2ex]
\label{eq:total-power-synchrotron-radiation}
P_0&=\frac{C_{\upgamma}}{2\pi}c(mc^2)^4\frac{\gamma^4}{R^2}\,,\quad \omega_c=\frac{3}{2}c\frac{\gamma^3}{R}\,.
\end{align}
\end{subequations}
Here $C_{\upgamma}$ is Sand's radiation constant:
\begin{equation}
C_{\upgamma}=\frac{4\pi}{3}\frac{r_c}{(mc^2)^3}\,,\quad r_c=\frac{e^2}{4\pi\varepsilon_0 mc^2}\,,
\end{equation}
with the classical electron radius $r_c$. The total radiated power integrated over the whole frequency range is denoted by $P_0$ and
$\omega_c$ is the critical frequency of synchrotron radiation. All physical constants are put in the prefactor of the spectrum in
\eqref{eq:power-spectrum-synchrotron}. The characteristic function $S_s(\xi)$ for synchrotron radiation is dimensionless and depends on
the dimensionless ratio $\omega/\omega_c$. The integrand of $S_s(\xi)$ is the modified Bessel's function $K_{5/3}(\zeta)$. A plot of the
characteristic function is shown in \figref{fig:characteristic-function}. It has a maximum at the numerical value $\xi\approx 0.285812$
and it is characterized by the following further properties:
\begin{equation}
\int_0^{\infty} S_s(\xi)\,\mathrm{d}\xi=1\,,\quad \int_0^1 S_s(\xi)\,\mathrm{d}\xi=\frac{1}{2}\,.
\end{equation}
The first of these results means that the total radiated power per unit frequency range is indeed given by $P_0/\omega_c$. The second result
shows that half of the power emitted is radiated by photons up to the critical frequency $\omega_c$. A plot of the function $S_s(\xi)$ is shown
in \figref{fig:characteristic-function}.

The following asymptotic expansions are valid for the power radiated:
\begin{equation}
\label{eq:synchrotron-power-small-frequencies}
\frac{\mathrm{d}P(\omega,t)}{\mathrm{d}\omega}=\frac{9\sqrt{3}}{4\sqrt[3]{2}\pi}\Gamma\left(\frac{2}{3}\right)\frac{P_0}{\omega_c}\left(\frac{\omega}{\omega_c}\right)^{1/3}\approx 1.33323\frac{P_0}{\omega_c^{4/3}}\omega^{1/3}\,,
\end{equation}
for $\omega\ll \omega_c$ with Euler's Gamma function $\Gamma(\xi)$ and
\begin{equation}
\frac{\mathrm{d}P(\omega,t)}{\mathrm{d}\omega}=\frac{9\sqrt{3}}{8\sqrt{2\pi}}\frac{P_0}{\omega_c}\frac{\sqrt{\omega/\omega_c}}{\exp(\omega/\omega_c)}\approx 0.777362\frac{P_0}{\omega_c^{3/2}}\frac{\sqrt{\omega}}{\exp(\omega/\omega_c)}\,,
\end{equation}
for $\omega\gg \omega_c$. Finally, the spectral photon flux giving the number of photons radiated per unit time and relative bandwidth is
then given by:
\begin{equation}
\frac{\dot{N}_{\upgamma}}{\mathrm{d}\omega/\omega}=\frac{1}{\hbar}\frac{\mathrm{d}P}{\mathrm{d}\omega}=\frac{P_0}{\hbar\omega_c}S_s\left(\frac{\omega}{\omega_c}\right)\,,
\end{equation}
with $\hbar=h/(2\pi)$ where $h$ is Planck's constant.

\subsection{Coherent synchrotron radiation}

\begin{figure}[b]
\centering
\includegraphics{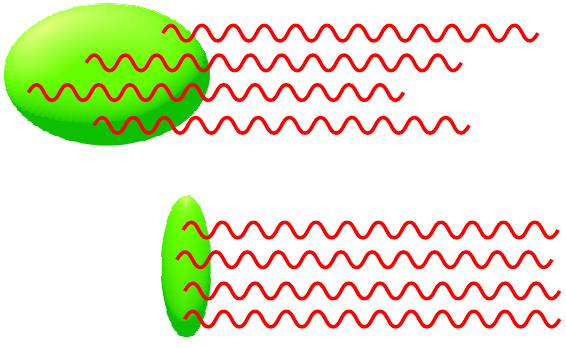}
\caption{Synchrotron radiation emitted by particles in a bunch whose length is much larger than the wavelength of the radiation (above)
and whose length lies in the order of magnitude of the radiation wavelength (below).}
\label{fig:coherent-radiation-bunch}
\end{figure}%
When the length of a particle bunch lies in the order of magnitude (or below) of the radiation wavelength then synchrotron radiation can be
emitted such that wave trains originating of different particles are in phase with each other (see \figref{fig:coherent-radiation-bunch}).
Then different wave trains can interfere constructively leading to a vast increase of the radiation intensity. Such kind of radiation is
called temporarily coherent or just coherent \cite{wiedemann:1999}.

The power spectrum of coherent synchrotron radiation emitted from a bunch can be obtained from the following equation \cite{wiedemann:1999}:
\begin{equation}
\label{eq:coherent-spectrum}
\left.\frac{\mathrm{d}P}{\mathrm{d}\omega}\right|_{\mathrm{CSR}}=N_{\mathrm{e}}(N_{\mathrm{e}}-1)\left.\frac{\mathrm{d}P}{\mathrm{d}\omega}\right|_{\mathrm{SR}}|F(\omega)|^2\,,
\end{equation}
where $N_{\mathrm{e}}$ is the number of radiating electrons, $\mathrm{d}P/\mathrm{d}\omega|_{\mathrm{SR}}$ is the single particle synchrotron
radiation power spectrum, and $F(\omega)$ is known as form factor of the bunch. The form factor is the Fourier transform of the particle density
distribution $\varrho(\mathbf{r})$ describing the bunch \cite{wiedemann:1999,Muller:2008zzh}:
\begin{equation}
F(\omega)=\int \mathrm{d}^3r\,\varrho(\mathbf{r})\exp\left(\mathrm{i}\frac{\omega}{c}\widehat{\mathbf{n}}\cdot \mathbf{r}\right)\,,
\end{equation}
where $\widehat{\mathbf{n}}$ is the unit vector pointing along the wave vector: $\mathbf{k}=k\widehat{\mathbf{n}}$ with $k=\omega/c$ and the frequency
$\omega$ of the wave. The form factor describes what wave numbers (i.e. frequencies or wavelengths) contribute to the coherent synchrotron
radiation spectrum. The coherent radiation spectrum given by \eqref{eq:coherent-spectrum} has certain interesting peculiarities. First of all,
it grows quadratically with the number of particles. This is due to the fact that amplitudes add up constructively whereby the amplitude linearly
depends on the particle number. The power then results from the amplitude squared.

Secondly, since the bunch length is the only physical length
scale in the form factor, its inverse appears in the spectrum. The power spectrum will significantly drop off for radiation wavelengths much smaller
than the bunch length. Thirdly, how fast this drop off takes place depends closely on the form factor of the bunch. This means that not only
the bunch length is crucial for the coherent synchrotron radiation spectrum but also the longitudinal particle distribution.

\subsection{Energy dependence of the radiated CSR power}

In CSRtrack simulations of the FLUTE chicane it was found that 1D and 3D simulations produce different results
for the phase space distribution after compression --- as long as the energy is low enough or the beam current high enough
\cite{naknaimueang:2013}. This especially was the case for the intended beam energy of approximately \unit[40]{MeV}. In the
framework of a 1D simulation the particle coordinates are projected on the longitudinal axis. This procedure is not followed in the 3D
calculation where the all coordinates are taken into account. The 1D and 3D calculations produce similar results before the bunches enter
the final chicane magnet. Therefore the main differences between the 1D and the 3D calculation must have the origin at the fourth magnet.
In this magnet the longitudinal bunch length becomes small enough such that the bunch produces a significant amount of coherent synchrotron
radiation and experiences space charge effects. This observation lead the authors to the conclusion that the difference in 1D and 3D simulations
for the FLUTE chicane is a measure for space charge and CSR effects. This is also due to the fact that the neglect of transverse coordinates
do not properly take these effects into account.

Besides, in \cite{naknaimueang:2013} it was also observed that for a hypothetical beam energy of \unit[300]{MeV} there was almost no difference
between the phase space distribution after compression obtained from 1D and 3D simulations. Based upon the previous conclusion one can infer
that for such high energies both CSR and space charge effects to not play that much of a role any more. For the space charge effects this
behavior is easy to understand since according to \cite{reiser:2008} they are suppressed by $1/\gamma^2$ where $\gamma$ is the Lorentz factor
(see also \secref{sec:space-charge-effects-flute}). However according to \eqref{eq:total-power-synchrotron-radiation} the total power of
synchrotron radiation grows proportional to $\gamma^4$. This originates from the behavior of the critical frequency $\omega_c$
which grows like $\gamma^3$ and for large Lorentz factor shifts the maximum of the synchrotron radiation spectrum to high frequencies.

However the behavior of the CSR spectrum with respect to the energy of the radiating particles is quite different. Note that contrary to the
normal synchrotron radiation spectrum, the CSR spectrum involves the form factor, i.e., the longitudinal shape of the bunch (see
\eqref{eq:coherent-spectrum}). Due to the form factor the radiated power spectrum drops off at frequencies larger than the bunch frequency
$\omega_b\equiv c/\sigma_s$. How fast this drop-off takes place, depends on the detail of the bunch shape. It occurs exponentially for a
Gaussian shape and only polynomially for a profile with a sharp edge \cite{Assmann:2013}. For simplicity we would like to assume that the
maximum frequency of the CSR spectrum equals $\omega_b$ and that all frequencies $\omega>\omega_b$ do not contribute to the radiated power.

To obtain the total radiated power, the spectrum should be integrated to the upper limit $\omega=\omega_b$ and not to infinitely high
frequencies:
\begin{equation}
\label{eq:modified-total-power}
P(t)=\int_0^{\omega_b} \mathrm{d}\omega\,\frac{\dot{N}_{\upgamma}}{\mathrm{d}\omega/\omega}=\frac{P_0}{\omega_c\hbar} \int_0^{\omega_b} \mathrm{d}\omega\,\frac{9\sqrt{3}}{8\pi}\left(\frac{\omega}{\omega_c}\right) \int_{\omega/\omega_c}^{\infty} \mathrm{d}\zeta\,K_{5/3}(\zeta)\,.
\end{equation}
In \appref{sec:computation-csr-power} we obtain the value of this integral for $\omega_b\ll \omega_c$. Its result is given by
\begin{equation}
\label{eq:modified-total-power-result}
P(t)=\frac{P_0}{\hbar}\frac{27\sqrt{3}}{16 \cdot \sqrt[3]{2}\pi}\Gamma\left(\frac{2}{3}\right)\left(\frac{\omega_b}{\omega_c}\right)^{4/3}+\mathcal{O}\left(\frac{\omega_b}{\omega_c}\right)^2\,.
\end{equation}
The frequency $\omega_b$ does not increase with the particle energy but depends only on the longitudinal bunch length. Since the critical
frequency $\omega_c$ grows with $\gamma^3$, for high energies $\omega_b/\omega_c\ll 1$ is guaranteed. The total power $P_0$ grows with $\gamma^4$
and this dependence is then cancelled by a factor $\gamma^{-4}$ coming from $\omega_c^{-4/3}$:
\begin{equation}
P(t)\sim \frac{P_0}{\omega_c^{4/3}}\sim \frac{\gamma^4}{\gamma^4}=1\,.
\end{equation}
This shows that the CSR power is independent from the particle energy at leading order and when $\omega_b\ll \omega_c$ holds. Since the particle
momentum is given by $p=\gamma mv$, i.e., it grows linearly with the Lorentz factor the back reaction of CSR on the bunch can be assumed to
decrease for high energies. The bunch becomes more stable with respect to perturbations whereas the radiated energy stays the same. This is
exactly the behavior that was observed with the tool CSRtrack in \cite{naknaimueang:2013}.

\subsection{Modification of the synchrotron radiation spectrum}

\begin{figure}[b]
\centering
\includegraphics[scale=1]{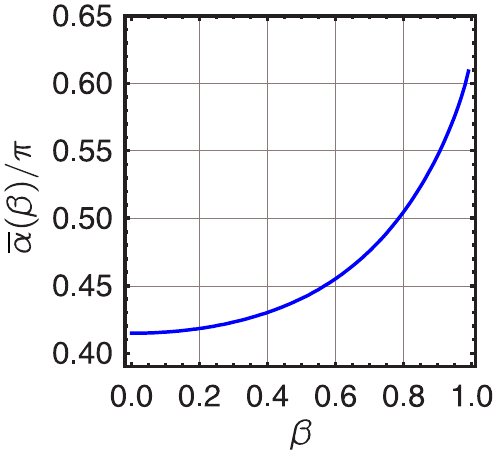}
\caption{Shown is $\overline{\alpha}(\beta)$, which is referred to in the text, as a function of $\beta=v/c$. The minimal value of
$\overline{\alpha}(\beta)$ (in the leftmost position) lies at approximately $0.41\pi$ and $\overline{\alpha}(\beta)$ steadily increases to about
$0.62\pi$.}
\label{fig:zeros-special-functions}
\end{figure}

In several papers it was noted that the synchrotron radiation spectrum for an electron moving through a dipole magnet once is different from
the radiation spectrum that emerges when an electron performs many revolutions inside a magnet \cite{bagrov:1982,saldin:1996}. The latter is
what was calculated in \cite{schwinger:1949} and what is usually referred to as synchrotron radiation spectrum. Since at FLUTE we intend
to generate CSR in the last bending magnet of the chicane but not via an electron performing many revolutions inside a circular
accelerator this statement deserves a further study. Equation (16) in \cite{bagrov:1982} gives the relativistic, angle-integrated spectrum
for frequencies $\omega\ll \omega_c$. It is calculated as an expansion in $\omega/\omega_0$ where $\omega_0=c\beta/R$ is the cyclotron frequency.
The result involves functions $F^0(\alpha,\beta)$ and $F^1(\alpha,\beta)$ depending on $\beta=v/c$ and the bending angle $\alpha$ of the radiating
particle inside the magnet. The function $F^0$ gives the contribution at zeroth order in $\omega^2/\omega_0^2$ and $F^1$ delivers a contribution
at first order in $\omega^2/\omega_0^2$. According to \cite{bagrov:1982} there is a criterion upon which can be decided whether or not the
radiation spectrum at small frequencies deviates from the standard synchrotron power spectrum. The equation $F^1(\alpha,\beta)=0$ defines a function
$\overline{\alpha}=\overline{\alpha}(\beta)$, whose plot has been reproduced in \figref{fig:zeros-special-functions}.

As long as the bending angle for a certain $\beta$ is larger than $\overline{\alpha}(\beta)$, the synchrotron radiation spectrum can be assumed to
coincide with \eqref{eq:coherent-spectrum} also for $\omega\mapsto 0$. However this is not the case if the bending angle is smaller than
$\overline{\alpha}(\beta)$. For a particle energy of \unit[41]{MeV} one obtains a limiting bending angle of $\overline{\alpha}\approx 112^{\circ}$.
Since in the FLUTE chicane we have bending angles ranging from $10.2^{\circ}$ to $11.5^{\circ}$ (for bunch charges of \unit[1]{pC} to \unit[3]{nC},
see \cite{Assmann:2013}) we can expect the synchrotron radiation spectrum to be modified for small frequencies. However this change is supposed to
occur for $\omega\lesssim \omega_0$, where $\omega_0 \approx \unit[2.98 \cdot 10^8]{Hz}$.\footnote{This is also evident from the paragraph directly
below Eq.~(II.5) in \cite{schwinger:1949}. In the latter equation an approximation was used that is only valid for $\omega\gg \omega_0$.} This lies
several orders of magnitude below the THz radiation regime, which we are interested in at FLUTE. Hence using the model above, the changes that are
expected to occur for the synchrotron radiation spectrum when an electron moves through a bending magnet only once, can be safely neglected at FLUTE.

\section{Conclusions and outlook}
\setcounter{equation}{0}

To summarize, analytical studies for bunch compression at the future linear accelerator FLUTE were performed whose results were compared
to the simulation output of the tools Astra and CSRtrack. The calculations were done for two typical bunches with the extremal charges of
\unit[1]{pC} and \unit[3]{nC} that had been simulated from the cathode to the entrance of the bunch compressor. Neglecting both space
charge and CSR effects, the final bunch profiles obtained from mere path length differences agree very well with the simulation results.
As a cross check, the problem was then treated within the transfer matrix formalism as well. First order perturbation theory in the
momentum spread gives a result for the final bunch length that deviates from the simulation results by several percent. For this reason
considering second order terms is mandatory to give a good agreement. Besides, in this context we obtained some second order coefficients
for dipole magnets, fringe fields, and drifts.

To consider space charge effects, a simple model was introduced where the bunch is described by a homogeneously charged cylinder.
The latter generates electric and also magnetic fields when moving. The equations of motion for a single electron at the surface of the
cylinder were obtained and solved numerically. Within this model, the space charge effects are overestimated for a bunch charge of
\unit[3]{nC}, whereas for \unit[1]{pC} there is a reasonable agreement with the simulations. It can be deduced that space charge effects
are negligible for bunch compression at FLUTE. In a future analysis the space charge effects could be considered with the
help of the more complicated Vlasov equation what was done in, e.g., \cite{Huang:2013aba}.

Concerning the backreation of bunches with their own CSR it was proven that the radiated CSR power does not scale with the Lorentz factor
of the bunch but it stays constant. Hence, for high energies a bunch is not sensitive to CSR effects any more, which agrees with recent
CSRtrack simulations referred to in the current article. A next step could be to compute the energy loss of a typical bunch
at FLUTE analytically according to \cite{Derbenev:1995} and to compare with the simulation results.

Another issue is that the synchrotron radiation spectrum is different for electrons moving through a bending magnet only once in comparison
to circulating electrons. We were able to demonstrate that the spectral differences occur for frequencies that lie several orders of 
magnitude below the THz range, whereby this effect does not play any role for FLUTE.

The paper demonstrates how powerful the combination of analytical methods and simulations is to investigate bunch compression. The techniques
presented shall provide a framework for further analytical compression studies. These can be used for future investigations of FLUTE or they
may be modified accordingly for other purposes.

\section{Acknowledgments}

It is a pleasure to thank M.~Fitterer, S.~Hillenbrand, V.~Judin, A.-S.~M\"{u}ller, S. Naknaimueang, S.~Marsching, M.~Nasse, A.~Papash, R.~Rossmanith,
M.~Schuh, and M.~Weber for helpful discussions. Furthermore the authors are indebted to M. Oyamada and M.~Schwarz for reading the paper and giving 
helpful comments. This work was mainly performed with
financial support within the program ``Accelerator Research and Development'' of the \textit{Hermann von Helmholtz-Gemeinschaft Deutscher
Forschungszentren}. One of us (M.S.) acknowledges additional support from the \textit{Deutsche Akademie der Naturforscher Leopoldina} within 
Grant No. LPDS 2012-17 to complete this article.

\cleardoublepage
\begin{appendix}
\numberwithin{equation}{section}

\section{Parametric representation of particle trajectory in the FLUTE chicane}
\label{sec:parametric-representation-trajectory}
\setcounter{equation}{0}

In what follows, find the electron trajectory for the FLUTE chicane used in \secref{ssec:electron-trajectory} (for $x'=0$)
and in \secref{ssec:transverse-components-magnetic-jitter}. The right-hand interval limits of $\mathbf{r}_1$, $\dots$, $\mathbf{r}_9$
are understood to correspond to $l_1$, $\dots$, $l_9$.
\begin{subequations}
\label{eq:parametric-representation-piecewise}
\begin{equation}
\label{eq:parametric-representation-piecewise-sub}
\mathbf{r}(l)=\left\{\begin{array}{lcl}
\mathbf{r}_1(l) & \text{for} & 0 \leq l \leq l_1\,, \\
\mathbf{r}_2(l-l_1) & \text{for} & l_1 \leq l \leq \sum_{i=1}^2 l_i\,, \\
\mathbf{r}_3\left(l-\sum_{i=1}^2 l_i\right) & \text{for} & \sum_{i=1}^2 l_i \leq l \leq \sum_{i=1}^3 l_i\,, \\
\mathbf{r}_4\left(l-\sum_{i=1}^3 l_i\right) & \text{for} & \sum_{i=1}^3 l_i \leq l \leq \sum_{i=1}^4 l_i\,, \\
\mathbf{r}_5\left(l-\sum_{i=1}^4 l_i\right) & \text{for} & \sum_{i=1}^4 l_i \leq l \leq \sum_{i=1}^5 l_i\,, \\
\mathbf{r}_6\left(l-\sum_{i=1}^5 l_i\right) & \text{for} & \sum_{i=1}^5 l_i \leq l \leq \sum_{i=1}^6 l_i\,, \\
\mathbf{r}_7\left(l-\sum_{i=1}^6 l_i\right) & \text{for} & \sum_{i=1}^6 l_i \leq l \leq \sum_{i=1}^7 l_i\,, \\
\mathbf{r}_8\left(l-\sum_{i=1}^7 l_i\right) & \text{for} & \sum_{i=1}^7 l_i \leq l \leq \sum_{i=1}^8 l_i\,, \\
\mathbf{r}_9\left(l-\sum_{i=1}^8 l_i\right) & \text{for} & \sum_{i=1}^8 l_i \leq l \leq \sum_{i=1}^9 l_i\,, \\
\end{array}
\right.
\end{equation}
\begin{align}
\mathbf{r}_1(l)&=\begin{pmatrix}
x_0 \\
z_0 \\
\end{pmatrix}+\begin{pmatrix}
\sin x' \\
\cos x' \\
\end{pmatrix}l\,,\quad l\in \left[0,\frac{z_1-z_0}{\cos x'}\right]\,, \displaybreak[0]\\[2ex]
\mathbf{r}_2(l)&=\mathbf{r}_1\left(\frac{z_1-z_0}{\cos x'}\right)+R\begin{pmatrix}
\cos x'-\cos(l/R+x') \\
\sin(l/R+x')-\sin x' \\
\end{pmatrix}\,,\quad l \in [0,R\alpha]\,, \displaybreak[0]\\[2ex]
\mathbf{r}_3(l)&=\mathbf{r}_2(R\alpha)+\begin{pmatrix}
\cos\delta \\
\sin\delta \\
\end{pmatrix}l\,,\quad l\in \left[0,\frac{L_{\mathrm{space}}}{\sin\delta}\right]\,, \\[2ex]
\mathbf{r}_4(l)&=\mathbf{r}_3\left(\frac{L_{\mathrm{space}}}{\sin\delta}\right)+R\begin{pmatrix}
\sin(l/R+\delta)-\sin\delta \\
\cos\delta-\cos(l/R+\delta) \\
\end{pmatrix}\,,\quad l\in [0,R\alpha]\,, \displaybreak[0]\\[2ex]
\mathbf{r}_5(l)&=\mathbf{r}_4(R\alpha)+\begin{pmatrix}
\sin x' \\
\cos x' \\
\end{pmatrix}l\,,\quad l\in \left[0,\frac{L_{\mathrm{drift}}}{\cos x'}\right]\,, \displaybreak[0]\\[2ex]
\mathbf{r}_6(l)&=\mathbf{r}_5\left(\frac{L_{\mathrm{drift}}}{\cos x'}\right)+R\begin{pmatrix}
\cos(l/R-x')-\cos x' \\
\sin(l/R-x')+\sin x' \\
\end{pmatrix}\,,\quad l\in [0,R(x'+\epsilon)]\,, \displaybreak[0]\\[2ex]
\mathbf{r}_7(l)&=\mathbf{r}_6(R[x'+\epsilon])+\begin{pmatrix}
-\sin\epsilon \\
\cos\epsilon \\
\end{pmatrix}l\,,\quad l\in \left[0,\frac{L_{\mathrm{space}}}{\cos\epsilon}\right]\,, \displaybreak[0]\\[2ex]
\mathbf{r}_8(l)&=\mathbf{r}_7\left(\frac{L_{\mathrm{space}}}{\cos\epsilon}\right)+R\begin{pmatrix}
\cos\epsilon-\cos(\epsilon-l/R) \\
\sin\epsilon-\sin(\epsilon-l/R) \\
\end{pmatrix}\,,\quad l\in [0,R(x'+\epsilon)]\,, \displaybreak[0]\\[2ex]
\mathbf{r}_9(l)&=\mathbf{r}_8(R[x'+\epsilon])+\begin{pmatrix}
\sin x' \\
\cos x' \\
\end{pmatrix}l\,,\quad l\in [0,z_2]\,. \\[2ex]
\alpha&=\arcsin\left(\frac{L_{\mathrm{mag}}}{R}+\sin x'\right)-x'\,,\quad \delta=\frac{\pi}{2}-(\alpha+x')\,, \\[2ex]
\epsilon&=\arcsin\left(\frac{L_{\mathrm{mag}}}{R}-\sin x'\right)\,.
\end{align}
\end{subequations}

\subsection{Parametric representation of trajectory in hypothetical sector magnet chicane}
\label{ssec:chicane-sector-dipoles}

A parametric representation of the trajectory for a particle in a chicane consisting of sector dipole magnets
(see \secref{ssec:sector-chicane}) is given by \eqref{eq:parametric-representation-piecewise-sub} with the following
piecewise functions:
\begin{subequations}
\begin{align}
\mathbf{r}_1(l)&=\begin{pmatrix}
x_0 \\
z_0+l \\
\end{pmatrix}\,,\quad l\in [0,z_1-z_0]\,, \displaybreak[0]\\[2ex]
\mathbf{r}_2(l)&=\mathbf{r}_1(z_1-z_0)+\begin{pmatrix}
R \\
0 \\
\end{pmatrix}+R'\begin{pmatrix}
-\cos(l/R') \\
\sin(l/R') \\
\end{pmatrix}\,,\quad l\in [0,R'\varepsilon]\,, \displaybreak[0]\\[2ex]
\mathbf{r}_3(l)&=\mathbf{r}_2(R'\varepsilon)+\begin{pmatrix}
\sin\varepsilon \\
\cos\varepsilon \\
\end{pmatrix}l\,,\quad l\in [0,L_{\mathrm{space}}']\,, \displaybreak[0]\\[2ex]
\mathbf{r}_4(l)&=\mathbf{r}_3(L_{\mathrm{space}}')+R''\begin{pmatrix}
\cos(l/R''-\varepsilon)-\cos\varepsilon \\
\sin(l/R''-\varepsilon)+\sin\varepsilon \\
\end{pmatrix}\,,\quad l\in [0,R''\varepsilon]\,, \displaybreak[0]\\[2ex]
\mathbf{r}_5(l)&=\mathbf{r}_4(R''\varepsilon)+\begin{pmatrix}
0 \\
l \\
\end{pmatrix}\,,\quad l\in [0,L_{\mathrm{drift}}]\,, \displaybreak[0]\\[2ex]
\mathbf{r}_6(l)&=\mathbf{r}_5(L_{\mathrm{drift}})+R''\begin{pmatrix}
\cos(l/R'')-1 \\
\sin(l/R'') \\
\end{pmatrix}\,,\quad l\in [0,R''\varepsilon]\,, \displaybreak[0]\\[2ex]
\mathbf{r}_7(l)&=\mathbf{r}_6(R''\varepsilon)+\begin{pmatrix}
-\sin\varepsilon \\
\cos\varepsilon \\
\end{pmatrix}l\,,\quad l\in [0,L_{\mathrm{space}}']\,, \displaybreak[0]\\[2ex]
\mathbf{r}_8(l)&=\mathbf{r}_7(L_{\mathrm{space}}')+R'\begin{pmatrix}
\cos\varepsilon-\cos(l/R'-\varepsilon) \\
\sin\varepsilon+\sin(l/R'-\varepsilon) \\
\end{pmatrix}\,,\quad l\in [0,R'\varepsilon]\,, \displaybreak[0]\\[2ex]
\mathbf{r}_9(l)&=\mathbf{r}_8(R'\varepsilon)+\begin{pmatrix}
0 \\
l \\
\end{pmatrix}\,,\quad l\in [0,z_2-z_1]\,, \displaybreak[0]\\[2ex]
R'&=R+\Delta R\,, \displaybreak[0]\\[2ex]
\varepsilon&=\arctan\left\{\frac{R\sin\alpha+\sin\alpha\left[\sqrt{R'^2-\Delta R^2\sin^2\alpha}-(R+\Delta R\cos\alpha)\right]}{\Delta R+R\cos\alpha+\cos\alpha\left[\sqrt{R'^2-\Delta R^2\sin^2\alpha}-(R+\Delta R\cos\alpha)\right]}\right\}\,, \displaybreak[0]\\[2ex]
L_{\mathrm{space}}'&=\frac{2L_{\mathrm{space}}-\Delta R\sin(2\alpha)-2R'\sin\varepsilon}{2\cos\alpha\cos(\alpha-\varepsilon)}+R'\tan\alpha\,, \displaybreak[0]\\[2ex]
R''&=\frac{1}{\cos(\alpha-\varepsilon)}\Big\{R\big[\sin(2\alpha)+\cot\varepsilon\big]\tan(\alpha)-R\cot\varepsilon\sin^2\alpha\tan\alpha \notag \\
&\phantom{{}={}}\hspace{0.5cm}+\sin\alpha\big[L_{\mathrm{space}}-R'/\sin\varepsilon+\cot\varepsilon(R'-R+R\cos\alpha-L_{\mathrm{space}}\tan\alpha)\big]\Big\}\,.
\end{align}
\end{subequations}
Here $\alpha$ is the bending angle and $\Delta R=\Delta p/(eB)$ with the momentum spread $\Delta p$, the magnetic field $B$, and the
elementary charge $e$. For the reference trajectory $\Delta R=0$ has to be set.

\section{Transfer matrices for the FLUTE chicane}
\label{sec:transfer-matrices-flute}
\setcounter{equation}{0}

In the current section we list the transfer matrices that are referred to in \secref{sec:transfer-matrix-formalism}. Besides, some
general remarks on the transfer matrix formalism are given.
The drift transfer matrix is the simplest and it is given by \cite{Iselin:1992}:
\begin{equation}
\label{eq:matrix-drift}
R_{\mathrm{drift}}(L)=\begin{pmatrix}
1 & L & 0 & 0 & 0 & 0 \\
0 & 1 & 0 & 0 & 0 & 0 \\
0 & 0 & 1 & L & 0 & 0 \\
0 & 0 & 0 & 1 & 0 & 0 \\
0 & 0 & 0 & 0 & 1 & L/(\beta^2\gamma^2) \\
0 & 0 & 0 & 0 & 0 & 1 \\
\end{pmatrix}\,,
\end{equation}
where $\beta=v/c$ and $\gamma$ is the Lorentz factor of the reference particle; the length of the drift is $L$. Note that terms
suppressed by $\beta^2\gamma^2$ are related to velocity differences of particles. We see that such a term appears in the element $R_{56}$
of the drift. However one should keep in mind that for electrons in the FLUTE chicane it holds that $\gamma\approx 80$ rendering such
contributions highly suppressed.

Now, the transfer matrix for a sector magnet reads \cite{Iselin:1992}:
\begin{subequations}
\label{eq:matrix-sector-dipole}
\begin{equation}
R_{\mathrm{sec}}(L,h)=\begin{pmatrix}
\mathrm{c}(k_xL) & \mathrm{s}(k_xL)/k_x & 0 & 0 & 0 & d(k_x,L)h/\beta \\
-k_x\mathrm{s}(k_xL) & \mathrm{c}(k_xL) & 0 & 0 & 0 & \mathrm{s}(k_xL)h/(\beta k_x) \\
0 & 0 & \mathrm{c}(k_yL) & \mathrm{s}(k_yL)/k_y & 0 & 0 \\
0 & 0 & -k_y\mathrm{s}(k_yL) & \mathrm{c}(k_yL) & 0 & 0 \\
-\mathrm{s}(k_xL) h/(\beta k_x) & -d(k_x,L)h/\beta & 0 & 0 & 1 & L/(\beta^2\gamma^2)-h^2J_1(L)/\beta^2 \\
0 & 0 & 0 & 0 & 0 & 1 \\
\end{pmatrix}
\end{equation}
with the functions
\begin{align}
\label{eq:definition-trigonometric-functions}
\mathrm{s}(x)&\equiv \sin(x)\,,\quad \mathrm{c}(s)\equiv \cos(x)\,, \\[2ex]
d(k_x,L)&\equiv \frac{1-\cos(k_xL)}{k_x^2}\,,\quad J_1(L)\equiv \frac{L}{k_x^2}-\frac{\sin(k_xL)}{k_x^3}\,,
\end{align}
and the quantities
\begin{equation}
\quad k_x\equiv h\sqrt{1-n}\,,\quad k_y\equiv h\sqrt{n}\,.
\end{equation}
\end{subequations}
The traveling length of the particle within the dipole is given by $L$. The curvature of the reference trajectory is denoted as $h\equiv 1/R$
and the dimensionless parameter $n$ is related to the gradient of the magnetic field. It appears when expanding the $y$-component
of the dipole magnetic flux density in $x$-direction:
\begin{subequations}
\begin{align}
B_y(x,0,t)&=B_y(0,0,t)\left[1-nhx+o^2h^2x^2+\hdots\right]\,, \\[2ex]
\label{eq:gradient-of-magnetic-field}
n&=\left.-\frac{1}{hB_y}\frac{\partial B_y}{\partial x}\right|_{\substack{x=0 \\ y=0}}\,,\quad
o=\left.\frac{1}{2!h^2B_y}\frac{\partial^2B_y}{\partial x^2}\right|_{\substack{x=0 \\ y=0}}\,.
\end{align}
\end{subequations}
\begin{figure}[b]
\centering
\subfloat[sector dipole with curvature radius $R_1$ and $R_2$]{\label{fig:sector-bend}\includegraphics[scale=1]{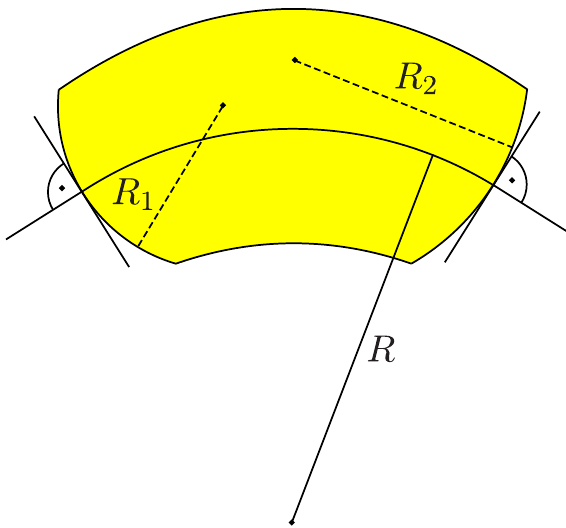}}\hspace{1cm}
\subfloat[right-turn (above) and left-turn reference trajectory (below)]{\label{fig:sector-bend-sign-conventions}\includegraphics[scale=1]{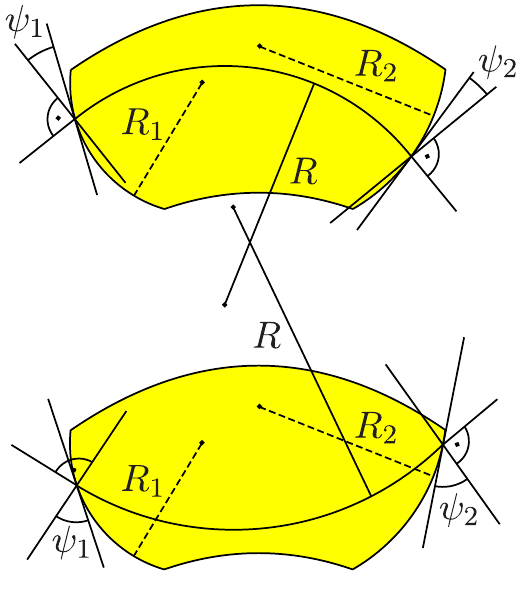}}
\caption{The left panel shows a sector dipole with curvature radii $R_1$ and $R_2$, where the particle both enters and exists the dipole
perpendicularly to the surface. In the right panel the particle enters the sector dipole under the angle $\psi_1$ and exists under the
angle $\psi_2$. The angles are defined to be enclosed by the tangent at the entering and exit point, respectively, and the corresponding
lines that are orthogonal to the trajectories at the respective points. Similar figures can be found in \cite{Brown:1984}.}
\label{fig:sector-bend-properties}
\end{figure}%
For a magnetic field $\mathbf{B}$ in vacuum and no electric field we have that $\boldsymbol{\nabla}\times \mathbf{B}=0$. Therefore, $\mathbf{B}$
can be derived from a scalar potential $\varphi$ via $\mathbf{B}=-\boldsymbol{\nabla}\varphi$. It is a common procedure to assume that this
potential is antisymmetric with respect to the median plane $y=0$: $\varphi(x,y,t)=-\varphi(x,-y,t)$. This simplifies the calculation and
disregarding this assumption has not shown to lead to any new insights \cite{Brown:1984}. From this symmetry follows that $B_x(x,0,t)=B_t(x,0,t)=0$
in the median plane where only $B_y(x,0,t)\neq 0$ and orthogonal to that plane. Hence every particle travelling in that plane will remain in the
plane and the whole magnetic field expanded around the reference trajectory can be expressed via the derivatives of \eqref{eq:gradient-of-magnetic-field}.

An illustration of the geometrical quantities that appear in the context of the sector dipole magnet is given by \figref{fig:sector-bend}. Note
that the curvature radii $R_1$ und $R_2$ of the sector magnet do not appear at first order perturbation theory. Besides, our conventions in
\eqref{eq:definition-trigonometric-functions} and \eqref{eq:gradient-of-magnetic-field} differ from what is used in
\cite{Brown:1984,Iselin:1992}.

The magnetic fringe fields of dipole magnets can be modeled by a further transfer matrix that is given by \cite{Iselin:1992}
\begin{equation}
\label{eq:matrix-fringe}
R_{\mathrm{fringe}}(\psi_i,h)=\begin{pmatrix}
1 & 0 & 0 & 0 & 0 & 0 \\
h\tan(\psi_i) & 1 & 0 & 0 & 0 & 0 \\
0 & 0 & 1 & 0 & 0 & 0 \\
0 & 0 & -h\tan(\overline{\psi_i}) & 1 & 0 & 0 \\
0 & 0 & 0 & 0 & 1 & 0 \\
0 & 0 & 0 & 0 & 0 & 1 \\
\end{pmatrix}\,,
\end{equation}
where the index $i\in\{1,2\}$ marks the entrance angle $\psi_1$ and the exit angle $\psi_2$, respectively. These angles are enclosed
by the tangent along the dipole surface at the entrance or exit point and the line running perpendicularly to the particle trajectory
at these points (see \figref{fig:sector-bend}). Furthermore the angle in the component $R_{43}$ of \eqref{eq:matrix-fringe} is
modified by a contribution that is linked to the profile of the $y$-component of the magnetic fringe field:
\begin{equation}
\label{eq:fringe-field-correction}
\overline{\psi_i}=\psi_i-hgI_1\left(\frac{1+\sin^2\psi_i}{\cos\psi_i}\right)\,,\quad
I_1=\int^{+\infty}_{-\infty} \frac{B_y(z)[B_0-B_y(z)]}{gB_0^2}\,\mathrm{d}z\,,
\end{equation}
where in \cite{Brown:1984} the division by $\cos\psi_i$ is stated, but it is missing in \cite{Iselin:1992}. For $\psi_i\ll \pi/2$, which
is especially the case for the FLUTE chicane, the division by this expression does not lead to drastic modifications.

\begin{figure}[b!]
\includegraphics[scale=0.5]{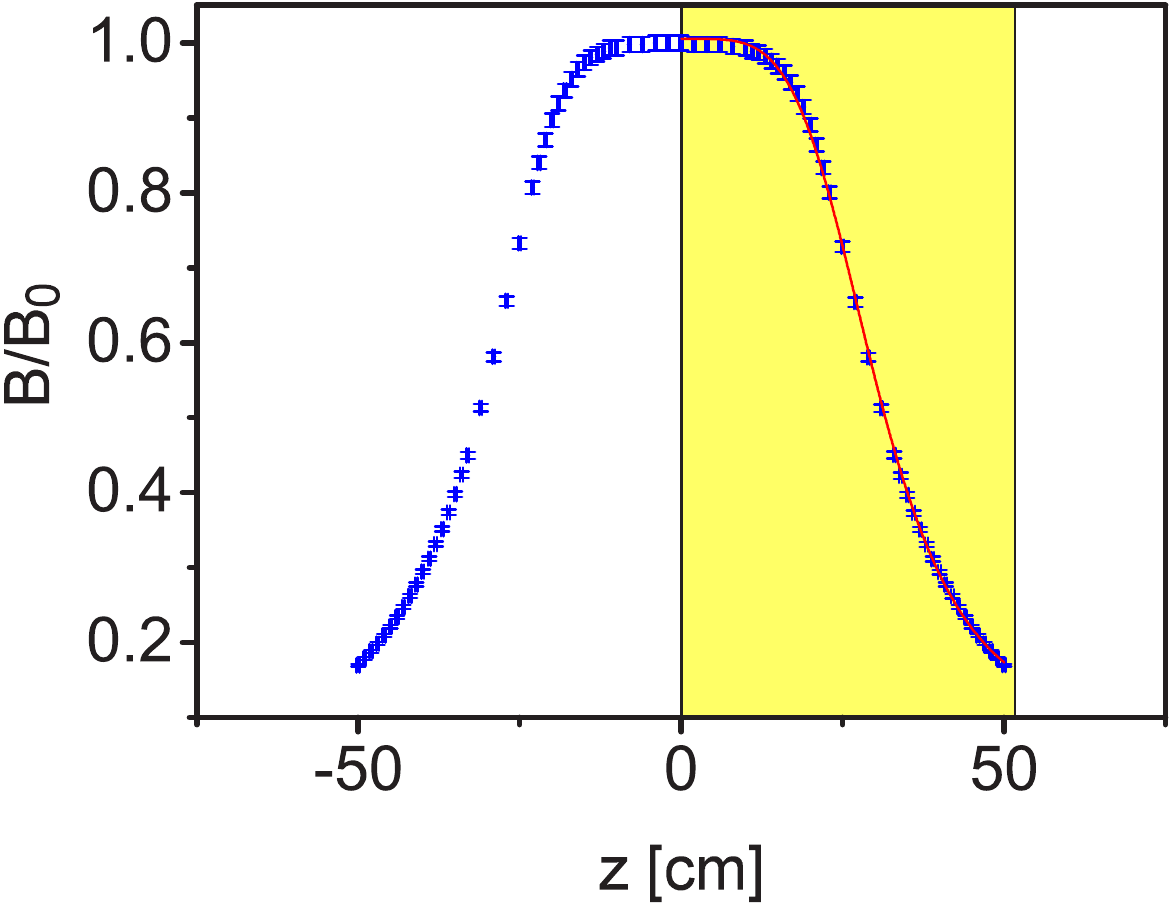}
\caption{Normalized fringe field $B/B_0$ of a dipole magnet produced by the company ``GMW Associates'' as a function of the distance $z$
from the magnet center. The distance is measured in the plane parallel to the magnetic poles. The crosses denote measured values and the
plain red line is a fitted function given in \eqref{eq:fringe-field-fit-function}. The horizontal and vertical errors bars are depicted
in blue. The electric current of the magnet is \unit[5.0]{A} and the gap $g$ is \unit[2.0]{cm}. The magnetic poles are quadratic with a
side length of approximately \unit[4.9]{cm}.}
\label{fig:fringe-field-dipole-gmw}
\end{figure}%
In \eqref{eq:fringe-field-correction}, $g$ is the full gap height of the dipole magnet and $I_1$ is the first fringing field integral.
The function $B_y(z)$ describes the magnetic fringe field on the median plane as a function of the perpendicular distance $z$ to the
entrance/exit face of the magnet. The value $B_0$ corresponds to the limiting value of the fringe field inside the magnet, i.e., in a
sufficient distance to the magnet entrance and exit, respectively.

Figure \ref{fig:fringe-field-dipole-gmw} shows measured data of the fringe field of a dipole magnet that were obtained at
KIT~\cite{Koppenhoefer:2013}. A logistic fit function well obeys the data (at one edge chosen as indicated by the yellow region
in \figref{fig:fringe-field-dipole-gmw}):
\begin{subequations}
\label{eq:fringe-field-fit-function}
\begin{align}
\left.\frac{B(z)}{B_0}\right|_{\mathrm{measured}}&=A_2+\frac{A_1-A_2}{1+(z/z_0)^q}\,, \\[1ex]
A_1&=1.00558\,,\quad A_2=0.0833\,,\quad z_0=30.2581\,,\quad q=4.42642\,.
\end{align}
\end{subequations}
Since the magnetic field has to vanish for $z\mapsto \infty$, there is undoubtedly an offset in the measured data. This is given by
the value $A_2$. Hence, we shift the $B$-axis appropriately and normalize the result again leading to:
\begin{equation}
\label{eq:fringe-field-fit-function-2}
\left.\frac{B'(z)}{B_0'}\right|_{\mathrm{adapted}}=\frac{1}{1+(z/z_0)^q}\,,
\end{equation}
with the values of $z_0$ and $q$ given above. The fringe field integral in \eqref{eq:fringe-field-correction} is computed numerically\footnote{The
integration is done over the whole range of the positive $z$-axis assuming \eqref{eq:fringe-field-fit-function-2} and the result is
multiplied by 2 to take the contribution for negative $z$ into account.} with the result $I_1\approx 7.45$.
In \cite{Brown:1984} it is stated for typical dipole magnets $I_1$ ranges from 0.3 to~1.0. However the experimental data indicates
that the fringe field for this particular dipole magnet has nonvanishing values even for distances that are much larger than the
typical dimensions of the dipole magnet (being several centimeters).

The upshot is that the fringe field profile may play an important role for dipole magnets. As soon as the dipole magnets for the
FLUTE bunch compressor will be available it may be useful to measure the profile to estimate its effect on bunch compression.
A logistic fit function according to \eqref{eq:fringe-field-fit-function} was shown to be appropriate for this purpose.

\section{Space charge effects for a cylindric bunch}
\label{sec:space-charge-cylindrical-bunch}
\setcounter{equation}{0}

To derive the equations of motion for electrons within a cylindric bunch in \secref{sec:space-charge-effects} the following
formulas are needed. The basis vectors $\widehat{\mathbf{b}}$ and $\widehat{\mathbf{n}}$ can be expressed by the new basis vectors
$\widehat{\mathbf{e}}_r$ and $\widehat{\mathbf{e}}_{\varphi}$ (and vice versa) as follows:
\begin{subequations}
\begin{align}
\widehat{\mathbf{e}}_r&=\widehat{\mathbf{n}}\cos\varphi+\widehat{\mathbf{b}}\sin\varphi\,,\quad
\widehat{\mathbf{e}}_{\varphi}=-\widehat{\mathbf{n}}\sin\varphi+\widehat{\mathbf{b}}\cos\varphi\,, \\[1ex]
\widehat{\mathbf{b}}&=\widehat{\mathbf{e}}_r\sin\varphi+\widehat{\mathbf{e}}_{\varphi}\cos\varphi\,,\quad
\widehat{\mathbf{n}}=\widehat{\mathbf{e}}_r\cos\varphi-\widehat{\mathbf{e}}_{\varphi}\sin\varphi\,.
\end{align}
\end{subequations}
The derivatives of the basis vectors $\{\widehat{\mathbf{e}}_r,\widehat{\mathbf{e}}_{\varphi},\widehat{\mathbf{t}}\}$ with respect to $t$ are
given by:
\begin{subequations}
\begin{align}
\dot{\widehat{\mathbf{e}}}_r&=-\widehat{\mathbf{n}}\dot{\varphi}\sin\varphi+\cos\varphi|\dot{\mathbf{r}}|(\widehat{\mathbf{b}}\tau-\widehat{\mathbf{t}}\kappa)+\widehat{\mathbf{b}}\dot{\varphi}\cos\varphi+\sin\varphi(-\widehat{\mathbf{n}}|\dot{\mathbf{r}}|\tau) \notag \\
&=-\widehat{\mathbf{t}}|\dot{\mathbf{r}}|\kappa\cos\varphi+\dot{\varphi}(-\widehat{\mathbf{n}}\sin\varphi+\widehat{\mathbf{b}}\cos\varphi)+|\dot{\mathbf{r}}|\tau(\widehat{\mathbf{b}}\cos\varphi-\widehat{\mathbf{n}}\sin\varphi) \notag \\
&=-\widehat{\mathbf{t}}|\dot{\mathbf{r}}|\kappa\cos\varphi+(\dot{\varphi}+|\dot{\mathbf{r}}|\tau)\widehat{\mathbf{e}}_{\varphi}\,,
\end{align}
\begin{align}
\dot{\widehat{\mathbf{e}}}_{\varphi}&=-\widehat{\mathbf{n}}\dot{\varphi}\cos\varphi-\sin\varphi|\dot{\mathbf{r}}|(\widehat{\mathbf{b}}\tau-\widehat{\mathbf{t}}\kappa)-\widehat{\mathbf{b}}\dot{\varphi}\sin\varphi+\dot{\widehat{\mathbf{b}}}\cos\varphi \notag \\
&=\widehat{\mathbf{t}}|\dot{\mathbf{r}}|\kappa\sin\varphi-\dot{\varphi}(\widehat{\mathbf{n}}\cos\varphi+\widehat{\mathbf{b}}\sin\varphi)-|\dot{\mathbf{r}}|\tau(\widehat{\mathbf{n}}\cos\varphi+\widehat{\mathbf{b}}\sin\varphi) \notag \\
&=\widehat{\mathbf{t}}|\dot{\mathbf{r}}|\kappa\sin\varphi-\widehat{\mathbf{e}}_r(\dot{\varphi}+|\dot{\mathbf{r}}|\tau)\,,
\end{align}
\begin{equation}
\dot{\widehat{\mathbf{t}}}=\widehat{\mathbf{n}}|\dot{\mathbf{r}}|\kappa=\widehat{\mathbf{e}}_r|\dot{\mathbf{r}}|\kappa\cos\varphi-\widehat{\mathbf{e}}_{\varphi}|\dot{\mathbf{r}}|\kappa\sin\varphi\,,
\end{equation}
\begin{align}
\ddot{\widehat{\mathbf{e}}}_r&=-\widehat{\mathbf{t}}\left(\frac{\mathrm{d}|\dot{\mathbf{r}}|}{\mathrm{d}t}\kappa\cos\varphi+|\dot{\mathbf{r}}|\dot{\kappa}\cos\varphi-\dot{\varphi}|\dot{\mathbf{r}}|\kappa\sin\varphi\right)-\dot{\widehat{\mathbf{t}}}|\dot{\mathbf{r}}|\kappa\cos\varphi+\widehat{\mathbf{e}}_{\varphi}\left(\ddot{\varphi}+\frac{\mathrm{d}|\dot{\mathbf{r}}|}{\mathrm{d}t}\tau+|\dot{\mathbf{r}}|\dot{\tau}\right) \notag \\
&\phantom{{}={}}+\dot{\widehat{\mathbf{e}}}_{\varphi}\left(\dot{\varphi}+|\dot{\mathbf{r}}|\tau\right) \notag \\
&=-\widehat{\mathbf{t}}\left(\frac{\mathrm{d}|\dot{\mathbf{r}}|}{\mathrm{d}t}\kappa\cos\varphi+|\dot{\mathbf{r}}|\dot{\kappa}\cos\varphi-2\dot{\varphi}|\dot{\mathbf{r}}|\kappa\sin\varphi-|\dot{\mathbf{r}}|^2\tau\kappa\sin\varphi\right) \notag \\
&\phantom{{}={}}+\widehat{\mathbf{e}}_{\varphi}\left(\ddot{\varphi}+\frac{\mathrm{d}|\dot{\mathbf{r}}|}{\mathrm{d}t}\tau+|\dot{\mathbf{r}}|\dot{\tau}+|\dot{\mathbf{r}}|^2\kappa^2\sin\varphi\cos\varphi\right)-\widehat{\mathbf{e}}_r\left[\left(\dot{\varphi}+|\dot{\mathbf{r}}|\tau\right)^2+|\dot{\mathbf{r}}|^2\kappa^2\cos^2\varphi\right]\,,
\end{align}
\begin{align}
\ddot{\widehat{\mathbf{t}}}&=\widehat{\mathbf{e}}_r\left(\frac{\mathrm{d}|\dot{\mathbf{r}}|}{\mathrm{d}t}\kappa\cos\varphi+|\dot{\mathbf{r}}|\dot{\kappa}\cos\varphi-|\dot{\mathbf{r}}|\kappa\dot{\varphi}\sin\varphi\right)+\dot{\widehat{\mathbf{e}}}_r|\dot{\mathbf{r}}|\kappa\cos\varphi \notag \\
&\phantom{{}={}}-\widehat{\mathbf{e}}_{\varphi}\left(\frac{\mathrm{d}|\dot{\mathbf{r}}|}{\mathrm{d}t}\kappa\sin\varphi+|\dot{\mathbf{r}}|\dot{\kappa}\sin\varphi+|\dot{\mathbf{r}}|\kappa\dot{\varphi}\cos\varphi\right)-\dot{\widehat{\mathbf{e}}}_{\varphi}|\dot{\mathbf{r}}|\kappa\sin\varphi \notag \\
&=\widehat{\mathbf{e}}_r\left(\frac{\mathrm{d}|\dot{\mathbf{r}}|}{\mathrm{d}t}\kappa\cos\varphi+|\dot{\mathbf{r}}|\dot{\kappa}\cos\varphi-|\dot{\mathbf{r}}|\kappa\dot{\varphi}\sin\varphi+\left(\dot{\varphi}+|\dot{\mathbf{r}}|\tau\right)|\dot{\mathbf{r}}|\kappa\sin\varphi\right) \notag \\
&\phantom{{}={}}-\widehat{\mathbf{e}}_{\varphi}\left[\frac{\mathrm{d}|\dot{\mathbf{r}}|}{\mathrm{d}t}\kappa\sin\varphi+|\dot{\mathbf{r}}|\dot{\kappa}\sin\varphi+|\dot{\mathbf{r}}|\kappa\dot{\varphi}\cos\varphi-|\dot{\mathbf{r}}|\kappa\cos\varphi\left(\dot{\varphi}+|\dot{\mathbf{r}}|\tau\right)\right]
-\widehat{\mathbf{t}}|\dot{\mathbf{r}}|^2\kappa^2\,.
\end{align}
\end{subequations}
From these results the velocity and the acceleration vector that are used in \eqref{eq:equations-of-motion-compact} can be computed:
\begin{subequations}
\begin{align}
\dot{\mathbf{r}}_b&=\dot{r}\widehat{\mathbf{e}}_r+r\dot{\widehat{\mathbf{e}}}_r+\dot{z}\widehat{\mathbf{t}}+z\dot{\widehat{\mathbf{t}}} \notag \\
&=\widehat{\mathbf{t}}\left(\dot{z}-r|\dot{\mathbf{r}}|\kappa\cos\varphi\right)+\widehat{\mathbf{e}}_{\varphi}\left[r\left(\dot{\varphi}+|\dot{\mathbf{r}}|\tau\right)-z|\dot{\mathbf{r}}|\kappa\sin\varphi\right]+\widehat{\mathbf{e}}_r\left(\dot{r}+z|\dot{\mathbf{r}}|\kappa\cos\varphi\right)\,,
\end{align}
\begin{align}
\ddot{\mathbf{r}}_b&=\ddot{r}\widehat{\mathbf{e}}_r+2\dot{r}\dot{\widehat{\mathbf{e}}}_r+r\ddot{\widehat{\mathbf{e}}}_r+\ddot{z}\widehat{\mathbf{t}}+2\dot{z}\dot{\widehat{\mathbf{t}}}+z\ddot{\widehat{\mathbf{t}}} \notag \\
&=\widehat{\mathbf{t}}\left[-r\left(\frac{\mathrm{d}|\dot{\mathbf{r}}|}{\mathrm{d}t}\kappa\cos\varphi+|\dot{\mathbf{r}}|\dot{\kappa}\cos\varphi-2\dot{\varphi}|\dot{\mathbf{r}}|\kappa\sin\varphi-|\dot{\mathbf{r}}|^2\tau\kappa\sin\varphi\right)+\ddot{z}\right. \notag \\
&\phantom{{}={}\widehat{\mathbf{t}}\Big[}\left.-|\dot{\mathbf{r}}|^2\kappa^2z-2\dot{r}|\dot{\mathbf{r}}|\kappa\cos\varphi\right] \notag \\
&\phantom{{}={}}+\widehat{\mathbf{e}}_{\varphi}\left[r\left(\ddot{\varphi}+\frac{\mathrm{d}|\dot{\mathbf{r}}|}{\mathrm{d}t}\tau+|\dot{\mathbf{r}}|\dot{\tau}+|\dot{\mathbf{r}}|^2\kappa^2\sin\varphi\cos\varphi\right)-2\dot{z}|\dot{\mathbf{r}}|\kappa\sin\varphi\right. \notag \\
&\phantom{{}={}+\widehat{\mathbf{e}}_{\varphi}\Big[}\left.-z\left(\frac{\mathrm{d}|\dot{\mathbf{r}}|}{\mathrm{d}t}\kappa\sin\varphi+|\dot{\mathbf{r}}|\dot{\kappa}\sin\varphi-|\dot{\mathbf{r}}|^2\kappa\tau\cos\varphi\right)+2\dot{r}\left(\dot{\varphi}+|\dot{\mathbf{r}}|\tau\right)\right] \notag \\
&\phantom{{}={}}+\widehat{\mathbf{e}}_r\left[-r\left((\dot{\varphi}+|\dot{\mathbf{r}}|\tau)^2+|\dot{\mathbf{r}}|^2\kappa^2\cos^2\varphi\right)+2\dot{z}|\dot{\mathbf{r}}|\kappa\cos\varphi\right. \notag \\
&\phantom{{}={}+\widehat{\mathbf{e}}_r\Big[}\left.+z\left(\frac{\mathrm{d}|\dot{\mathbf{r}}|}{\mathrm{d}t}\kappa\cos\varphi+|\dot{\mathbf{r}}|\dot{\kappa}\cos\varphi+|\dot{\mathbf{r}}|^2\kappa\tau\sin\varphi\right)+\ddot{r}\right]\,.
\end{align}
\end{subequations}
Finally, the following cross product is needed to obtain the Lorentz force:
\begin{align}
\label{eq:cross-product-velocity-magnetic-field}
\dot{\mathbf{r}}\times \mathbf{B}&=(\dot{\mathbf{r}}_r\widehat{\mathbf{e}}_r+\dot{\mathbf{r}}_{\varphi}\widehat{\mathbf{e}}_{\varphi}+\dot{\mathbf{r}}_t\widehat{\mathbf{t}})\times (B_r\widehat{\mathbf{e}}_r+B_{\varphi}\widehat{\mathbf{e}}_{\varphi}+B_t\widehat{\mathbf{t}}) \notag \\
&=(\dot{\mathbf{r}}_rB_{\varphi}-\dot{\mathbf{r}}_{\varphi}B_r)\widehat{\mathbf{t}}+(\dot{\mathbf{r}}_tB_r-\dot{\mathbf{r}}_rB_t)\widehat{\mathbf{e}}_{\varphi}+(\dot{\mathbf{r}}_{\varphi}B_t-\dot{\mathbf{r}}_tB_{\varphi})\widehat{\mathbf{e}}_r\,.
\end{align}
The general differential equations are given below together with the remaining coefficients that are not needed in
\secref{sec:space-charge-effects}:
\begin{subequations}
\label{eq:space-charge-eq-general-1}
\begin{align}
\eta_1\left(\varrho'+\eta_2f\zeta\cos\varphi\right)&+\varrho''+\eta_3\varrho'+\left[\eta_4(\eta_3f+g)\zeta+\eta_5f\zeta'\right]\cos\varphi-\varrho(\varphi'^2+\eta_6f^2\cos^2\varphi) \notag \\
&=\eta_7\left[-\eta_8(\widetilde{E}^{\mathrm{int}}_{\varrho}-\widetilde{v}\widetilde{B}^{\mathrm{int}}_{\varphi})-\left(\eta_9f\varrho\cos\varphi-\eta_{10}\zeta'\right)(\widetilde{B}^{\mathrm{int}}_{\varphi}+\widetilde{B}^{\mathrm{ext}}_{\varphi})\right. \notag \\
&\phantom{{}={}\eta_7\Big[}\left.-\left(\eta_{12}\varrho\varphi'-\eta_{13}f\zeta\sin\varphi\right)\widetilde{B}^{\mathrm{int}}_t\right]\,, \\[2ex]
\eta_8&=\frac{1}{r_0}\,,\quad \eta_{12}=\frac{\sqrt{2K}\widetilde{v}}{r_0}\,,\quad \eta_{13}=\frac{\sqrt{2K}\widetilde{v}}{r_0}\frac{L}{R}\,.
\end{align}
\end{subequations}
\begin{subequations}
\label{eq:space-charge-eq-general-2}
\begin{align}
\chi_1(\varrho\varphi'-\chi_2f\zeta\sin\varphi)&+\varrho\left[\varphi''+\chi_3\varphi'+\chi_4f^2\sin(2\varphi)\right]+2\varrho'\varphi'-\left[\chi_5(\chi_3f+g)\zeta+\chi_6f\zeta'\right]\sin\varphi \notag \\
&=\chi_7\left[-\chi_8\left(\widetilde{E}^{\mathrm{int}}_{\varphi}+\widetilde{v}\widetilde{B}^{\mathrm{int}}_{\varrho}\right)+\left(\chi_9f\varrho\cos\varphi-\chi_{10}\zeta'\right)(\widetilde{B}^{\mathrm{int}}_{\varrho}+\widetilde{B}^{\mathrm{ext}}_{\varrho})\right. \notag \\
&\phantom{{}={}\chi_7\Big[}\left.+\left(\chi_{11}f\zeta\cos\varphi+\chi_{12}\varrho'\right)\widetilde{B}^{\mathrm{int}}_t\right]\,, \\[2ex]
\chi_8&=\frac{1}{r_0}\,,\quad \chi_{11}=\frac{\sqrt{2K}\widetilde{v}}{r_0}\frac{L}{R}\,,\quad \chi_{12}=\frac{\sqrt{2K}\widetilde{v}}{r_0}\,.
\end{align}
\end{subequations}
\begin{subequations}
\label{eq:space-charge-eq-general-3}
\begin{align}
\psi_1(\zeta'-\psi_2f\varrho\cos\varphi)&+\zeta''+\psi_3\zeta'-\psi_4f^2\zeta-\psi_5f\varrho'\cos\varphi+\varrho\left[\psi_5f\varphi'\sin\varphi-\psi_6(\psi_3f+g)\cos\varphi\right] \notag \\
&=\psi_7\left[-\psi_8\widetilde{E}^{\mathrm{int}}_t-(\psi_9\rho'+\psi_{10}f\zeta\cos\varphi)(\widetilde{B}^{\mathrm{int}}_{\varphi}+\widetilde{B}^{\mathrm{ext}}_{\varphi})\right. \notag \\
&\phantom{{}={}\psi_7\Big[}\left.+(\psi_9\rho \varphi'-\psi_{10}f\zeta\sin\varphi)(\widetilde{B}^{\mathrm{int}}_{\varrho}+\widetilde{B}^{\mathrm{ext}}_{\varrho})\right]\,, \\[2ex]
\psi_8&=\frac{1}{L}\,.
\end{align}
\end{subequations}

\section{Computation of radiated CSR power}
\label{sec:computation-csr-power}
\setcounter{equation}{0}

Equation (\ref{eq:modified-total-power}) is a double integral. This can be evaluated by substituting $\zeta=\omega/\omega_c$ and a
successive partial integration (see, e.g., \cite{schwinger:1949}):
\begin{align}
\label{eq:modified-total-power-computation}
P(t)&=\frac{P_0}{\hbar}\frac{9\sqrt{3}}{8\pi} \int^{\omega_b/\omega_c}_0 \mathrm{d}\xi\,\xi \int_{\xi}^{\infty} \mathrm{d}\zeta\,K_{5/3}(\zeta) \notag \\
&=\frac{P_0}{\hbar}\frac{9\sqrt{3}}{8\pi} \left\{\left.\frac{1}{2}\xi^2 \int_{\xi}^{\infty} \mathrm{d}\zeta\,K_{5/3}(\zeta)\right|_0^{\omega_b/\omega_c}-\frac{1}{2}\int_0^{\omega_b/\omega_c} \mathrm{d}\xi\,\xi^2\left[K_{5/3}(\infty)-K_{5/3}(\xi)\right]\right\} \notag \\
&=\frac{P_0}{\hbar}\frac{9\sqrt{3}}{8\pi}\left[\frac{1}{2}\left(\frac{\omega_b}{\omega_c}\right)^2 \int_{\omega_b/\omega_c}^{\infty} \mathrm{d}\zeta\,K_{5/3}(\zeta)+\frac{1}{2}\int_0^{\omega_b/\omega_c} \mathrm{d}\xi\,\xi^2K_{5/3}(\xi)\right] \notag \\
&=\frac{P_0}{2\hbar}\left[\frac{\omega_b}{\omega_c}S_s\left(\frac{\omega_b}{\omega_c}\right)+\frac{9\sqrt{3}}{8\pi} \int_0^{\omega_b/\omega_c} \mathrm{d}\xi\,\xi^2K_{5/3}(\xi)\right]\,.
\end{align}
In the limit $\omega_b\ll \omega_c$ it is ensured that the integration variable in the second term solely runs over small values. Because
of this, we can evaluate the integral by expanding the integrand. However in the first term the integration runs to infinity and so
an expansion of the integrand for a small integration variable is not valid for the whole integration domain. Therefore, it makes
sense to bring the first term in a different shape and to use the recurrence relation $K_{5/3}(x)=-2K_{2/3}'(x)-K_{1/3}(x)$ of the
modified Bessel's functions first:
\begin{align}
xS_s(x)&=\frac{9\sqrt{3}}{8\pi} x^2 \int_x^{\infty} \mathrm{d}\zeta\,K_{5/3}(\zeta)=\frac{9\sqrt{3}}{8\pi} x^2\left\{-2\left[K_{2/3}(\infty)-K_{2/3}(x)\right]-\int_x^{\infty} \mathrm{d}\zeta\,K_{1/3}(\zeta)\right\} \notag \\
&=\frac{9\sqrt{3}}{8\pi} x^2\left[2K_{2/3}(x)-\int_0^{\infty} \mathrm{d}\zeta\,K_{1/3}(\zeta)+\int_0^x \mathrm{d}\zeta\,K_{1/3}(\zeta)\right] \notag \\
&=\frac{9\sqrt{3}}{8\pi} x^2\left[2K_{2/3}(x)-\frac{\pi}{\sqrt{3}}+\int_0^x \mathrm{d}\zeta\,K_{1/3}(\zeta)\right]\,.
\end{align}
Performing a Taylor expansion of the latter result for $x\ll 1$ leads to
\begin{equation}
\label{eq:expansion-1}
xS_s(x)=\frac{9\sqrt{3}}{4\cdot 2^{1/3}\pi}\Gamma\left(\frac{2}{3}\right)x^{4/3}+\mathcal{O}(x^2)\,.
\end{equation}
The second term in \eqref{eq:modified-total-power-computation} can be directly Taylor-expanded with the result
\begin{equation}
\label{eq:expansion-2}
\int^x_0 \mathrm{d}\xi\,\xi^2K_{5/3}(\xi)=\frac{\Gamma(2/3)}{2^{1/3}}x^{4/3}+\mathcal{O}(x^{10/3})\,.
\end{equation}
Inserting the expansions of Eqs.~(\ref{eq:expansion-1}), (\ref{eq:expansion-2}) in \eqref{eq:modified-total-power-computation}
leads to the final result of \eqref{eq:modified-total-power-result}. As an independent cross check, the same
result can also be obtained by using the expansion of \eqref{eq:synchrotron-power-small-frequencies} directly.

\end{appendix}

\newpage



\end{document}